\documentclass[twoside,11pt]{article}
\usepackage[utf8]{inputenc}
\usepackage[english]{babel}
\usepackage{amsmath}
\usepackage{mathtools}
\usepackage{bbm}
\usepackage{bm}
\usepackage{xcolor}
\usepackage{setspace}
\usepackage{algorithm}
\usepackage{algpseudocode}
\usepackage[inline]{enumitem}
\usepackage{caption}
\usepackage{subcaption}
\usepackage{multirow}
\usepackage{soul}
\usepackage{todonotes}
\usepackage{longtable}
\usepackage{lscape}
\usepackage{xcite}
\usepackage{booktabs,tabularx}

\usepackage{threeparttable, tablefootnote}

\usepackage{xcite}

\usepackage{xr}
\makeatletter
\newcommand*{\addFileDependency}[1]{% argument=file name and extension
  \typeout{(#1)}% latexmk will find this if $recorder=0 (however, in that case, it will ignore #1 if it is a .aux or .pdf file etc and it exists! if it doesn't exist, it will appear in the list of dependents regardless)
  \@addtofilelist{#1}% if you want it to appear in \listfiles, not really necessary and latexmk doesn't use this
  \IfFileExists{#1}{}{\typeout{No file #1.}}% latexmk will find this message if #1 doesn't exist (yet)
}
\makeatother

\usepackage{tikz,array,calc}
\usetikzlibrary{decorations.pathreplacing}

\usepackage{jmlr2e}

\usepackage{xr}
\ShortHeadings{Hybrid Censored Quantile Regression Forest to Assess the Heterogeneous Effects}{HCQRF}
\firstpageno{1}

\def\bbeta{\boldsymbol \beta}
\def\bz{\bm z}
\def\bZ{\bm Z}
\def\bx{\bm x}
\def\bX{\bm X}

\begin{document}

\title{Hybrid Censored Quantile Regression Forest to Assess the Heterogeneous Effects}

\author{\name Huichen Zhu\email zhuhuichenecho@gmail.com\\ \addr Department of Statistics\\ The Chinses University of Hong Kong
\AND 
\name Yifei Sun \email ys3072@cumc.columbia.edu\\
\addr Department of Biostatistics\\
Columbia University
\AND
\name Ying Wei \email yw2148@cumc.columbia.edu\\
\addr Department of Biostatistics\\
Columbia University}

% \author{\name Marina Meil\u{a} \email mmp@stat.washington.edu \\
%       \addr Department of Statistics\\
%       University of Washington\\
%       Seattle, WA 98195-4322, USA
%       \AND
%       \name Michael I.\ Jordan \email jordan@cs.berkeley.edu \\
%       \addr Division of Computer Science and Department of Statistics\\
%       University of California\\
%       Berkeley, CA 94720-1776, USA}

\editor{ }

\maketitle

\begin{abstract}%   <- trailing '%' for backward compatibility of .sty file
In many applications, heterogeneous treatment effects on a censored response variable are of primary interest, and it is natural to evaluate the effects at different quantiles (e.g., median).
% It is challenging that the effects may depend on a large amount of covariates and the structure of the effect is unknown. 
The large number of potential effect modifiers, the unknown structure of the treatment effects, and the presence of right censoring pose significant challenges.  
In this paper, we develop a hybrid forest approach called Hybrid Censored Quantile Regression Forest (HCQRF) to assess the heterogeneous effects varying with high-dimensional variables. The hybrid estimation approach takes advantage of the random forests and the censored quantile regression. We propose a doubly-weighted estimation procedure that consists of a redistribution-of-mass weight to handle censoring and an adaptive nearest neighbor weight derived from the forest to handle high-dimensional effect functions. We propose a variable importance decomposition to measure the impact of a variable on the treatment effect function. Extensive simulation studies demonstrate the efficacy and stability of HCQRF. The result of the simulation study also convinces us of the effectiveness of the variable importance decomposition. We apply HCQRF to a clinical trial of colorectal cancer. We achieve insightful estimations of the treatment effect and meaningful variable importance results. The result of the variable importance also confirms the necessity of the decomposition. 
\end{abstract}


\section{Introduction}

Precision medicine and individualized health care are a future direction of medicine to optimize health benefits for all. To achieve that, one needs to have a thorough understanding of heterogeneous treatment effects in a target population. More and more studies (e.g., \cite{kravitz2004evidence,kosorok2015adaptive}) suggest that the effect of a drug or an intervention could depend on individuals' health conditions, medical history, genetic profiles, and many other factors. Hence, it would be desirable to assess the heterogeneous treatment effect incorporating high-dimensional covariates. In this paper, we focus on censored time-to-event outcomes, a family of widely-considered outcomes in health analyses and risk modeling. 

Estimating the heterogeneous effects potentially varied with a large number of covariates for the censored data brings many challenges. Besides the curse of dimensionality, the effect function is likely to be an unknown function of covariates. The naive estimation approaches, such as the linear additive model with treatment-covariate interaction effects, may oversimplify the complexity of response-covariate associations in reality. Recent advancements in machine learning methods make it possible to model nonlinear and complex associations with high-dimensional covariates. Among them, random forest is a popular choice due to its competitive prediction accuracy and feature selection, which are important considerations for health applications. 

Random forest \citep{breiman2001random} is an ensemble of regression and classification trees constructed from recursive partitions. The partitions separate dissimilar subjects by a specific splitting criterion, resulting in homogeneous subgroups in terminal nodes. The nonparametric nature of the recursive partition framework makes random forest adept to high-dimensional covariates, and complex and nonlinear associations. It hence becomes a powerful tool for prediction and is widely used across many fields. Several random forest approaches were developed for time-to-event outcomes. 
\cite{ishwaran2008random} introduced the random survival forest to estimate the cumulative hazard function by averaging the Nelson-Aalen estimator of the terminal node where the individual locates in each tree of the random forest. \cite{zhu2012recursively} used extremely randomized trees and proposed an imputation procedure that recursively updates the censored observations. \cite{steingrimsson2018censoring} extended the censoring unbiased transformation to squared error loss functions and proposed a new recursive partition approach based on it. The target function in the above approaches is the conditional survival or conditional cumulative hazard function. \cite{hothorn2006survival} proposed a random forest to estimate the conditional mean of the log–survival time. \cite{li2020censored} extended the quantile forest \citep{athey2019generalized} with a loss function adept to the censored data to estimate the quantile of the survival time.
The aforementioned random forests are powerful tools for risk prediction, but they are not designed to assess the heterogeneous covariates effects on the censored outcome.

In this paper, we propose a forest-based estimation procedure to assess the heterogeneous effects of the covariates for right-censored data. 
The proposed hybrid censored random forest integrates the censored quantile regression with the random forest. The estimation procedure based on it enables us to estimate the heterogeneous effects of the covariates at a quantile level of interest. It is a doubly-weighted estimation framework, which comprises a redistribution-of-mass weight \citep{wang2009locally} to handle censoring and an adaptive nearest neighbor weight derived from the forest to handle high-dimensional effect functions. The proposed forest algorithm is based on a two-step splitting rule by evaluating the heterogeneity of the coefficients and goodness-of-fit of a quantile regression model. An inference-based criterion is used to choose the splitting variable. A loss-function-based criterion is used to select the splitting value.

The main contributions of the proposed approach are as follows. First, the proposed approach models the survival time directly by a censored quantile regression, which results in a relevant interpretation of the estimation and also provides an opportunity to assess the future risk based on the estimated model. 
Second, it expands the utility of the classical random forest to assess the heterogeneous effects of the covariates potentially varying with high-dimensional variables. The perspective of viewing a random forest as an adaptive kernel approach \citep{lin2006random,MeinshausenN2006,athey2019generalized} is widely adopted. However, the estimated nonparametric regression functions derived by those approaches cannot directly evaluate the heterogeneity of the covariate effects, which is of great need in many medical applications. The proposed hybrid approach is able to help us understand the effects varying by individuals and relieves the restriction of the specific structure assumption of effects functions in other estimation approaches.
Third, we propose a two-step splitting rule in the random forest, avoiding favoring variables with more possible splits.
Last but not least, the mutually advantageous conjunction of the quantile regression model and random forest allows us to tackle all the challenges in assessing the heterogeneous effects with high-dimension.

The rest of the paper is organized as follows. In Section \ref{sec:model}, we introduce the censored quantile regression model with heterogeneous coefficient functions. In Section \ref{sec:estimation}, we present the estimation procedure of the heterogeneous coefficient functions obtained by the proposed hybrid forest. Details about how the hybrid forest is built are given in Section \ref{sec:hybridforest}. In Section \ref{sec:simulation}, we conduct a variety of simulation experiments to evaluate the estimation performance of the proposed approach and compare it with other approaches. The proposed method is illustrated with the analysis of a randomized phase III study in Section \ref{sec:realdata} with some discussions.

\section{Censored Quantile Regression Model with Heterogeneous Coefficient Functions}\label{sec:model}

Let $T$ be a survival time outcome. We denote by $\bZ$ a $q$-dimensional vector of predictive variables including a constant $1$, whose effects on the quantiles of $T$ are of interest, and
denote by $\bX$ a $p$-dimensional vector of modifiers \citep{hastie1993varying}, which potentially modify the effects of $\bZ$. Defining $Q_\tau(T|\bX,\bZ)$ as the conditional quantile function of $T$ given $(\bX,\bZ)$ at the quantile level $\tau$, we consider the following quantile regression model
\begin{align}\label{eq:survival_main_model}
	Q_\tau(T|\bX,\bZ) = \bZ^\top\bbeta_\tau(\bX),
\end{align}
where $\bbeta_\tau(\bX)$ is the effect of $\bZ$ on the $\tau$th quantile of $T$, and is an unknown function of $\bX$.  Since $\bZ$ includes a constant 1,  the coefficient function $\bbeta_\tau(\bX) = (\beta_{\tau,0}(\bX),\bbeta_{\tau,1}(\bX))$ includes both the intercept function $\beta_{\tau,0}(\bX)$ and the slope functions $\bbeta_{\tau,1}(\bX)$.

The choice of $\bZ$ depends on specific applications. For example,  when the predictive variable is a binary indicator of treatment, $\bbeta_{\tau,1}(\bX)$ is the individualized quantile treatment effect (QTE) given the covariates $\bX$. When the predictive variable is a continuous dose variable, $\bbeta_{\tau,1}(\bX)$ outputs a heterogeneous dose-response relationship that could vary by $\bX$. There are also applications where the choice of predictive variables is informed by domain knowledge. In Model \eqref{eq:survival_main_model}, the dimension of $\bX$ can be high and grow with the sample size, while the dimension of $\bZ$ is low and fixed. This way, we could fully capture individualized heterogeneous quantile effect $\bbeta_\tau(\bX)$, and ensure its estimation is feasible.

In this paper, we assume that the survival time $T$ is subject to right censoring, and the censoring time, denoted by $C$, is independent of $T$ conditional on $(\bX,\bZ)$. In the presence of censoring, we observe the censored time $Y = \min(T, C)$ and the event indicator $\Delta = \mathbbm 1 (T\le C)$. 
% In the following paragraphs, we define $\{(y_i,\delta_i,\bx_i, \bz_i),~i = 1,2,\dots,N\}$ to be the independent and identically distributed random sample of $(Y,\Delta,\bX, \bZ)$. If $\delta_i = 1$, then $y_i = t_i$ where $t_i$ is the observed survival time, and if $\delta_i = 0$, then $y_i = c_i$ is the censoring time.  
Throughout the paper, we denote a censored random forest by $\mathbb T = \{{\mathcal T}_b:b = 1,2,\ldots, B\}$, where ${\mathcal T}_b$ refers to the $b$th tree in the forest.

%The varying-coefficient model for uncensored data was first proposed by \citep{hastie1993varying} for mean regression and extended to quantile regression by \citet{honda2004quantile} and \citet{cai2009nonparametric}.
%Kernel smoothing and spline approximation are the two most frequently used techniques in estimating the coefficient function of varying coefficient models. For example, \citet{yu1998local} studied the quantile regression estimation by kernel weighted local linear fitting; polynomial spline \citep{kim2007quantile} and B-spline  \citep{wang2009quantile} approximation are used to estimate the function coefficients of the quantile regression model. \citet{xie2015quantile} further extended the varying coefficient approach to regression quantiles under random censoring by exploiting the inverse of probability weighting method with local linear smoothing kernels. Both approaches are restricted to the dimension of the modifiers and can only be applied to the cases when the coefficient function is smooth. Thus, they are not feasible to estimate coefficients varying with a higher dimension of variables and without shape constrain of the function.

\section{Doubly-weighted \texorpdfstring{$\bbeta_\tau(\bX)$ }{TEXT} Estimation From A Censored Forest} 
\label{sec:estimation}

In Model \eqref{eq:survival_main_model}, if the coefficient function $\bbeta_\tau(\bX)$ is a constant, Model \eqref{eq:survival_main_model} reduces to a simple linear quantile regression of the survival time $T$ and its predictors $\bZ$. On the other hand, if $\bZ$ only includes the constant 1, Model \eqref{eq:survival_main_model} reduces to a nonparametric quantile regression model. The latter can also be estimated by existing censored random forests (i.e., \cite{li2020censored}). In this section, we are going to propose a doubly-weighted estimation approach to estimate $\bbeta_\tau(\bX)$ in Model \eqref{eq:survival_main_model}.

If the dimension of covariates $\bX$ is low, Model \eqref{eq:survival_main_model} shares the same form as a varying-coefficient model without censoring, and the estimation of $\bbeta_\tau(\bX)$ can be carried out through kernel smoothing or spline approximation \citep{hastie1993varying, honda2004quantile, cai2009nonparametric}. Those approaches are not scalable for high dimensions. To estimate $\bbeta_\tau(\bX)$ in Model \eqref{eq:survival_main_model} with high-dimensional $\bX$ and censored outcome, we propose a doubly-weighted hybrid estimation procedure in conjunction with censored quantile regression and random forest. In this section, we assume that a forest $\mathbb T$ is available.  

\subsection{Forest-based Weights} \label{sec:forestweights}
We start by estimating $\bbeta_\tau(\bX)$ using uncensored data.
Our approach to estimate the quantile coefficient function $\bbeta_\tau(\bx_0)$ for a given $\bx_0$ relies on a similarity weights $\{\omega_i(\bx_0),~i=1,\ldots,N\}$, where $\omega_i(\bx_0)$ measures the contribution of the $i$th sample in estimating $\bbeta_\tau(\bx_0)$.
%For example, when the dimension of $\bX$ is low, one may construct the weight using kernel smoothing. However, when the number of potential modifiers is large, the kernel weighting approach suffers from the curse of dimensionality. 
To accommodate the high-dimensional $\bX$,  we propose to use a random forest to construct the similarity weights $\omega_i(\bx_0)$ as in \cite{MeinshausenN2006} and \cite{athey2019generalized}. In what follows, $\omega_i(\bx_0)$ is termed forest-based weight. 
%The forest-based weights are obtained by averaging neighborhood data determined by a large number of trees, which is a prevalent view of a random forest as an adaptive weight generator \citep{MeinshausenN2006,athey2019generalized}. 
Suppose we have a random forest of $B$ trees, indexed by $b = 1,2,\ldots,B$. In each tree ${\mathcal T}_b$, we denote by $\mathcal N_b(\bx_0)$ the terminal node that contains $\bx_0$. The forest-based weights $\omega^{\mathbb T}_i(\bx_0)$ is then defined as, 
\begin{align}\label{eq:forestweight}
    \omega^{\mathbb T}_i(\bx_0) = \frac{1}{B}\sum_{b = 1}^B\frac{\mathbbm 1\{\bX_i\in \mathcal{N}_b(\bx_0)\}}{| \mathcal{N}_b(\bx_0)|},
\end{align}
where $|\mathcal{N}_b(\bx_0)|$ is the number of observations in the terminal node $\mathcal N_b(\bx_0)$. 
In each tree ${\mathcal T}_b$, if $\bX_i$ is contained in the terminal node $\mathcal N_b(\bx_0)$, it contributes towards estimating $\bbeta_\tau(\bx_0)$ with weight $\frac{I(\bX_i\in \mathcal{N}_b(\bx_0))}{\mid \mathcal{N}_b(\bx_0)\mid}$. Otherwise, its weight is zero. The weight $\omega^{\mathbb T}_i(\bx_0)$ is obtained by aggregating over $B$ trees.
It is easy to see that $\sum_{i = 1}^N\omega^{\mathbb T}_i(\bx_0) = 1$, and the weights $\omega^{\mathbb T}_i(\bx_0)$ define an adaptive neighborhood of $\bx_0$ identified by the random forest.

In the absence of censoring, one can estimate the quantile coefficient function $\bbeta_\tau(\bx_0)$ by minimizing weighted quantile loss 
\begin{equation} \label{eq:qloss1}
    \widehat{\bbeta}_\tau(\bx_0)=\arg\min_{\bbeta}\frac{1}{N}\sum_{i = 1}^N\omega^{\mathbb T}_i(\bx_0)\rho_\tau(T_i-\bZ_i^T\bbeta),
\end{equation} 
where $\rho_\tau(u) = u(\tau-\mathbbm 1\{u< 0\})$ is the quantile loss function.  

%We will introduce in detail how the random forest is built in Section \ref{sec:hybridforest}. 

\subsection{Redistribution Weights}\label{sec:redisweight}
In the presence of censoring, the quantile loss in \eqref{eq:qloss1} can no longer be used.
Instead, we consider a weighted quantile regression with redistribution weights suggested in \cite{wang2009locally} when censoring exists.
The weighted quantile loss can be written as
\begin{align}\label{eq:individualloss}
\widetilde\rho_{\tau}(Y_i-\bZ_i^\top\bbeta)
	 =  \left\{u_{i,\tau}\rho_\tau(Y_i-\bZ_i^\top\bbeta)+(1-u_{i,\tau})\rho_\tau(Y^{+\infty}-\bZ_i^\top\bbeta)\right\},
\end{align}  
where $Y^{+\infty}$ is any value sufficiently large to exceed all $\bZ_i^\top\bbeta$.  The redistribution weights $u_{i,\tau}$ in \eqref{eq:individualloss} are  defined as
\begin{align}\label{eq:redisweight}
	u_{i,\tau} = \begin{cases}
		1, & \Delta_i= 1 \text{ or } F(Y_i|\bX_i,\bZ_i)>\tau\\
		\frac{\tau-F(Y_i|\bX_i,\bZ_i)}{1-F(Y_i|\bX_i,\bZ_i)} & \Delta_i = 0 \text{ and } F(Y_i|\bX_i,\bZ_i)<\tau,
	\end{cases}
\end{align}
where $F(t|\bX,\bZ)$ is the conditional cumulative distribution function of $T$ given $\bX,\bZ$. The conditional cumulative function $F(t|\bX,\bZ)$ is unknown. In practice, we can use proper estimation approaches to obtain the estimated $u_{i,\tau}$.

The redistribution weights root from a unique property of the quantile regression, whose estimator is only determined by the signs of residuals. 
By design, the redistribution weight $u_{i,\tau} =1$ if $Y_i$ is either an observed event time or the conditional censoring probability given $\bX_i,\bZ_i$ is larger than $\tau$. In both cases, the sign of the residual $T_i-\bZ_i^\top\bbeta_{\tau}(\bX_i)$ is determined. Hence it contributes towards the quantile estimate as a full observation. However, when $Y_i$ is censored, and the conditional censoring probability is smaller than $\tau$, 	the sign of $T_i-\bZ_i^\top\bbeta_{\tau}(\bX_i)$ becomes undetermined. In this case, $u_{i,\tau} $ assign the probability mass $F(Y_i|\bX_i,\bZ_i)$ to a sufficiently-large value, and ``redistribute'' the remaining probability by 
$\frac{\tau-F(Y_i|\bX_i,\bZ_i)}{1-F(Y_i|\bX_i,\bZ_i)}$.

%that redistributes the mass of censored data lying under the quantile of interest to the uncensored ones to the right. After considering the redistribution weights for the censored data, the individual loss function becomes:
%In the loss function \eqref{eq:individualloss},
%The redistribution scheme takes advantage of the property that the quantile regression estimator is only determined by the signs of residuals. Thus, for uncensored data ($\Delta_i = 1$) and censored data ($\Delta_i = 0$) with $Y_i = C_i>\bZ_i^\top\bbeta_{\tau}(\bX_i)$,  the value of $\mathbbm 1\{T_i<\bZ_i^\top\bbeta_{\tau}(\bX_i)\}$ is directly observed. For censored data ($\Delta_i = 0$) with $Y_i = C_i< \bZ_i^\top\bbeta_{\tau}(\bX_i)$, the sign of $T_i-\bZ_i^\top\bbeta_{\tau}(\bX_i)$ is ambiguous. 
%Therefore, $\frac{\tau-F(C_i|\bX_i,\bZ_i)}{1-F(C_i|\bX_i,\bZ_i)}$ is assigned for the ambiguous situation, and 1 is assigned for the first situation. Based on \eqref{eq:redisweight}, it is easy to know that the redistribution weight is $u_{i,\tau}\in(0,\tau)\cup\{1\}$ for $ F(\cdot|\bX_i,\bZ_i)\in(0,1)$. \cite{leng2013quantile}, and \cite{xie2015quantile} employed inverse probability weighting (IPW) to handle censoring in the data. However, the estimator might not be stable if the estimated propensity is too small. The redistribution weight \eqref{eq:redisweight} can avoid this issue since it is within a range for all values of $F(\cdot|\bX_i,\bZ_i)$.

Combining the forest-based weights \eqref{eq:forestweight} together with the redistribution weight \eqref{eq:redisweight}, we could  estimate $\bbeta_\tau(\cdot)$ at a fixed $\bx_0$ and a fixed quantile level $\tau$ from a random forest by 
\begin{align}\label{eq:quanteffectest}
	\widehat{\boldsymbol{\beta}}_{\mathbb{T},\tau}(\bx_0) =\arg\min_{\boldsymbol{\beta}} L_{\mathbb{T},\tau}(\boldsymbol{\beta}, \bx_0)=
	\arg\min_{\boldsymbol{\beta}}
	\frac{1}{N}\sum_{i = 1}^N  \omega^{\mathbb T}_i(\bx_0)\widetilde\rho_{\tau}(Y_i-\bZ_i^\top\bbeta).
	%L_{\mathbb{T},\tau}(\boldsymbol{\beta},\bx_0),
\end{align}
%where $\mathbb T = \{{\mathcal T}_b:b = 1,2,\ldots\}$ could be any censored random forest and $\omega^{\mathbb T}_i(\bx_0)$ is the forest weight of $i$ the observation for $\bx_0$. 
In the above discussion, we assume a forest $\mathbb T$ is available. In what follows, we consider a censored quantile random forest algorithm to build the forest.

\begin{remark}
For censored outcomes, many researchers have considered modeling the quantiles under linear assumptions. \cite{powell1984least} and \cite{powell1986censored} proposed a linear quantile regression model under fixed censoring. \cite{portnoy2003censored} introduced a weighted estimation approach exploiting the idea of redistribution-of-mass. \cite{peng2008survival} proposed a martingale-based estimating procedure. Both \cite{portnoy2003censored}'s and \cite{huang2002varying}'s approaches relied on a global linear assumption at all quantile levels. \cite{de2019adapted} introduced a new loss function tackling the censoring based on the conditional distribution of the survival and the observed time. However, the computation of the adapted majorize-minimize algorithm is complex. \cite{leng2013quantile}, and \cite{xie2015quantile} employed inverse probability weighting (IPW) to handle the censoring. Such IPW approaches are often unstable when some estimated propensity scores are too small. We consider the redistribution weights suggested in \cite{wang2009locally} due to its relaxed linear assumption and simple minimizing algorithm. The redistribution weight is controlled as $u_{i,\tau}\in(0,\tau)\cup\{1\}$, and hence avoids the risk of variance inflation.
\end{remark}

%\begin{eqnarray}\label{eq:objectfun}
%	&&L_{\mathbb{T},\tau}(\boldsymbol{\beta}, \bx_0) = .
% 	&=& \frac{1}{N}\sum_{i = 1}^N  \omega^{\mathbb T}_i(\bx_0)  u_{i,\tau}\rho_\tau(Y_i-\bZ_i^\top\bbeta)+\omega^{\mathbb T}_i(\bx_0)(1-u_{i,\tau})\rho_\tau(Y^{+\infty}-\bZ_i^\top\bbeta).
%\end{eqnarray}

\section{Hybrid Censored Quantile Regression Random Forest}
\label{sec:hybridforest}
%In order to estimate the coefficient function $\bbeta_\tau(\bx)$ in Model \eqref{eq:survival_main_model}, we propose a doubly-weighted estimation procedure which consists of a redistribution-of-mass weight and a forest-based weight in Section \ref{sec:estimation}. The forest-based weight \eqref{eq:forestweight} is based on a censored random forest. In order to further improve the estimation procedure, 
In this section, we propose a hybrid censored random forest, where we incorporate the censored quantile regression in the partition algorithm.
Instead of the heterogeneity of the conditional mean or quantile \citep{MeinshausenN2006,athey2019generalized}, the proposed partition algorithm aims to identify and differentiate the heterogeneity of the function $\bbeta_\tau(\bX)$ in Model \eqref{eq:survival_main_model} in the covariate space. In each node, we intend to search for the most effective modifier of the coefficient function in Model \eqref{eq:survival_main_model} by the proposed splitting criterion.

Following the convention in random forest \citep{breiman2001random}, we denote 
$\mathfrak{S}_{\mathcal N} =\{s\}$ as the collection of all possible binary split rules determined by a single modifier in $\bX$ and a cut-off value, where $\mathcal N$ is denoted as a collection of some observations in a node. We denote by $|\mathcal N|$ the sample size of the node. We also denote by $\mathcal N_L^s$ and $\mathcal N_R^s$ the left and right child nodes respectively based on a split $s$. Consequently, $\mathcal N_L^s\cap  \mathcal N_R^s = \emptyset$ and  $\mathcal N_L^s\cup  \mathcal N_R^s = \mathcal  N$.

%, where our proposed splitting criterion accommodating the partition mechanism to focus on the heterogeneity of the function $\bbeta_\tau(\bX)$ in Model \eqref{eq:survival_main_model}. 

% \todo[inline]{We need to clearly explain conceptually what "hybrid" means, and how it distinguish from other methods}

% Indeed, we seek splits that improve the fitting of the tree as mush as possible. 

A greedy search over $\mathfrak{S}_{\mathcal N}$ might favor a continuous splitting variable, because it has many more possible splits \citep{loh2014fifty}. To avoid such selection bias and reduce the computation burden,
we consider a two-step splitting algorithm, where we first use a re-distributed rank statistics to select the {best} splitting variable that maximizes the marginal heterogeneity in $\bbeta_\tau(\bX)$. Once the splitting variable is chosen, we search for a split that optimizes the goodness-of-fit of Model \eqref{eq:survival_main_model}. 
Details of the algorithm are given in Algorithm \ref{alg:vcqrf}.

\subsection{Re-distributed Rank-score for Choosing Splitting}\label{sec:choosesplit}
When choosing an {optimal} splitting variable, we consider a screening procedure to search for the most effective modifier of $\bbeta_\tau(\bX)$ in a parent node $\mathcal N$. We use an indicator function $\eta_i(s)$ to indicate where the observation $(Y_i,\Delta_i,\bX_i,\bZ_i)$ locates according to the split $s$: $\eta_i(s) = 1$ if $\bX_i\in \mathcal N_L^s$ and $\eta_i(s) = 0$ otherwise. We propose to use a rank-based statistics to find the optimal splitting variable. We first consider when we can observe complete data. For Model \eqref{eq:survival_main_model} without censoring, there is no direct approach to identify the most effective modifier in an unknown function $\bbeta_\tau(\bX)$. Instead, we rely on a working model  $Q_\tau(T|X_k,\bZ,\bX\in\mathcal N) = \bZ^\top\bbeta_1+\bZ^\top\boldsymbol\gamma_kX_k$, for each $k = 1,\ldots,K$. 
The rank-based test is well defined for testing the null hypothesis $H_{0k}: \boldsymbol\gamma_k = \bm 0$. Even if the working model is misspecified, it is still natural to choose the variable with the largest rank-score statistics as a splitting variable \citep{loh2002regression}. The rank-based statistics will make the selection procedure simple and efficient. However, the regression rank score for censored quantile regression is not well defined. 

% We follow the idea proposed by \cite{sun2020rank} to define the regression rank score with redistribution weight \eqref{eq:redisweight} for censored data. 
The following outlines how we obtain the rank-based statistics to choose the optimal variable in the node $\mathcal N$ without censoring. We let $\widetilde\bZ_{\mathcal N}$ and $\widetilde\bX_{\mathcal N}$ be the design matrix of Model \eqref{eq:survival_main_model} and the matrix of modifiers correspondingly in the node $\mathcal N$, and $\widetilde\bX_{\mathcal N \cdot k}$ be the $k$th column in $\widetilde\bX_{\mathcal N}$. The quantile loss function in the node $\mathcal N$ is,	
\begin{align}\label{eq:weightloss}
	L_\tau(\mathcal N,\bbeta) = \sum_{i:\bX_i\in \mathcal N}	\rho_\tau(T_i-\bZ_i^T\bbeta),
\end{align} 
where function $\rho_\tau(\cdot)$ is defined in \eqref{eq:qloss1}. We let $\widehat{\boldsymbol a}_{\tau,\mathcal N}$ be a column vector with $i$th element being the regression rank-score \citep{hajek1965extension} for the quantile regression \citep{koenker_2005},
\begin{align}
    \label{eq:ranks}
    \widehat a_{i,\tau,\mathcal N} = \tau-\mathbbm 1\{T_i-\bZ_i^T\widehat \bbeta_{\mathcal N}<0\},
\end{align}
% \begin{align}
%     \label{eq:rankscensored}
%     \widehat a_{i,\tau,\mathcal N}^C = 1-u_{i,\tau}\mathbbm 1\{Y_i-\bZ_i^T\widehat \bbeta_{\mathcal N}<0\},
% \end{align}
where $\widehat \bbeta_{\mathcal N}$ is the sample minimizer of $L_\tau(\mathcal N,\bbeta)$ .  Then let $\Lambda_{\mathcal N k} = diag(\widetilde\bX_{\mathcal N  \cdot k})\widetilde\bZ_{\mathcal N}$ be the interaction matrix  between the predictive variables and the $k$th modifier and $diag(\widetilde\bX^{T}_{\mathcal N  \cdot k})$ is a $|\mathcal N|\times |\mathcal N|$ diagonal matrix with $\widetilde\bX^{T}_{\mathcal N  \cdot k}$ being the diagonal elements. The projection of the interaction matrix to the spanned column space of $\widetilde\bZ_{\mathcal N}$ is $P_{\mathbf Z}\Lambda_{\mathcal N k}$ where $P_{\mathbf Z} =  \widetilde\bZ_{\mathcal N}(\widetilde\bZ_{\mathcal N}^T\widetilde\bZ_{\mathcal N})^{-1}\widetilde\bZ_{\mathcal N}^T$. 

\sloppy
Then, the rank score statistics to evaluate the heterogeneity of the quantile coefficients induced by the $k$th modifier is 
\begin{align}
    \label{eq:rankscorecensored}
    \mathcal T_{k,\mathcal N} = S_{k,\mathcal N}^TQ_{k,\mathcal N}^{-1}S_{k,\mathcal N},
\end{align}
where $S_{k,\mathcal N} = (\Lambda_{\mathcal N k}-P_{\mathbf Z} \Lambda_{\mathcal N k})^T\widehat{\boldsymbol a}_{\tau,\mathcal N}$ and $Q_{k,\mathcal N} = (\Lambda_{\mathcal N k}-P_{\mathbf Z} \Lambda_{\mathcal N k})^T(\Lambda_{\mathcal N k}-P_{\mathbf Z} \Lambda_{\mathcal N k})$. Intuitively, $\widehat{\boldsymbol a}_{\tau,\mathcal N}$ only depends on the signs of $\{T_i-\bZ_i^T\widehat \bbeta_{\mathcal N},~i = 1,2,\ldots,N\}$ and hence it represents the relative positions of $\{T_i~i = 1,2,\ldots,N\}$ after adjusting for $\bZ_i$ at $\tau$th quantile in the node $\mathcal N$. A smaller value of the norm of $S_{k,\mathcal N}$ means that no variation in $\widehat{\boldsymbol a}_{\tau,\mathcal N}$ can be further explained by $(\Lambda_{\mathcal N k}-P_{\mathbf Z} \Lambda_{\mathcal N k})$. Thus, a larger value of $\mathcal T_{k,\mathcal N}$ implies that more variation in $\widehat{\boldsymbol a}_{\tau,\mathcal N}$ can be explained by the interaction $\Lambda_{\mathcal N k}$. In the presence of censoring, we follow \cite{sun2020rank} and replace $\widehat a_{i,\tau,\mathcal N}$ with $\widehat a_{i,\tau,\mathcal N}^C = \tau-u_{i,\tau}\mathbbm 1\{Y_i-\bZ_i^T\widehat \bbeta_{\mathcal N}<0\}$, where $u_{i,\tau}$ is the redistribution weight \eqref{eq:redisweight}. The expectation of the new regression rank score with censoring equals to the one without censoring. Thus, $\mathcal T_{k,\mathcal N}$ with $\widehat a_{i,\tau,\mathcal N}^C$ is an approximation of the rank test statistics for uncensored data, and hence we choose the modifier with the largest value of $\mathcal T_{k,\mathcal N}$ as the splitting variable.

% \subsection{Choosing Splitting Value}
% Once the splitting variable is chosen, we seek a cut-off value that improves the fitting of the trees as much as possible. In the presence of right censoring, we consider the quantile loss with redistribution weights \eqref{eq:individualloss} to \textcolor{orange}{minimize the in-sample prediction error}.
% We let $\widehat\bbeta_{\mathcal N}$ be the minimizer of $\widetilde L_\tau(\mathcal N,\bbeta)$ in the node $\mathcal N$.
% The optimal split is given by
% \begin{align}\label{eq:splitcri}
% 	s^* = \arg\max_{s\in\mathfrak S_{\mathcal N}} \frac{\widetilde L_\tau(\mathcal N,\widehat \bbeta_{\mathcal N})-\widetilde L(\mathcal N_L^s,\widehat \bbeta_{\mathcal N_L^s})-\widetilde L_\tau(\mathcal N_R^s,\widehat \bbeta_{\mathcal N_R^s})}{\widetilde L_\tau(\mathcal N,\widehat\bbeta_{\mathcal N})}.
% \end{align} 
%  Equivalently, the optimal split value is chosen by $s^* = \arg\min_{s \in\mathfrak S_{\mathcal N}} \{\widetilde L_\tau(\mathcal N_L^s,\widehat \bbeta_{\mathcal N_L^s})+\widetilde L_\tau(\mathcal N_R^s,\widehat \bbeta_{\mathcal N_R^s})\}$. For any arbitrary split $s$, the fraction in \eqref{eq:splitcri} is always larger than or equal to $0$.
% \blue{I still feel we should switch the order of (10) and the equation below. Shall we combine 4.1 and 4.2?}

Once the splitting variable is chosen, we seek a cut-off value that improves the fitting of the trees as much as possible. 
Denote by $\widehat\bbeta_{\mathcal N}$ the minimizer of $\widetilde L_\tau(\mathcal N,\bbeta)$.
The optimal split is given by
\begin{align}\label{eq:splitcri}
s^* = \arg\min_{s \in\mathfrak S_{\mathcal N}^*} \{\widetilde L_\tau(\mathcal N_L^s,\widehat \bbeta_{\mathcal N_L^s})+\widetilde L_\tau(\mathcal N_R^s,\widehat \bbeta_{\mathcal N_R^s})\},
\end{align} 
where $\mathfrak S_{\mathcal N}^*$ is the collection of all possible binary split rules determined by the selected splitting variable and a cut-off value in the node $\mathcal N$.
It can be shown that splitting generally improves the loss function, that is, for any split $s$, we have $\widetilde L_\tau(\mathcal N_L^s,\widehat \bbeta_{\mathcal N_L^s})+\widetilde L_\tau(\mathcal N_R^s,\widehat \bbeta_{\mathcal N_R^s})\le \widetilde L_\tau(\mathcal N,\widehat \bbeta_{\mathcal N})$.

\subsection{Variable Importance}\label{sec:VarImp}
Variable importance is often used as a reference to measure the prediction strength of each variable in the data set. In our framework, we are further interested in the effective modifiers of the quantile coefficient function. %Variable importance measures in random forests have been receiving increased attention recently. It is a powerful screening tool to rank or select variables. 
We propose two types of permutation-based variable importance. The first one evaluates the overall impact of a modifier in the conditional quantile of the response variable. The second one measures the impact of a modifier on the treatment effect when the predictive variable is a binary treatment variable.

 We first introduce how we obtain the variable importance to evaluate the overall prediction strength. We denote the original data set by $\mathcal D = \{(Y_i,\Delta_i,\bX_i,\bZ_i),~i = 1,2,\dots,N\}$. Once the random forest $\mathbb T$ has been grown, we follow the procedure of permutation importance \citep{breiman2001random} and calculate the variable importance for $k$th variable with the following steps,  
\begin{enumerate}
    \item estimate the quantile loss in $\mathcal D$ based on $\mathbb T$: $\widetilde L_\tau(\mathcal D,\mathbb T) = \sum_{i = 1}^N\widetilde\rho_\tau(Y_i- \bZ_i^\top\widetilde\bbeta_{\mathbb T,\tau,-i}(\bX_i))$, where $\widetilde \bbeta_{\mathbb T,\tau,-i}(\bX_i)$ is the estimated quantile coefficient estimated by \eqref{eq:quanteffectest} but with those trees where $\bX_i$ is in the out-of-bag samples;
    
    \item generate a new data set $\mathcal D^m_k = \{(Y_i,\Delta_i,\bX_i^{(k)},\bZ_i),~i = 1,2,\dots N\}$ by shuffling the $k$th modifier randomly, where $\bX_i^{(k)}$ is a vector of modifiers for $i$th observation with $k$th element being the value after permutation and other elements being the same as in $\bX_i$;
        
    \item calculate the quantile loss in the new data set $\mathcal D^m_k$ based on $\mathbb T$: $\widetilde L_\tau(\mathcal D^m_k,\mathbb T) = \sum_{i = 1}^N\widetilde\rho_\tau(Y_i- \bZ_i^\top\widetilde\bbeta_{\mathbb T,\tau,-i}(\bX_i^{(k)}))$, where $\widetilde\bbeta_{\mathbb T,\tau,-i}(\bX_i^{(k)})$ is the value of $\widetilde \bbeta_{\mathbb T,\tau,-i}(\cdot)$ at $\bX_i^{(k)}$;
 
    \item  repeat 2-3 $M$ times (e.g., $M = 100$) and the variable importance of $k$th modifier is: $\text{VarImp}_k = \frac{1}{M}\sum_{m = 1}^{M} \widetilde L_\tau(\mathcal D^m_k,\mathbb T)- \widetilde L_\tau(\mathcal D,\mathbb T)$.
    \end{enumerate}
With the above procedure, we can obtain the variable importance for each modifier $k = 1,2,\ldots,p$. The permutation step (step 2) breaks any association between $k$th modifier and the response. Thus, the difference of quantile losses before and after the permutation step measures the overall prediction strength of the $k$th modifier. However, sometimes, the overall importance is manifested by the prediction strength in the main effect $\beta_{\tau,0}(\bX)$ while the one in $\bbeta_{\tau,1}(\bX)$ is of our interest. Therefore, we propose the second variable importance, which measures the predictive strength of the $k$th modifier in $\bbeta_{\tau,1}(\bX)$ when the predictive variable is a binary treatment variable. 

If $Z$ is a binary variable, $\bbeta_\tau(\bX)$ then contains two functions, the intercept function  $\beta_{\tau,0}(\bX)$ and the slope function $\beta_{\tau,1}(\bX)$. It is of primary interest to identify effective modifiers in $\beta_{\tau,1}(\bX)$. 
We propose a variable decomposition which decomposes the variable importance score into two parts by different values of $Z$ to discriminate the importance for $\beta_{\tau,0}(\bX)$ and $\beta_{\tau,1}(\bX)$. The overall variable importance can be decomposed into the importance score for $Z=1$ and the one for $Z = 0$. We assume that the treatments are randomly assigned and hence define the variable importance given $Z = j$ as, 
 \begin{align}\label{eq:varimp_z0}
     &\text{VarImp}_{Z = j,k,\tau} = \\\nonumber
     %=  \sum_{i = 1}^N \left\{\left[\widetilde\rho_\tau(Y_i-Z_i^\top\widetilde\bbeta_{\mathbb T,\tau,-i}(\bX_i^{(k)}))-\widetilde\rho_\tau(Y_i-Z_i^\top\widetilde\bbeta_{\mathbb T,\tau,-i}(\bX_i))\right]\nu_{i,0} \right\},
     &  \frac{1}{\sum_{i = 1}^N\mathbbm 1\{Z_i=j\}}\sum_{i = 1}^N\left\{\left[\widetilde\rho_\tau(Y_i-Z_i^\top\widetilde\bbeta_{\mathbb T,\tau,-i}(\bX_i^{(k)}))-\widetilde\rho_\tau(Y_i-Z_i^\top\widetilde\bbeta_{\mathbb T,\tau,-i}(\bX_i))\right]\mathbbm 1\{Z_i=j\}\right\}.
 \end{align}
 When $Z=0$, $\text{VarImp}_{Z = 0,l,\tau}$ is the importance score of $k$th modifier in $\beta_{\tau,0}(\bX)$. 
When $Z = 1$, we have $Q_\tau(T|\bX,Z = 1) = \beta_{\tau,0}(\bX) + \beta_{\tau,1}(\bX)$. The variable importance $\text{VarImp}_{Z = 1,l,\tau}$ contains both the importance score of $k$th modifier for both $\beta_{\tau,0}(\bX)$ and $\bbeta_{\tau,1}(\bX)$. Therefore, intuitively, the variable importance for $\bbeta_{\tau,1}(\bX)$ can be calculated by $|\text{VarImp}_{Z = 1,l,\tau}-\text{VarImp}_{Z = 0,l,\tau}|$.  
  
If $\beta_{\tau,1}(\bX)$ does not vary with the change of $k$th modifier, the difference of the variable importance scores between two groups, $\text{VarImp}_{Z = 1,l,\tau}-\text{VarImp}_{Z = 0,l,\tau}$ should be close to $0$. A value of $\text{VarImp}_{Z = 1,l,\tau}-\text{VarImp}_{Z = 0,l,\tau}$ being away from 0 possibly indicates that $\beta_{\tau,1}(\bX)$ changes with $k$th modifier. Both positive and negative values of the difference provide evidence for the interaction effect of the treatment and the modifier. Thus, we use the absolute value of the difference between the two importance scores $|\text{VarImp}_{Z = 1,l,\tau}-\text{VarImp}_{Z = 0,l,\tau}|$ to evaluate the impact of a modifier on the treatment effect. 
% We will explore more about the negative value of $\text{VarImp}_{Z = 1,l,\tau}-\text{VarImp}_{Z = 0,l,\tau}$ in Section \ref{sec:simulation} and Section \ref{sec:realdata}.

 \subsection{Computation Algorithm}\label{sec:alg}
%Algorithm \ref{alg:vcqrf} is our proposed algorithm. 

% \todo[inline]{I found it very hard to understand the computation algorithm here. }
The proposed ensemble procedure is summarized in Algorithm \ref{alg:vcqrf}. 
% In the algorithm, the function SUBSAMPLE draws a subsample with rate $r$ from $\mathcal D$. The function INITIALIZEQUEUE initializes a queue with a single element. The function NOTNULL checks whether if the queue is null. POP returns and removes the oldest element of a queue $\mathcal Q$ when $\mathcal Q$ is not empty, and returns null otherwise. The function SPLITSUCCEEDED checks whether all observations in the parent node are censored and whether the number of observations in each child node after split $s$ is smaller than or equal to a pre-specified number NINSPLIT. ADDTOQUEUE appends the latter element in the parenthesis to the former element. GETLEFTCHILD and GETRIGHTCHILD get the left and right child nodes of $\mathcal N$ based on split $s$. 

\begin{algorithm}[!ht]
\caption{Hybrid censored quantile regression forest}
\label{alg:vcqrf}
\begin{algorithmic}
\State The algorithm involves the following pre-specified tuning parameters:  the number of trees $B$, the sub-sampling $r$ rate, the number of candidate splitting variables in each split $mtry$, and the minimal node size for splitting MINSPLIT.
\Procedure{HCQRF}{Data $\mathcal D$}
    % \State weight vector $\omega^{\mathbb T}_i(\bx_0) \gets \mathbf 0_{|\mathcal D|}$, which is a vector of zeros of length $|\mathcal D|$
    \For{$b = 1$ to $B$}
    \State Randomly draw a subset from data $\mathcal D$ with sampling rate $r$ to grow the tree ${\mathcal T}_b$
    % which is denoted as $\mathcal I_b$
    % and its corresponding index is $I_b$
    % \State Root node $\mathcal N_0\gets \mathcal I_b$
 
    \State Initialize a tree structure ${\mathcal T}_b$
    \State Initialize a queue $\mathcal Q$  with the root node as the first element; $\mathcal Q$ maintains the current sequence of nodes for splitting
    
    \While{$\mathcal Q$ is not empty} %\hfill{ $\triangleright ~\mathcal Q$ contains the nodes to be split;}
    %\State \hfill{$\triangleright$ empty $\mathcal Q$ indicates the termination of the tree growing}
        \State Remove the first element in $\mathcal Q$, and assign it to  
        $\mathcal N $, the current node to split.
        \If{$\mathcal N$ reached the candidacy for splitting }
        %        The oldest element in $\mathcal Q$ and remove the oldest element in $\mathcal Q$;  $\mathcal N\gets$ is the current node for splitting. 
        \State Randomly draw $mtry$ candidate splitting variables
        \State Identify the optimal split $s$ for $\mathcal N $ from the $mtry$ variables as in Section \ref{sec:hybridforest}

        \State Split $\mathcal N$ into two child nodes $\mathcal N_L$ and $\mathcal N_R$ by the split $s$
        \State Append $\mathcal N_L$ and $\mathcal N_R$ to $\mathcal Q$
        \State Append the split $s$ to $\mathcal T_b$
        \EndIf 
    \EndWhile
    
    % \State $\mathcal N_b(\bx_0)\gets \textit{NEIGHBORHOOD}(\bx_0,{\mathcal T}_b,\mathcal D)$.
    % \For{all \{$i$: $\bx_i\in \mathcal D$ AND $\bx_i  \in \mathcal N_b(\bx_0)$\}}
    % \State $\omega^{\mathbb T}_i(\bx_0) = \omega^{\mathbb T}_i(\bx_0)+1/(B|\mathcal N_b(\bx_0)|)$
    % \EndFor
    \EndFor
    
% \noindent\textbf{output} $\widehat{\bbeta}(\bx_0)$, the minimized of \eqref{eq:objectfun} with redistribution weights weights $u_{i,\tau}$ and the forest-based weights $\omega^{\mathbb T}_i(\bx_0)$.
\EndProcedure
\State \textbf{output} Forest $\mathbb T = \{\mathcal T_b,~b = 1,2,\ldots, B\}$.
% and the corresponding sub-sampling index $\mathbb I = \{I_b,~b = 1,2,\ldots, B\}$.
\end{algorithmic}
The candidacy of a node $\mathcal N$ for splitting: the number of observations of $\mathcal N$ is larger than the pre-specified number MINSPLIT.
% The function INITIALIZEQUEUE initializes a queue with a single element. The function NOTNULL checks whether if the queue is null. POP returns and removes the oldest element of a queue $\mathcal Q$ when $\mathcal Q$ is not empty, and returns null otherwise. The function SPLITSUCCEEDED checks whether all observations in the parent node are censored and whether the number of observations in each child node after split $s$ is smaller than or equal to a pre-specified number NINSPLIT. ADDTOQUEUE appends the latter element in the parenthesis to the former element. GETLEFTCHILD and GETRIGHTCHILD get the left and right child nodes of $\mathcal N$ based on split $s$.
\end{algorithm}

\subsection{Computation Specification}\label{sec:compspec}
In the simulation and real data analysis, when calculating the redistribution weights \eqref{eq:redisweight}, we need to obtain the conditional cumulative distribution function of the survival time $T$ given $(\bX,\bZ)$: $F(t|\bX,\bZ) = \mathbb P(T\le t|\bX,\bZ)$. In practice, $F(\cdot |\bX,\bZ)$ is unknown. A variety of attempts have been made to estimate it. To name a few, \citet{portnoy2003censored} suggested estimating $F(\cdot |\bX,\bZ)$ through fitting an entire quantile regression process under the global linearity assumption of the conditional quantile functions. \cite{mckeague2001median} tried to fit a semiparametric model (e.g., Cox proportional hazard model) to obtain an approximation of $F(\cdot |\bX,\bZ)$. \citet{wang2009locally} proposed a fully nonparametric approach based on a local Kaplan-Meier estimator to estimate $F(\cdot |\bX,\bZ)$. The parametric and semiparametric approaches endure strong linearity assumptions, which is sometimes not practical. Furthermore, the nonparametric approach is only feasible when the dimension of covariates is small. Later, \cite{wang2013variable} proposed to perform a global dimension reduction formulation to facilitate the local weight estimation for multivariate covariates. Albeit it makes the nonparametric adapted to multivariate covariates, the dimension reduction formulation may lead to the loss of information. The kernel-based nonparametric approach is restricted to the types of covariates (e.g., when both continuous and categorical covariates exist). In order to relieve the restrictions of the previous approaches, we propose to estimate $F(\cdot|\bX,\bZ)$ by the random survival forest (RSF, \cite{ishwaran2008random}). By taking advantage of the random forests, the estimator by RSF is able to accommodate both continuous and categorical covariates and bypasses the restriction of types of covariates. The estimated conditional cumulative distribution function of $T$, denoted by $\widehat F(\cdot|\bX,\bZ)$, takes the place of $F(\cdot|\bX,\bZ)$ in \eqref{eq:redisweight}.

We also point out other computing specifications as follows, 
\begin{enumerate}
\item We set the total number of trees in the forest as $B =500$, the minimal number of samples in a terminal node is MINSPLIT$ = 20$, the subsampling rate is $r = 80\%$, and the number of randomly selected splitting variables at each split is $mtry = p/3$.

\item When optimizing the loss function in \eqref{eq:quanteffectest}, we suppose that among $N$ observations, the first $N^*$ survival times are censored, and the remaining are observed. When estimating the quantile coefficient, we append $(Y^{+\infty},\bX_1,\bZ_1,1-u_{1,\tau}),(Y^{+\infty},\bX_2,\bZ_2,1-u_{2,\tau}),\ldots, (Y^{+\infty},\bX_{N^*},\bZ_{N_{\mathcal N}^*},1-u_{N^*,\tau})$ to $\{(Y_i,\bX_i,\bZ_i,u_{i,\tau}):i = 1,2,\dots,N\}$, where $Y^{+\infty}$ is a sufficiently large number and when implemented in computation, we let $Y^{+\infty} = 10\max\{Y_1,Y_2,\ldots,Y_n\}$.  Then we can simply use the function \verb+rq+ in \verb+R+ package \verb+quantreg+ with the corresponding weight $u_{i,\tau}$ or $1-u_{i,\tau}$ to obtain the estimated quantile coefficient.
\item The random survival forest to estimate the conditional distribution $F(\cdot|\bX,\bZ)$ was implemented by the function \verb+rfsrc+ in the R package \verb+randomForestSRC+ with default settings.
\end{enumerate}

\section{Simulation}\label{sec:simulation}

% In the rest of the paper,  we  abbreviate the proposed hybrid censored quantile regression forest as HCQRF for notional convenience.  
In this section, we present simulation studies to evaluate the finite sample performance of the proposed HCQRF with comparisons of the following alternative methods. 
\vspace{5pt}
	\begin{itemize}[noitemsep,topsep=0.5pt,leftmargin = *]
	\item[] \textbf{Censored Quantile Regression Forest \citep[CQRF,][]{li2020censored}}: CQRF is an extension of generalized random forest \citep[grf,][]{athey2019generalized} but tailored for censored data. It treats both $\bX$ and $\bZ$ equally as splitting variables, and is designed to estimate the conditional quantile function $Q_\tau(T|\bX = \bx, \bZ = \bz)$.  When $Z$ is binary, one can estimate $\widehat \beta_{\mathbb T,\tau,1}(\bx_0) = \widehat Q_\tau(T|\bX = \bx_0,Z = 1)-\widehat Q_\tau(T|\bX = \bx_0,Z = 0)$. 
 \vskip 3pt
\item[] \textbf{HCQRF-complete (HCQRF-c)}:  We apply the proposed HCQRF on the complete data without censoring (i.e., we use the simulated data before introducing censoring such that all the $T_i$'s are fully observed).  
% The estimation of the quantile coefficient function follows the equation \eqref{eq:quanteffectest}.% \ref{sec:estimation}.
	\vskip 3pt
\item[] \textbf{grf-complete (grf-c)}:  We also construct CQRF from the complete data, but use the proposed double-weighed approach to estimate the coefficient function $\beta_{\tau}(\bX)$ from the resulting random forest. Since the double-weighting estimation scheme by \eqref{eq:quanteffectest} is valid when $\bZ$ is not included  as the splitting variables, and only $\bX$ are used as splitting variables. We denote this approach as grf-c, as it is in the essence of generalized random forest, which views random forest as a way to generate adaptive kernels \eqref{eq:forestweight} \citep{lin2006random, scornet2016random}. 
\end{itemize}
\vspace{5pt}
We summarize the key differences across those approaches in Table \ref{tbl:comparisonapproaches}. Among those comparison methods, CQRF is a popular existing random forest approach for censored quantiles. By comparing HCQRF and CQRF, we evaluate the estimation performance in the context of the latest literature. On the other hand, HCQRF-c serves as a benchmark approach. We compare HCQRF to HCQRF-c to assess the impact of censoring. By design, the main difference between HCQRF-c to grf-c is the choice of splitting rules. Although grf-c is not an existing approach, we include it as one of the comparison methods, and compare HCQRF-c to grf-c to illustrate the value and need of the proposed splitting rules when it comes to estimating heterogeneous treatment effects.  Specifically, in Sections \ref{sec:various_functions} and \ref{sec:complicated}, we evaluate their performance in estimating the coefficient functions $\boldsymbol{\beta}_\tau(\bX)$ under various scenarios.  Since CQRF can only estimate $\boldsymbol{\beta}_\tau(\bX)$ for binary $Z$, we compare its  performance in estimating the conditional quantile function $Q_\tau(T| \bX = \bx,Z = z)$ with the proposed HCQRF, and the results are presented in Section \ref{sec:condquantile}. Finally, we present a simulation study in Section \ref{sec:vi} to assess the effectiveness of the proposed variable importance decomposition in feature selection.

\subsection{Simulation Settings} %Once the modifiers and the predictive variables $(\bx_i,\bz_i),~ i = 1,2,\ldots,N$ are generated, 
We design the simulation scenarios with the following survival model 
\begin{align} \label{eq:simmodel}
	T_i = \beta_0(\bX_i)+\bZ_i^\top\bbeta_1(\bX_i)+\varepsilon_i
\end{align}
where 
 $\beta_0(\bX_i)$ and $\bbeta_1(\bX_i)$ are functions of effect modifiers $\bX_i$, and $\varepsilon_i$ is an error term.   In the Sections \ref{sec:various_functions} and  \ref{sec:complicated}, we outline the specifics of the distributions of $\bX_i$ and $\bZ_i$, the functions $\beta_0(\bX)$ and $\beta_1(\bX)$, and the distribution of $\varepsilon_i$ under each scenario.  In the meantime, we generate both completely random and covariate-dependent censoring time.
 The censoring rate in the following simulation scenarios is approximately 25\% at $\tau = 0.5$. Thus, the observed time is $Y_i = \min\{T_i,C_i\}$, and the censoring status is $\Delta_i = \mathbbm 1\{T_i\le C_i\}$.

 %The censoring time $c_i$ is generated variously according to different scenarios and hence the observed time is $y_i = \min\{t_i,c_i\}$. The event indicator is $\delta_i = \mathbbm 1\{t_i\le c_i\}$.  We divide the data set $\mathcal D = \{(y_i,\bx_i,\bz_i,\delta_i),~i = 1,2,\dots,N\}$ into training set $\mathcal D_{train}$ with sample size $N_1$ and test set $\mathcal D_{test}$ with sample size $N_2$.

In each Monte Carlo repetition, we generated a data set, $\mathcal D_1 = (Y_i,\Delta_i,\bX_i,\bZ_i)_{i = 1}^{N_1}$, on which the random forest is built based.  We consider two sample sizes,  $N_1 = 500$ and $N_1 = 1000$. We evaluate the estimation performance based on a new data set, $\mathcal D_2 = (\bX_i^*,\bZ_i^*)_{i = 1}^{N_2}$, which is generated independently in each Monte Carlo repetition following the same generation mechanism. We let $N_2 = 400$.
We assessed the performance in estimation $\bbeta_\tau(\bX)$ by the mean squared error (MSE), and mean absolute error (MAE) of both the estimated quantile coefficient function and the estimated quantile functions. Specifically, for each Monte Carlo simulation, we calculate
		\begin{align}
		&\text{MSE} = 	\frac{1}{N_2}\sum_{\bX_i^*\in\mathcal D_2}\left(\widehat\bbeta_{ \tau}(\bX_i^*)-\bbeta_{ \tau}(\bX_i^*)\right)^2,\nonumber\\
% 		&MSE_{\widehat Q_\tau} = \frac{1}{N_2}\sum_{j= 1}^{N_2}\left(\widehat Q_\tau(T_j|\bx^*_j,\bz^*_j)-Q_\tau(T_j|\bx^*_j,\bz^*_j)\right)^2\\
% 		&\text{relative MSE} =  	\frac{1}{N_2}\sum_{j= 1}^{N_2}\left(\frac{\widehat\bbeta_{ \tau}(\bx^*_j)-\bbeta_{ \tau}(\bx^*_j)}{\bbeta_\tau(\bx^*_j)}\right)^2,\\
		&\text{MAE} =  	\frac{1}{N_2}\sum_{\bX_i^*\in\mathcal D_2}\left|\widehat\bbeta_{ \tau}(\bX_i^*)-\bbeta_{ \tau}(\bX_i^*)\right|,
	\end{align}
where $\widehat\bbeta_{\tau}(\bX_i^*)$ is the estimated quantile coefficient function and $\bbeta_{\tau}(\bX_i)$ is the true one. In the Supplementary Materials, we also demonstrate the relative mean squared error (RMSE) and relative mean absolute error (RMAE) as additional scale-free assessments of the estimation performance.

\begin{landscape}
\begin{table}
\begin{center}
\begin{threeparttable}
\caption{Summary of the different approaches to be compared with.}\label{tbl:comparisonapproaches}
\begin{tabular}{p{0.12\textwidth}p{0.11\textwidth}p{0.25\textwidth}p{0.11\textwidth}p{0.23\textwidth}p{0.18\textwidth}}
 \cline{1-6}
 Method&Outcome used& Splitting rule & Splitting variables & Estimate $\bbeta_\tau(\bx_0)$ &Estimate  $Q_\tau(\bx_0, \bz)^e$ \\
 \cline{1-6}
% &\tikzmark{e}{}&
HCQRF& Censored$^a$ $(Y,\Delta)$&  Comparing \ul{conditional} quantiles of $T$ given $\bZ$  between the left and right child nodes
& $\bX$
& By minimizing the objective function in \eqref{eq:quanteffectest} with random forest weight \eqref{eq:forestweight} 
& $\bz^T\widehat\bbeta_{\tau,\mathbb T}(\bx_0)$ \\

%\vspace{0.2in}\\

CQRF & Censored $(Y)$ & Comparing the \ul{marginal} quantiles of $Y$ between the left and right child nodes$^c$
& $\bX$ and $\bZ$
& Unable to estimate $\bbeta_\tau(\bx_0)$ directly$^d$
& $\widehat Q_\tau(\bx_0,\bz)$ \\

%\vspace{0.2in}\\

HCQRF-c& Completely observed  $(T)^b$& The same as HCQRF & $\bX$ & The same as HCQRF& $\bz^T\widehat\bbeta_{\tau,\mathbb T}(\bx_0)$\\

%\vspace{0.2in}\\

grf-c& Completely observed  $(T)$& Comparing the \ul{marginal} quantiles of $T$ between the left and right child nodes
& $\bX$
& The same as HCQRF$^f$ &  $\bz^T\widehat\bbeta_{\tau,\mathbb T}(\bx_0)$\\
 \cline{1-6}
\end{tabular}

%\sidelink{a}{b}{}{A}
%\sidelink{c}{d}{}{B}
%\sidelink{e}{f}{}{C}

\footnotesize
\begin{tablenotes}
\item[a] The censored data $(Y_i,\Delta_i,\bX_i,\bZ_i),~i = 1,2,\ldots,N$ is used.
\item[b] The complete data $(T_i,\bX_i,\bZ_i),~i = 1,2,\ldots,N$ is used.
\item[c] The splitting rule is directly applied on  censored data without adjustment of the censoring.
\item[d] If the predictive variable is a binary variable, the coefficient function can estimated by $\widehat \beta_{\mathbb T,\tau,0}(\bx_0) = \widehat Q_\tau(T|\bX = \bx_0,Z = 0)$ and $\widehat \beta_{\mathbb T,\tau,1}(\bx_0) = \widehat Q_\tau(T|\bX = \bx_0,Z = 1)-\widehat Q_\tau(T|\bX = \bx_0,Z = 0)$.
\item[e] We abbreviate $\widehat Q(T|\bX = \bx_0,\bZ = \bz_0)$ as $\widehat Q(\bx_0,\bz_0)$.
\item[f] CQRF applying on the complete data is in the essense of grf \citep{athey2019generalized}, which is not able to estimate $\bbeta_\tau(\bx_0)$ directly. We can obtain the estimated $\bbeta_\tau(\bx_0)$ by following the same double-weighting estimation scheme \eqref{eq:quanteffectest} with forest weight \eqref{eq:forestweight} generated by grf.
% \item[g] The conditional quantile can be estimated by minimizing the adapted quantile loss function proposed by \cite{de2019adapted}, $\widehat Q_\tau(T|\bX = \bx_0,\bZ = \bz_0) = \arg\min_q\frac{1}{N}\sum_{i = 1}^N \omega_i^{\mathbb T}(\bx_0,\bz_0)\widetilde\rho_\tau(Y_i-q)$, where the forest weight $\omega_i^{\mathbb T}(\bx_0,\bz_0)$ is generated by the same scheme as \eqref{eq:forestweight} given $(\bx_0,\bz_0)$.
\end{tablenotes}
\end{threeparttable}
\end{center}
\end{table}
\end{landscape}

\subsection{Estimation Accuracy}\label{sec:various_functions}

\begin{figure}[!ht]
    \centering
    \caption{Heat map of the true $\beta_{1}(\bX)$ for \textbf{Scenario \ref{ex:tree_binary}}, \textbf{Scenario \ref{ex:boosting}} and \textbf{Scenario \ref{ex:cosine_continuous}}. The colors refer to the value of quantile coefficient in the corresponding grid.}
    \label{fig:heatmapbetatrue}
\begin{subfigure}{0.33\textwidth}
\caption{Scenario \ref{ex:tree_binary}: $\beta_1(x_1,x_2)$}
\label{fig:heatmapbeta_tree_binary_true1}
\includegraphics[width = \textwidth]{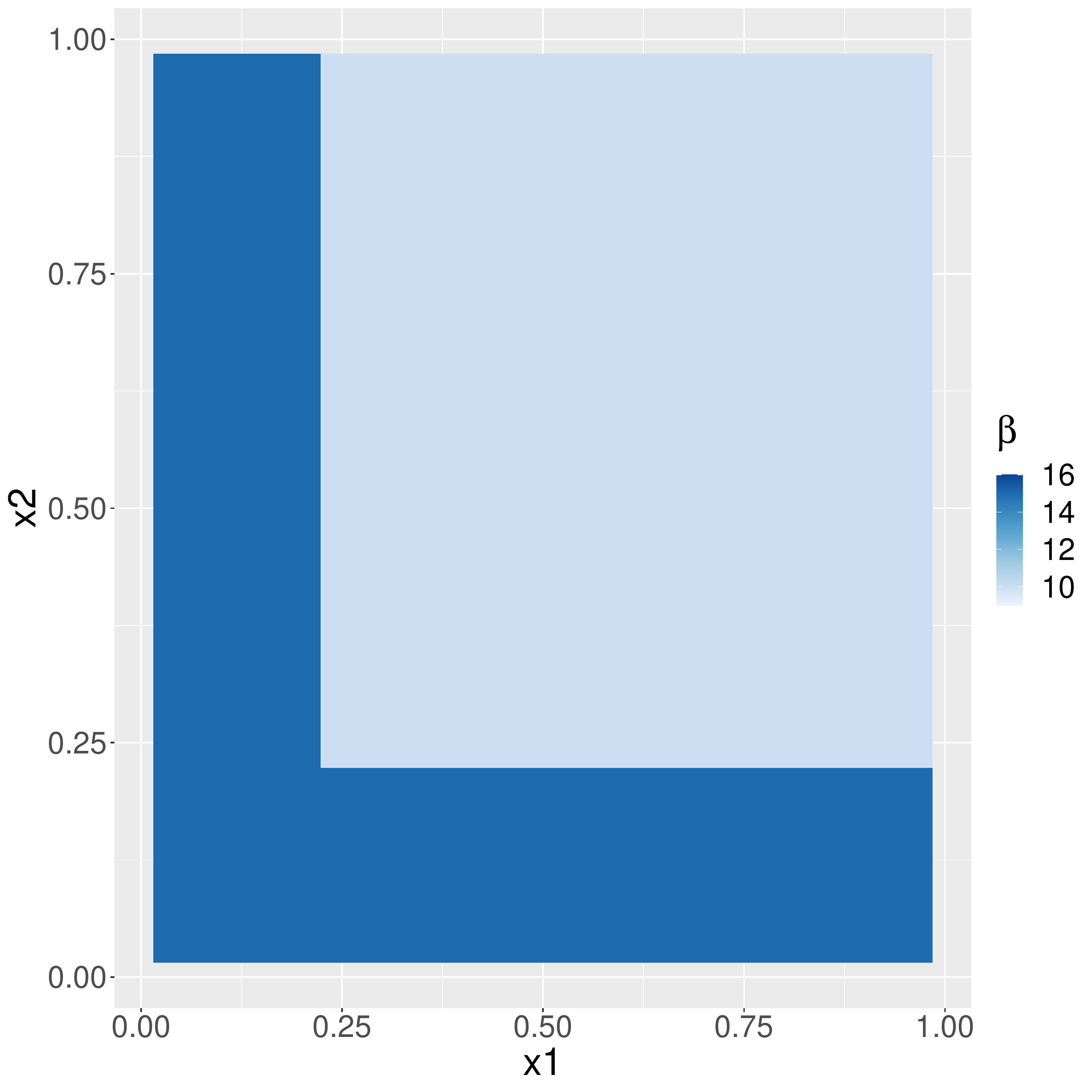}
\end{subfigure}\hfill
\begin{subfigure}{0.33\textwidth}
\caption{Scenario \ref{ex:boosting}: $\beta_1(x_1,x_2)$}
\label{fig:heatmapbeta_boosting_true1}
\includegraphics[width = \textwidth]{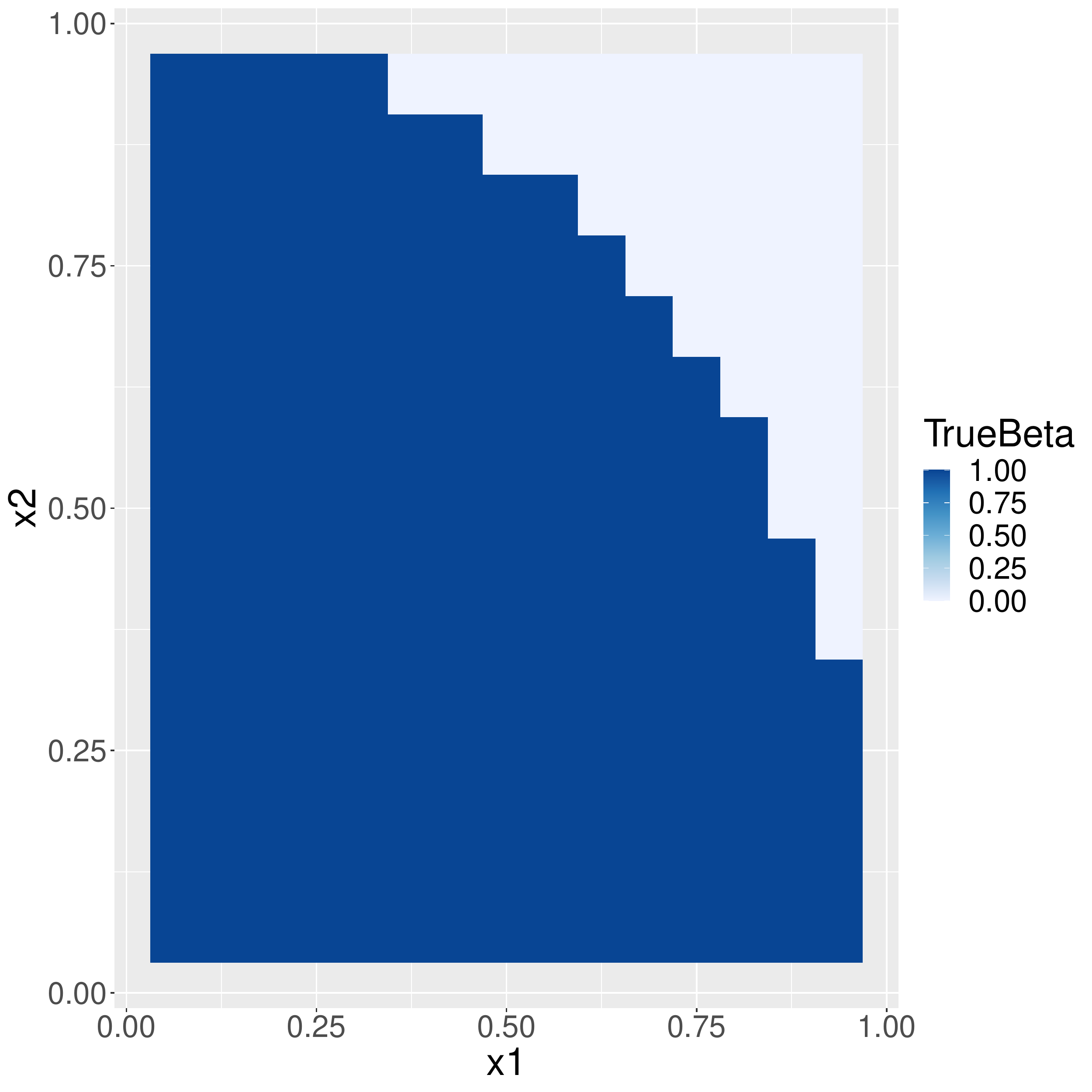}
\end{subfigure}\hfill
\begin{subfigure}{0.33\textwidth}
\caption{Scenario \ref{ex:cosine_continuous}: $\beta_1(x_1)$}
\label{fig:heatmapbeta_cosine_continuous_true1}
\includegraphics[width = \textwidth]{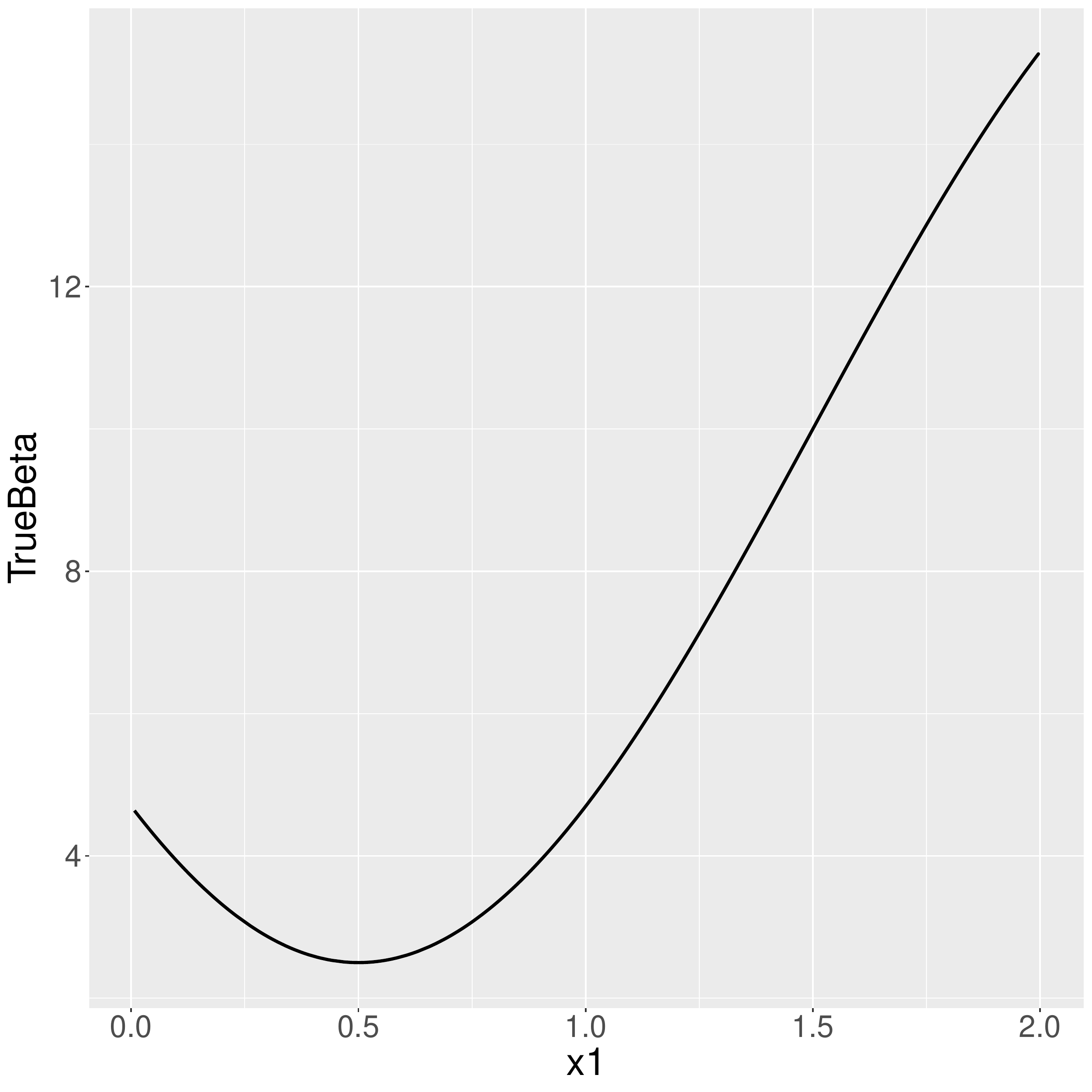}
\end{subfigure}
\end{figure}

In this subsection, we present a simulation study to assess and compare HCQRF, CQRF, HCQRF-c and grf-c in estimating the coefficient functions $\bbeta(\tau)$ under various scenarios.  We start with a simple scenario (\textbf{Scenario \ref{ex:tree_binary}}), where $Z$ is binary, and its quantile coefficient  is a piece-wise constant function of $\bX$.  In \textbf{Scenarios \ref{ex:boosting}} and \textbf{\ref{ex:cosine_continuous}}, we considered continuous $Z$ and with non-linear and continuous coefficient functions. Figure \ref{fig:heatmapbetatrue} displays the $\beta_1(\bX)$ under the three scenerios.

%We further 

%have three continuous predictors $\bZ$, whose quantile coefficients $ \bbeta_\tau(\bX)$ vary across several \textbf{non-linear} subregions of the modifiers $\bX$ as shown in Figure \ref{fig:heatmapbeta_boosting_true1}. Finally in \textbf{Scenario \ref{ex:cosine_continuous}}, we consider  $\bbeta(\bX)$ as continuous functions of $\bX$, which is shown in Figure \ref{fig:heatmapbeta_cosine_continuous_true1}. 

% In \textbf{Scenario \ref{ex:cosine_continuous_new}}, the two predictive variables are negatively correlated with identical coefficient functions. If we use the traditional recursive partition algorithm without considering distinguishing $\bX$ and $\bZ$ in Model \eqref{eq:survival_main_model}, the algorithm may ignore the importance of the predictive variables. The proposed splitting criterion which endures a model assumption is a possible solution to overcome it.

% In \textbf{Scenario \ref{ex5}}, the coefficient is a typical varying-coefficient model where there is only one modifier $x_{i1}$. 

\begin{example} \label{ex:tree_binary} The predictive variable $\bZ_i$ is a  Bernoulli random variable with success probability $0.5$, and its  coefficient functions are 
\begin{align*}
	&\beta_0 = 5,\\
	&\bbeta_1(\bX_i) = 15-5\mathbbm 1\{X_{i1}>0.2\}\times\mathbbm 1\{X_{i2}>0.2\}.
\end{align*} 
The error term in Model (\ref{eq:simmodel}), $\varepsilon_i$, was generated by $\varepsilon_i = \eta_i-\Phi^{-1}(\tau)$ with $\eta_i\sim N(0,0.25)$. The modifiers $\bX_i$ were uniform random variables generated from $U(0,1)^{p}$.
The censoring time $c_i$ was generated independently from $U(0,50)$.

%The quantile function of standard normal distribution, $\Phi^{-1}(\tau)$, is 0 at $\tau = 0.5$. The predictive variable $\bz_i$ is a binary variable which was generated from a Bernoulli distribution with success probability $0.5$. The modifiers $\bx_i$ was generated from $U(0,1)^p$.  
\end{example}

\begin{figure}[!ht]
    \centering
    \caption{Heat map of the true and estimated $\beta_{1}(\bX)$ for \textbf{Scenario \ref{ex:tree_binary}} with $N_1=500$. The estimated quantile coefficients are averaged over 500 Monte Carlo repetitions. The ranges of $\bX_1$ and $\bX_2$ are divided into 16 equally spaced intervals. The colors refer to the value of quantile coefficient in the corresponding grid.}
    %The heat maps for the true quantile coefficients are in the first row. The heat maps for the estimate quantile coefficients by HCQRF-c are in the second row. The heat maps for the estimated quantile coefficients by grf-c are in the third row.}
    \label{fig:heatmapbeta_tree_binary}
\begin{subfigure}{0.25\textwidth}
\caption{$\widehat\beta_1(\bx)$ by CQRF}
\label{fig:heatmapbeta_tree_binary_cqrf}
\includegraphics[width = \textwidth]{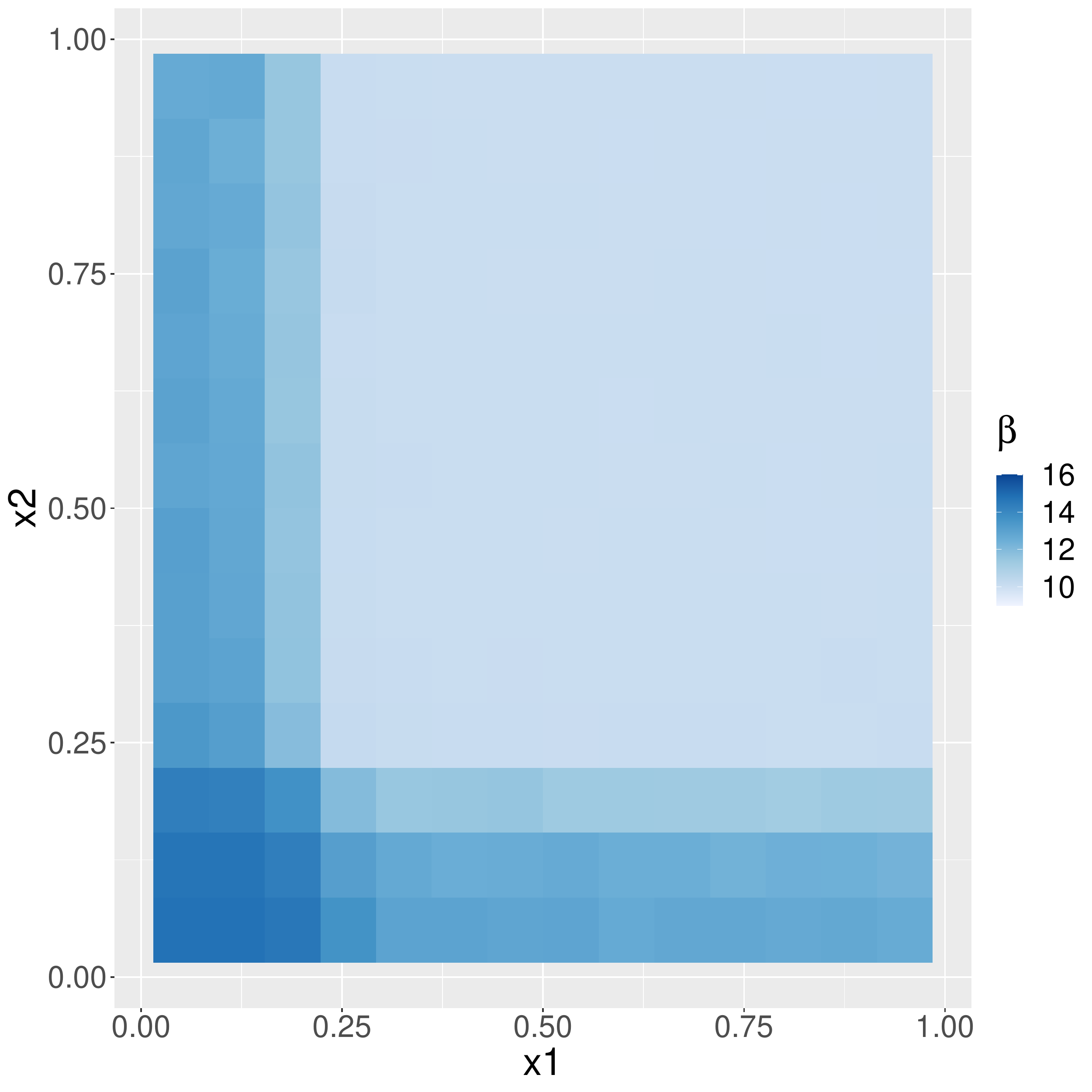}
\end{subfigure}\hfill
\hfill
\begin{subfigure}{0.25\textwidth}
\caption{$\widehat\beta_1(\bx)$ by HCQRF}
\label{fig:heatmapbeta_tree_binary_hcqrf}
\includegraphics[width = \textwidth]{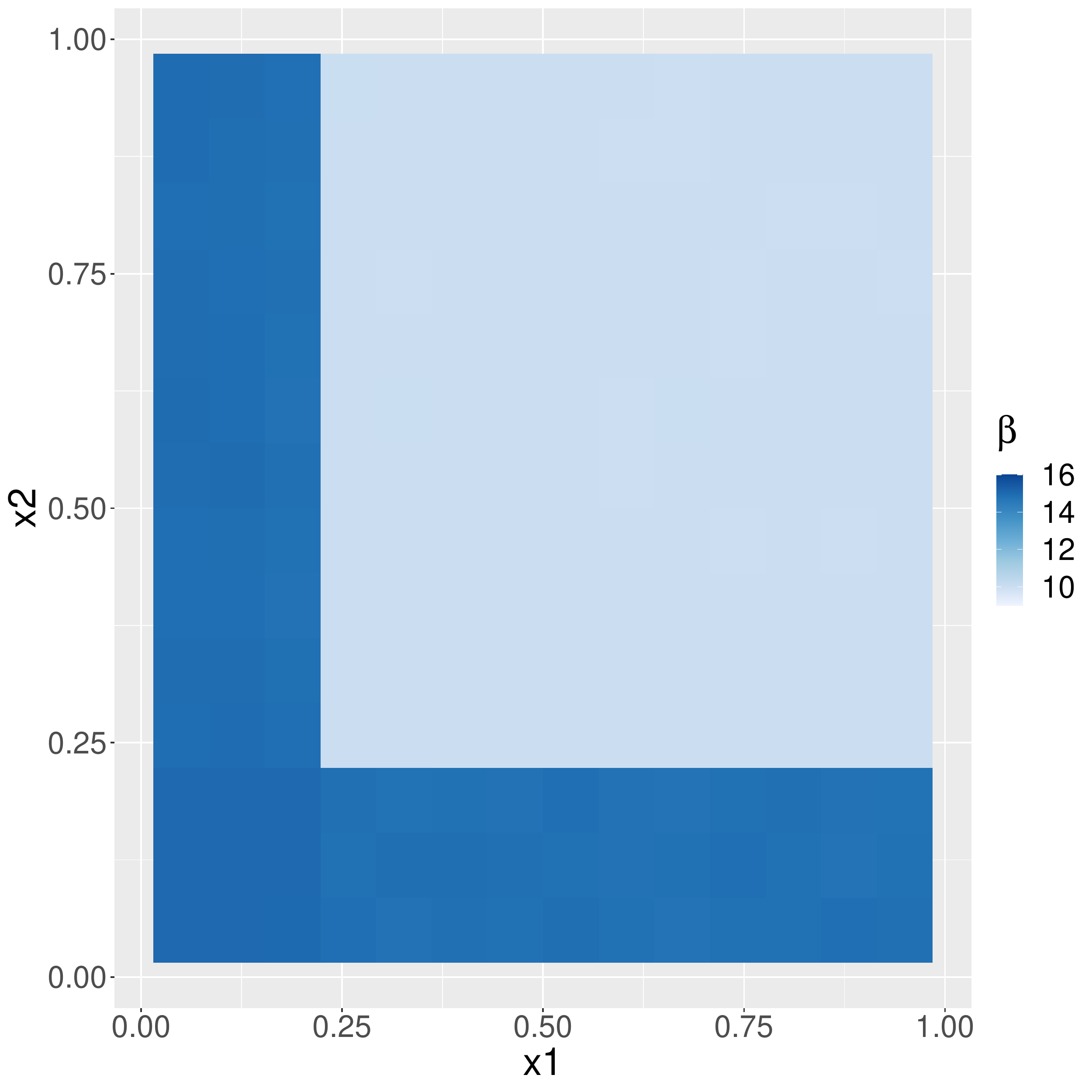}
\end{subfigure}\hfill
\begin{subfigure}{0.25\textwidth}
\caption{$\widehat\beta_1(\bx)$ by HCQRF-c}
\label{fig:heatmapbeta_tree_binary_hcqrfc}
\includegraphics[width = \textwidth]{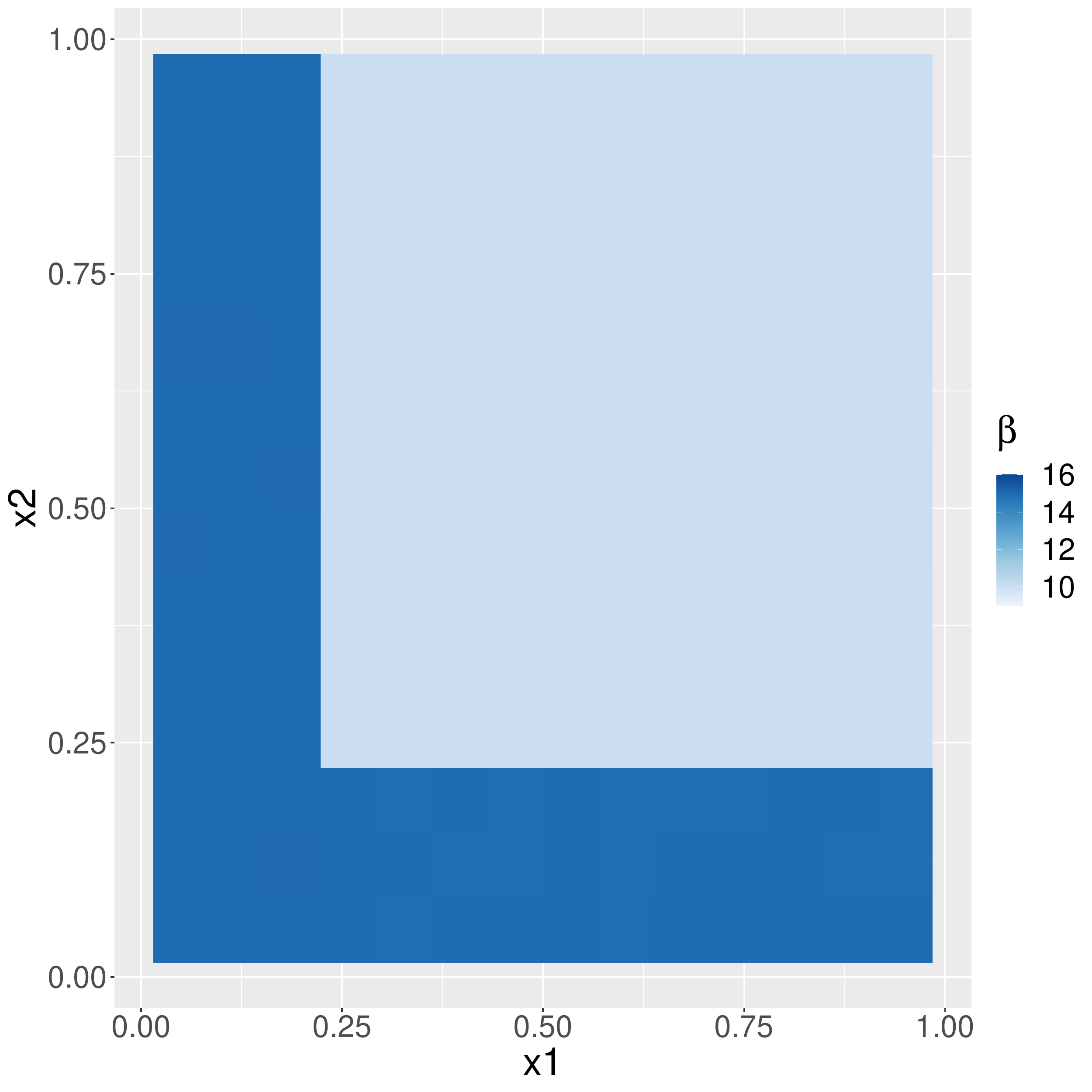}
\end{subfigure}\hfill
\begin{subfigure}{0.25\textwidth}
\caption{$\widehat\beta_1(\bx)$ by grf-c}
\label{fig:heatmapbeta_tree_binary_grf}
\includegraphics[width = \textwidth]{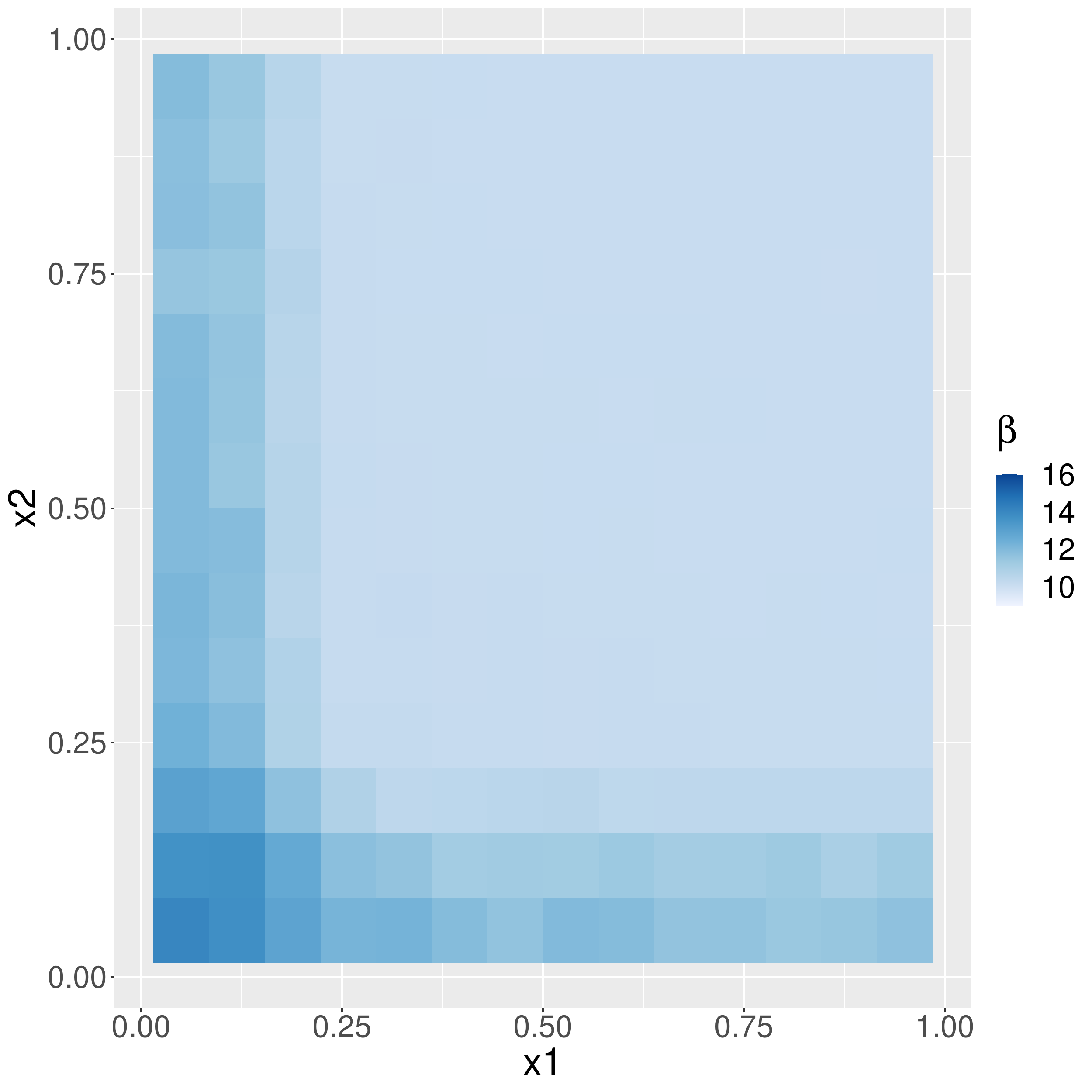}
\end{subfigure}
\end{figure}

Table \ref{tbl:tree_binary} summarizes the simulation results of \textbf{Scenario \ref{ex:tree_binary}} at sample sizes at $N_1 = 500$ and $1000$. In this simple scenario, the conditional quantile is a piece-wise constant function. The estimation performance of $\beta_0(\bX)$ by HCQRF is close to but slightly worse than that of CQRF, while the estimation of $\beta_1(\bX)$ by HCQRF outperforms the one of CQRF. With the complete data, the estimation performance of HCQRF-c is always better than the one of grf-c. In some cases,   grf-c is  even worse than HCQRF with the censored data, especially for $\beta_1(\bX)$. Figure \ref{fig:heatmapbeta_tree_binary} presents the heat maps for both true and estimated quantile coefficient $\beta_1(x_1, x_2)$ (averaged over $500$ Monte Carlo repetitions). It is clear that HCQRF-c does the best job of capturing the true function, followed by HCQRF and CQRF. It suggests the need of incorporating $Z$ into the splittings in the presence of heterogeneous treatment effects. The post-construction weighted adjustment (as in CQRF and grf-c) alone is insufficient to restore the coefficient function.  As expected, Table \ref{tbl:tree_binary} also shows that the measurements of the estimation performance diminish with the increase of the sample size.

%We can tell from the figure that the estimations around the partition line $\mathbbm 1\{X_1<0.5\}\times\mathbbm 1\{X_2<0.5\}$ by grf-c performs poorly. It indicates that the splitting criterion of grf which only considers the marginal quantiles of the survival time leads us to a non-satisfactory estimation. 

\begin{table}[!ht]

\begin{center}
\begin{threeparttable}
% \centering
\caption{Summary of the MSE and MAE across different approaches at $\tau = 0.5$  and under \textbf{Scenario \ref{ex:tree_binary}} based on 500 simulation runs.}
	\label{tbl:tree_binary}
\begin{tabular}{cccccc}
\hline\hline	 
&Method & $\beta_0$ & $\beta_1$ &$\beta_0$ & $\beta_1$\\
\hline
  && \multicolumn{2}{c}{$N_1 = 500$}&\multicolumn{2}{c}{$N_1 = 1000$}\\
\cline{3-6}

\multirow{4}{*}{MSE}&HCQRF & 0.005&0.213&0.004&0.009\\
&CQRF&0.002&4.140&0.002&1.907 \\
 &HCQRF-c  & 0.004&0.040&0.003&0.007\\
&grf-c &  0.001&6.194&0.001&5.661\\
\cline{2-6} 
\multirow{4}{*}{MAE}&HCQRF& 0.056&0.134&0.049&0.074\\
&CQRF&0.039&0.970&0.034&0.482 \\
&HCQRF-c&0.053&0.083&0.046&0.065 \\
&grf-c&0.028&1.412&0.023&1.303 \\
% \cline{3-7}
% &\multirow{3}{*}{Relative MSE}&HCQRF &1E-4&0.008&1E-3&0.002\\
% &&CQRF&9E-5&0.005&7E-5&0.001\\
%  &&HCQRF-c  &1E-4&0.004&1E-3&0.001\\
% &&grf-c &5E-5&0.036&4E-5&0.027\\
% \cline{3-7}
% &\multirow{3}{*}{Relative MAE}&HCQRF &0.010&0.024&0.009&0.010\\
% &&CQRF&0.008&0.019&0.009&0.007\\
%  &&HCQRF-c  &0.010&0.014&0.005&0.066\\
% &&grf-c & 0.006&0.084&0.007&0.006\\

\hline\hline
\end{tabular}
\begin{tablenotes}\footnotesize
\item MSE: mean squared error, MAE: mean absolute error.
\end{tablenotes}
\end{threeparttable}
\end{center}
\end{table}

\begin{figure}[!ht]
\centering
\caption{Heat map of the true and estimated quantile coefficients for \textbf{Scenario \ref{ex:boosting}} with $N_1=500$ . The estimated quantile coefficients are averaged based on 500 Monte Carlo repetitions. The ranges of $\bx_1$ and $\bx_2$ are divided into 16 equally spaced intervals. The colors refer to the value of quantile coefficient in the corresponding grid. The heat maps for the true quantile coefficients are in the first row. The heat maps for the estimate quantile coefficients by HCQRF are in the second row. The ones by HCQRF-c are in the third row. The ones by grf-c are in the fourth row.}
\label{fig:heatmapbeta_boosting}
\begin{subfigure}[b]{0.25\textwidth}
	\centering
		\caption{$\beta_0$}
	\includegraphics[width=\textwidth]{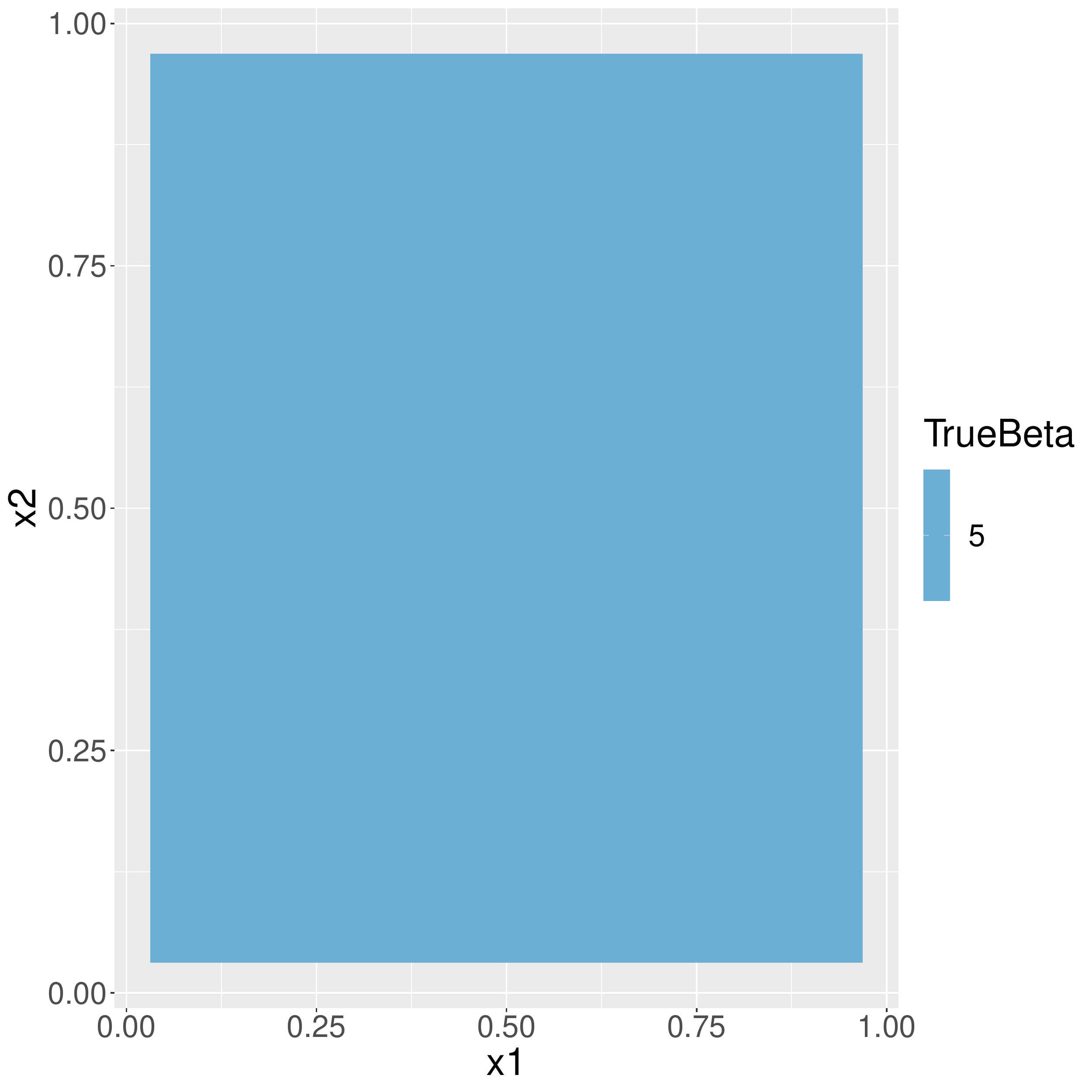}
	\label{fig:beta0}
\end{subfigure}\hfill
\begin{subfigure}[b]{0.25\textwidth}
	\centering
	\caption{$\beta_1(\bx)$}
	\includegraphics[width=\textwidth]{Figures/Boosting_continuous/truebeta2HM.pdf}
	\label{fig:beta1}
\end{subfigure}\hfill
\begin{subfigure}[b]{0.25\textwidth}
	\centering
	\caption{$\beta_2(\bx)$}
	\includegraphics[width=\textwidth]{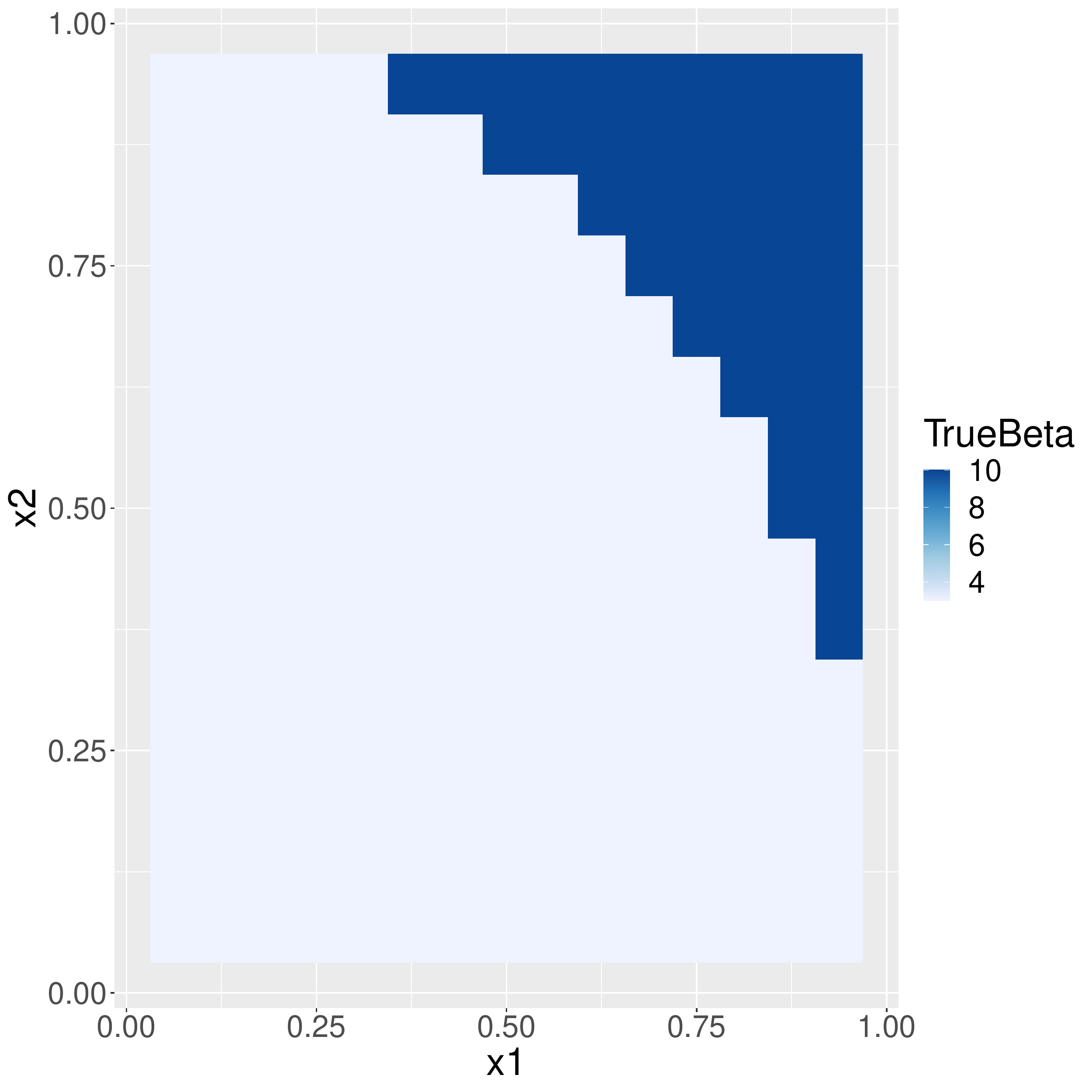}
	\label{fig:beta2}
\end{subfigure}\hfill
\begin{subfigure}[b]{0.25\textwidth}
\centering
\caption{$\beta_3(\bx)$}
\includegraphics[width=\textwidth]{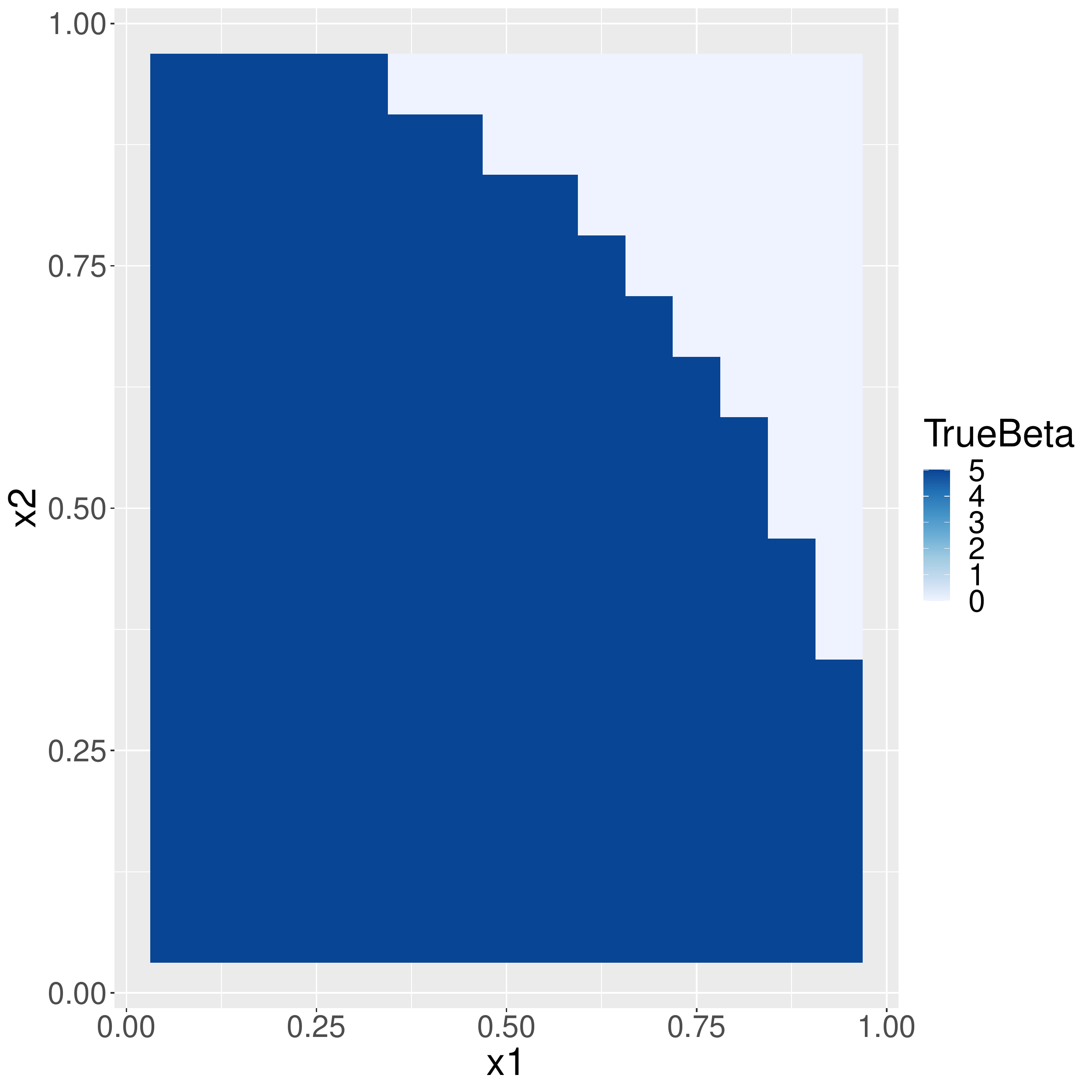}
\label{fig:beta3}
\end{subfigure}\vfill

\begin{subfigure}[b]{0.25\textwidth}
\centering
\includegraphics[width=\textwidth]{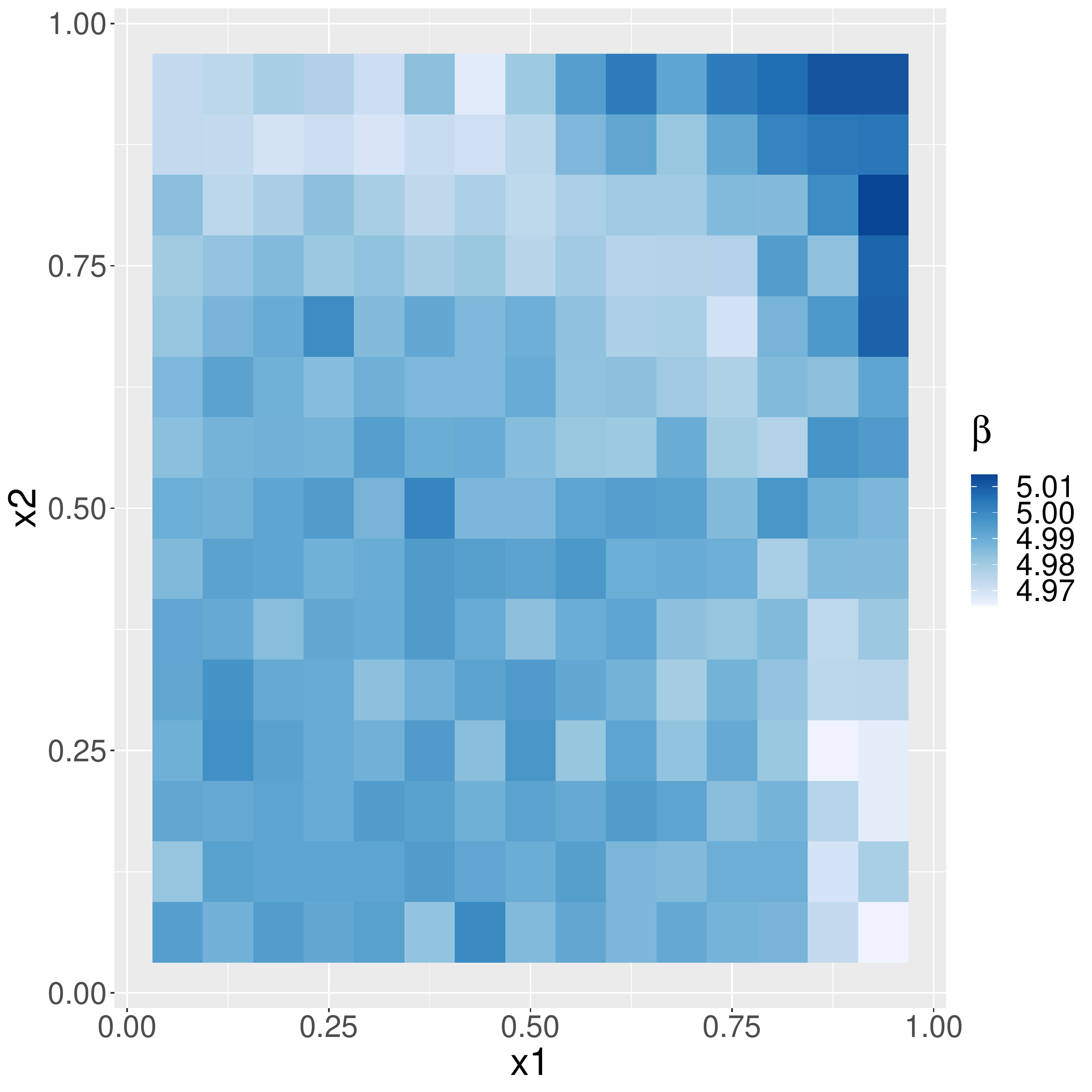}
\end{subfigure}\hfill
\begin{subfigure}[b]{0.25\textwidth}
\centering
\includegraphics[width=\textwidth]{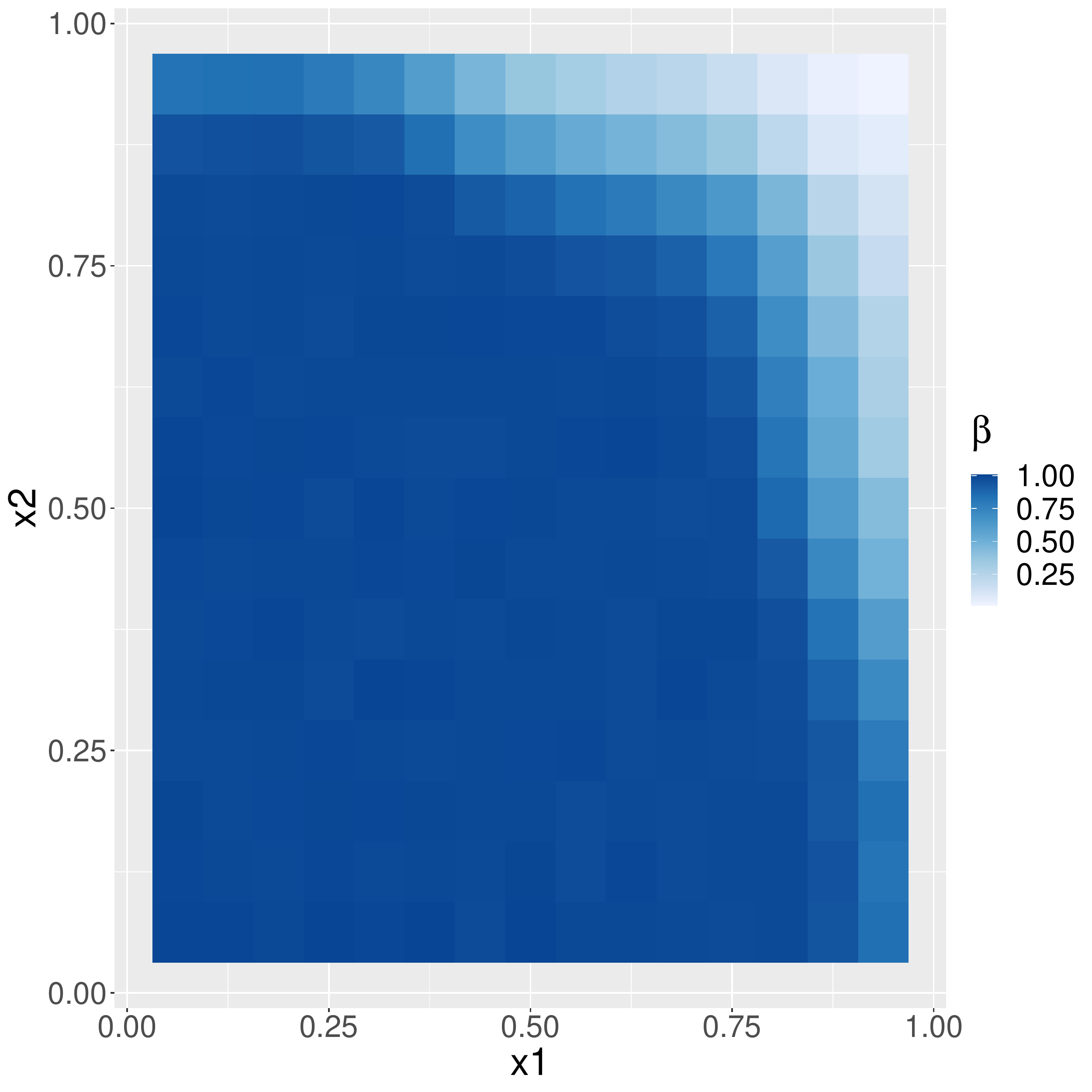}
\end{subfigure}\hfill
\begin{subfigure}[b]{0.25\textwidth}
\centering
\includegraphics[width=\textwidth]{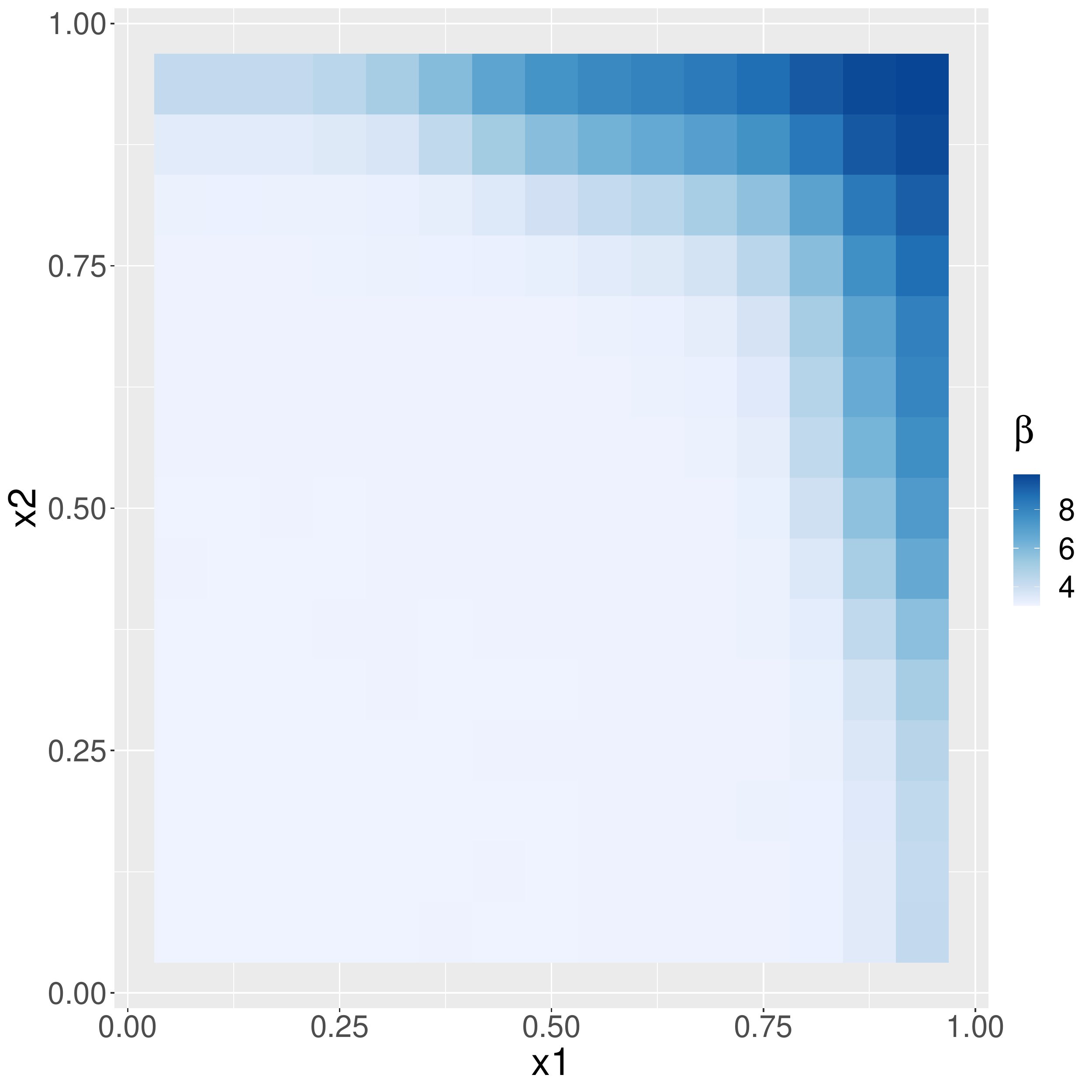}
\end{subfigure}\hfill
\begin{subfigure}[b]{0.25\textwidth}
\centering
\includegraphics[width=\textwidth]{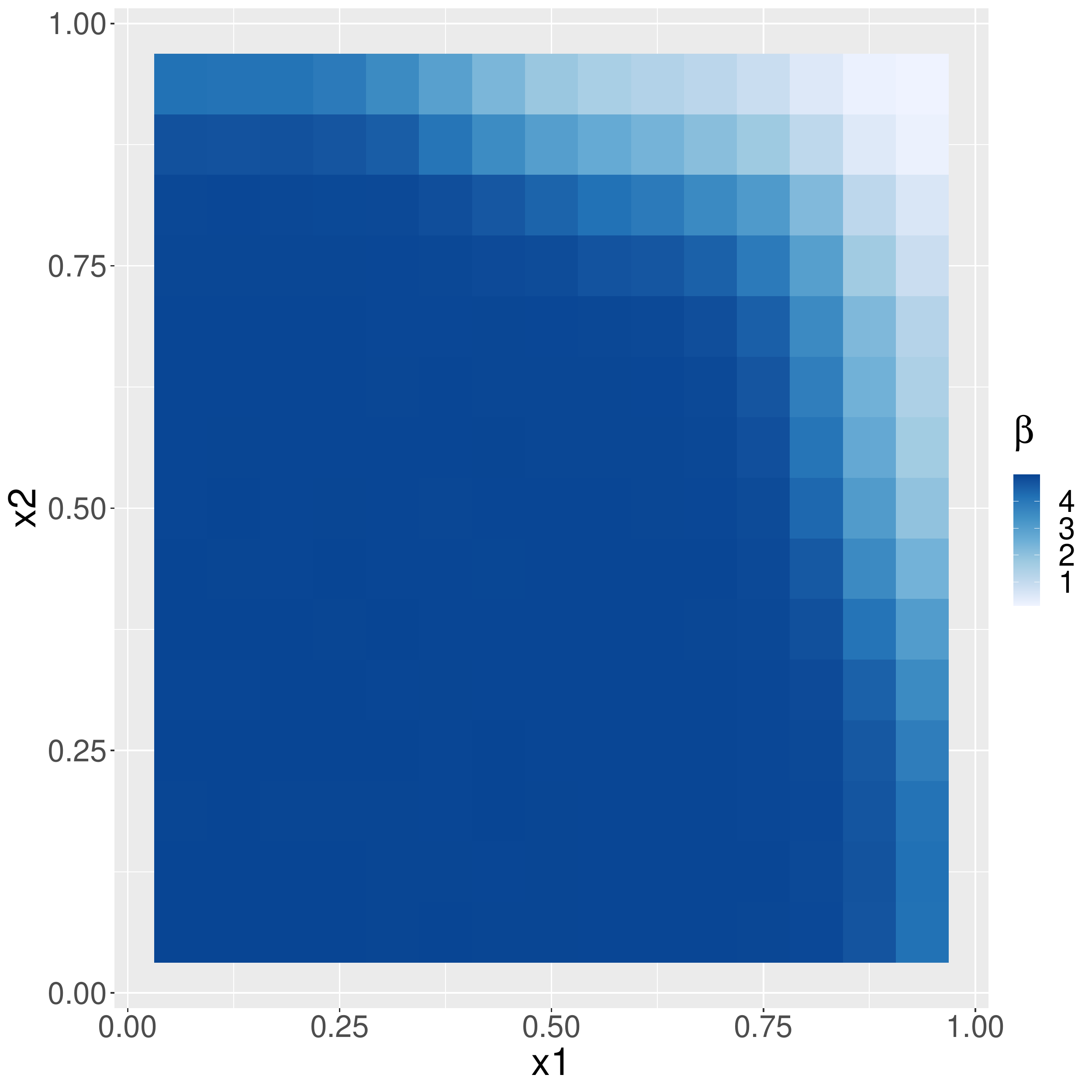}
\end{subfigure}\vfill

\begin{subfigure}[b]{0.25\textwidth}
\centering

\includegraphics[width=\textwidth]{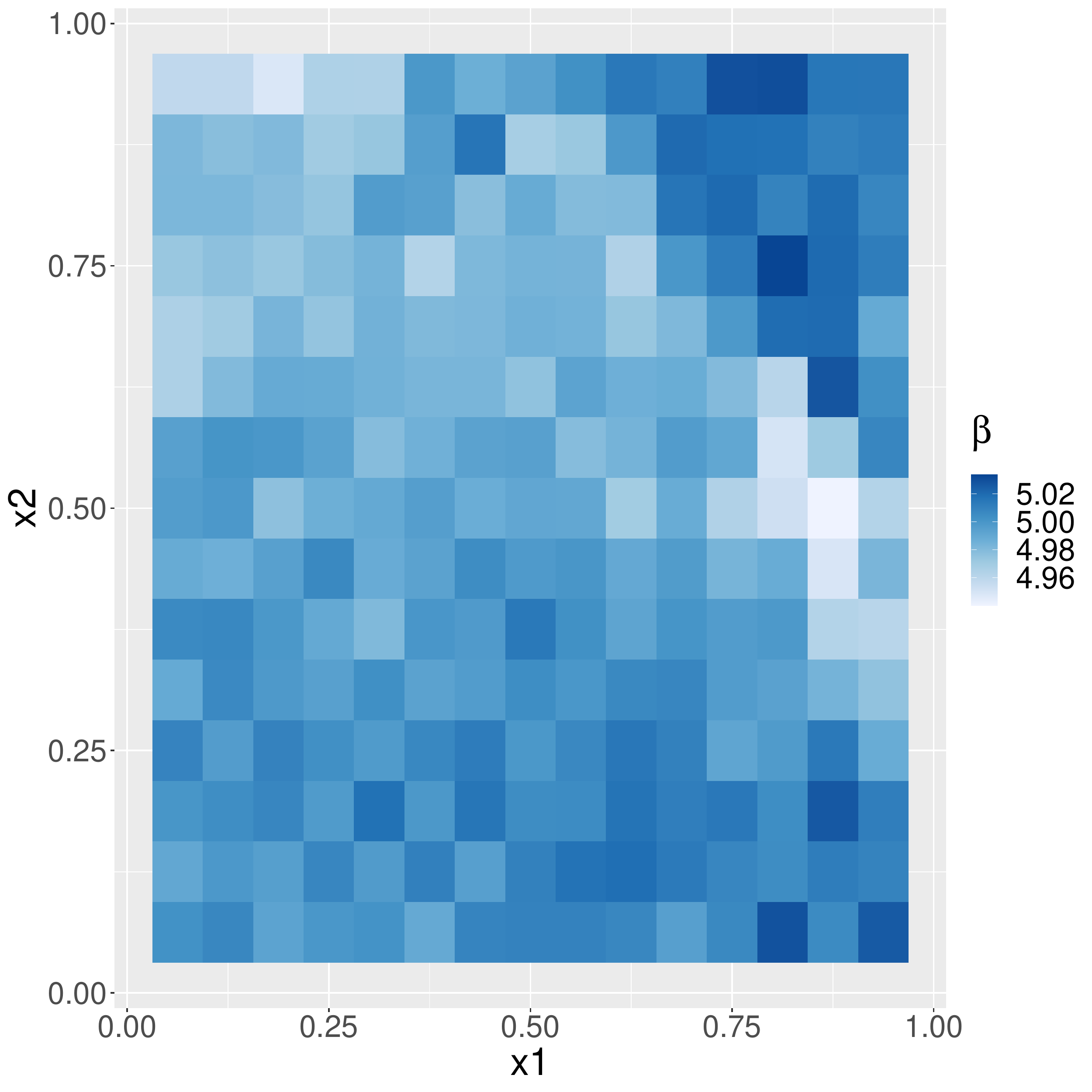}

\end{subfigure}\hfill
\begin{subfigure}[b]{0.25\textwidth}
	\centering

	\includegraphics[width=\textwidth]{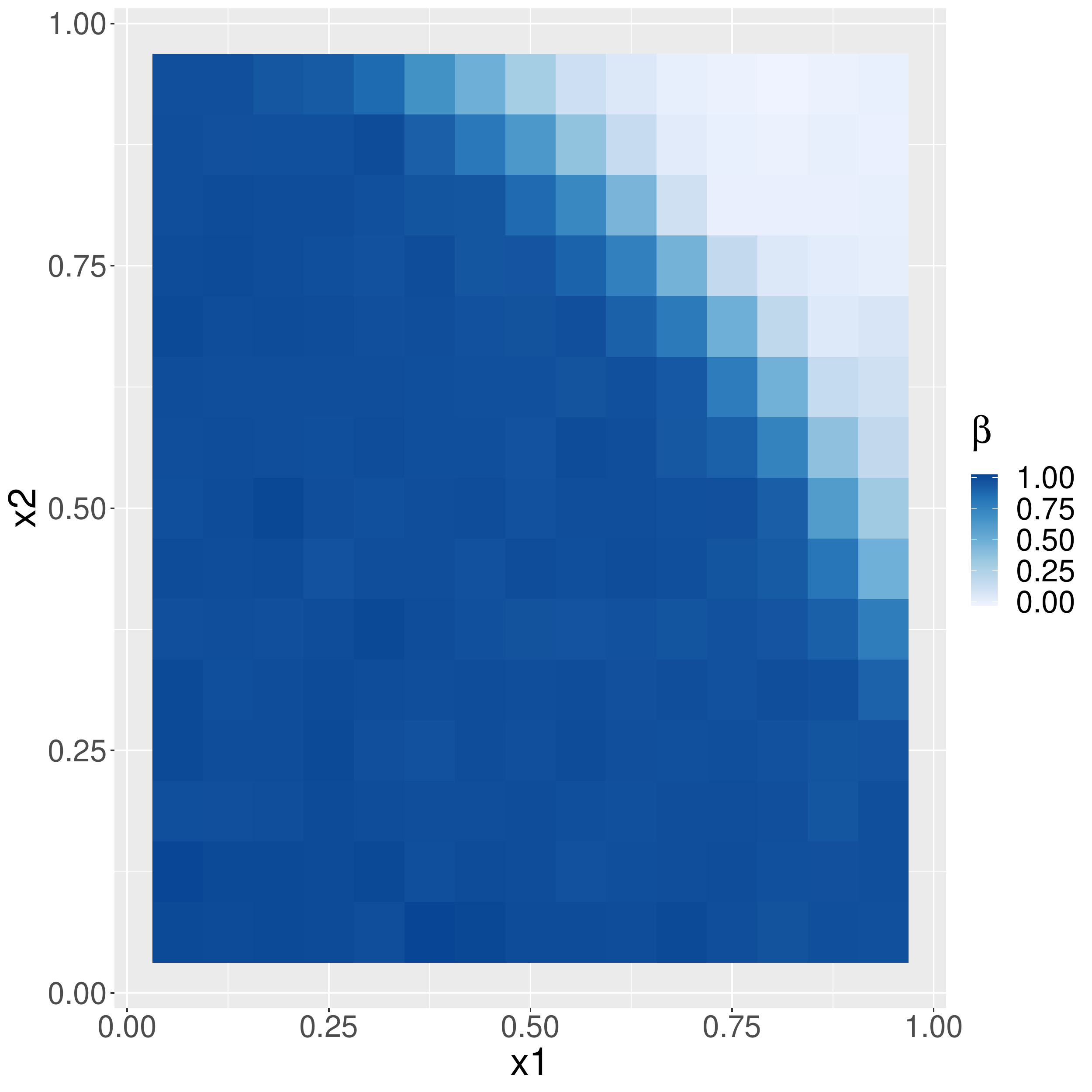}

\end{subfigure}\hfill
\begin{subfigure}[b]{0.25\textwidth}
	\centering

	\includegraphics[width=\textwidth]{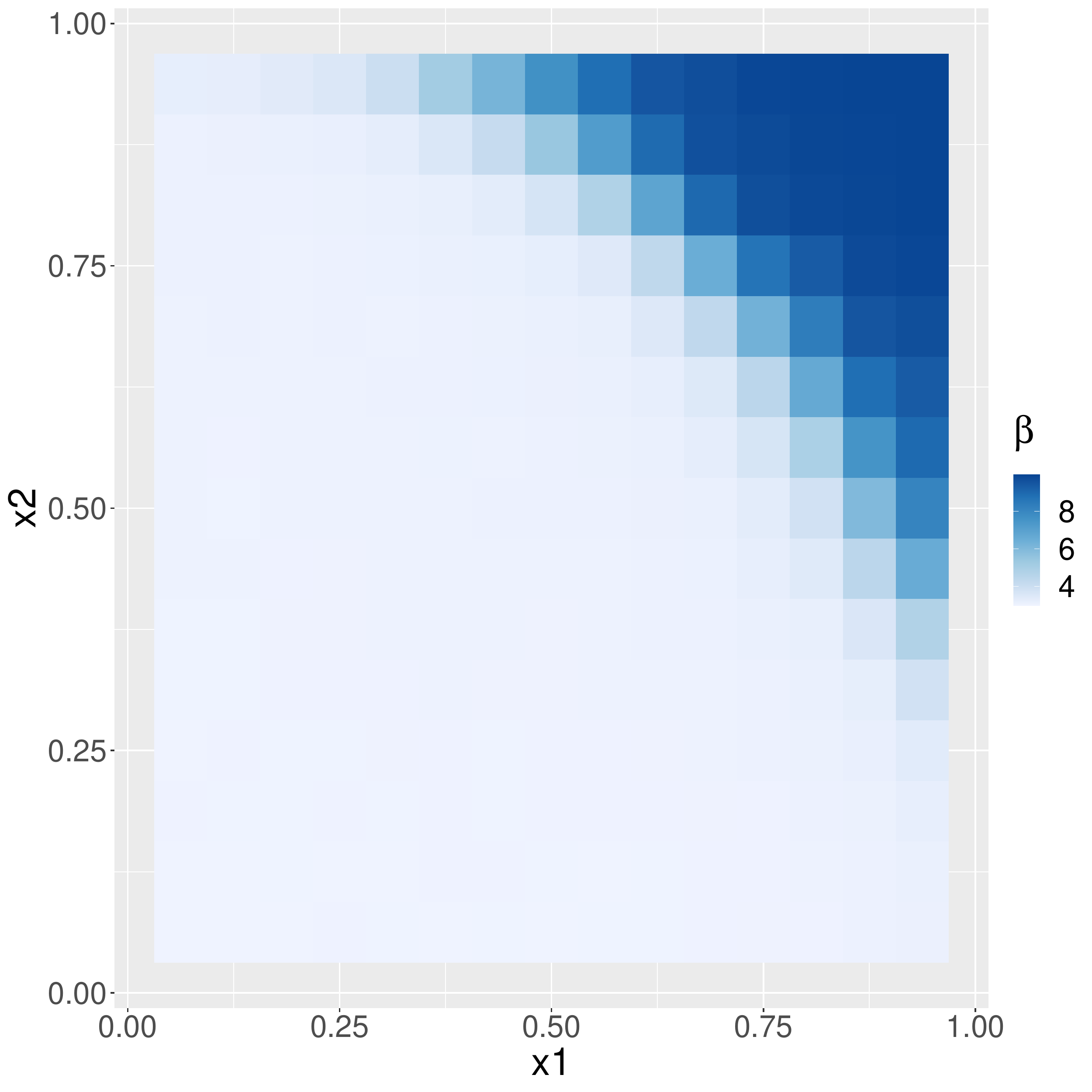}

\end{subfigure}\hfill
\begin{subfigure}[b]{0.25\textwidth}
	\centering

	\includegraphics[width=\textwidth]{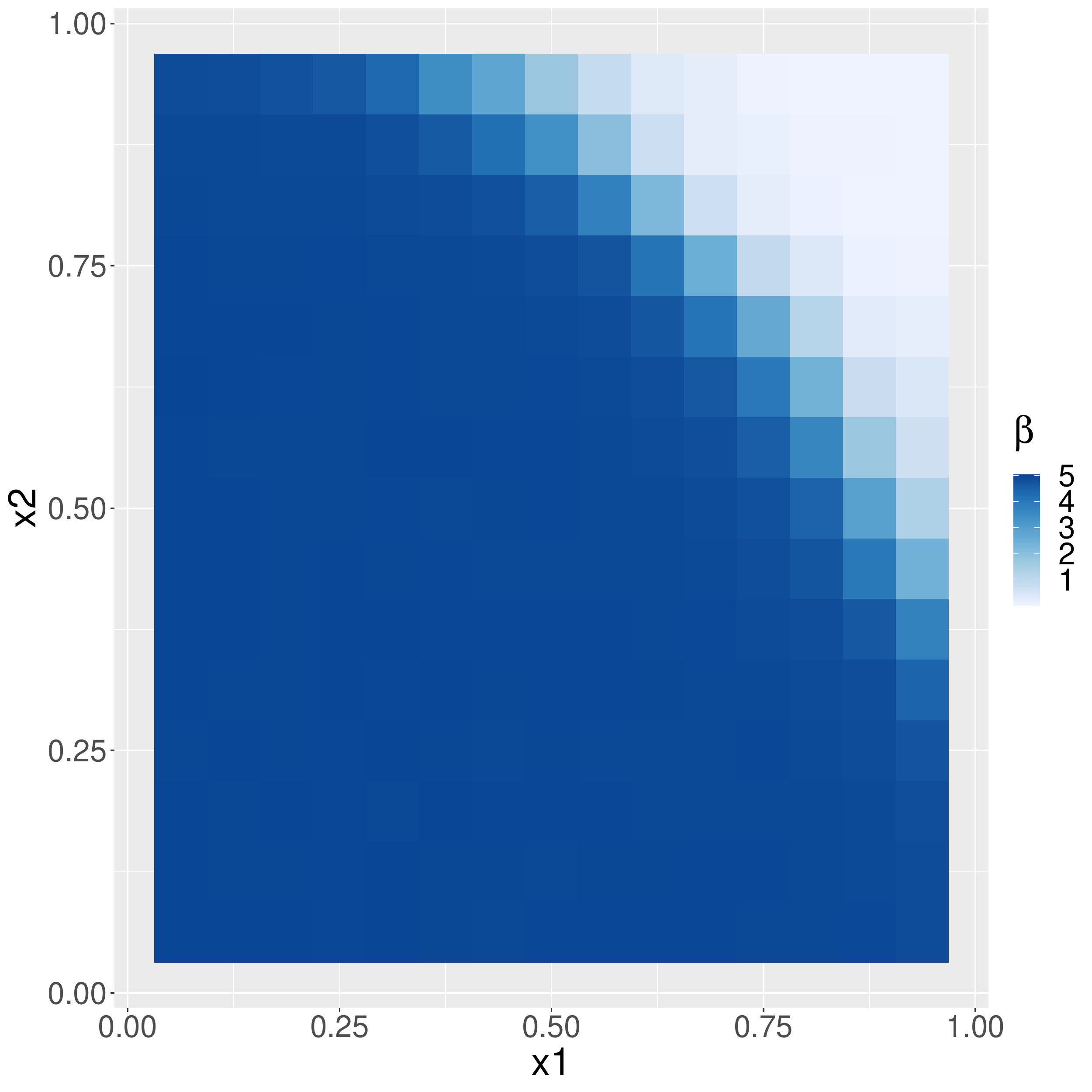}

\end{subfigure}\vfill
\begin{subfigure}[b]{0.25\textwidth}
\centering

\includegraphics[width=\textwidth]{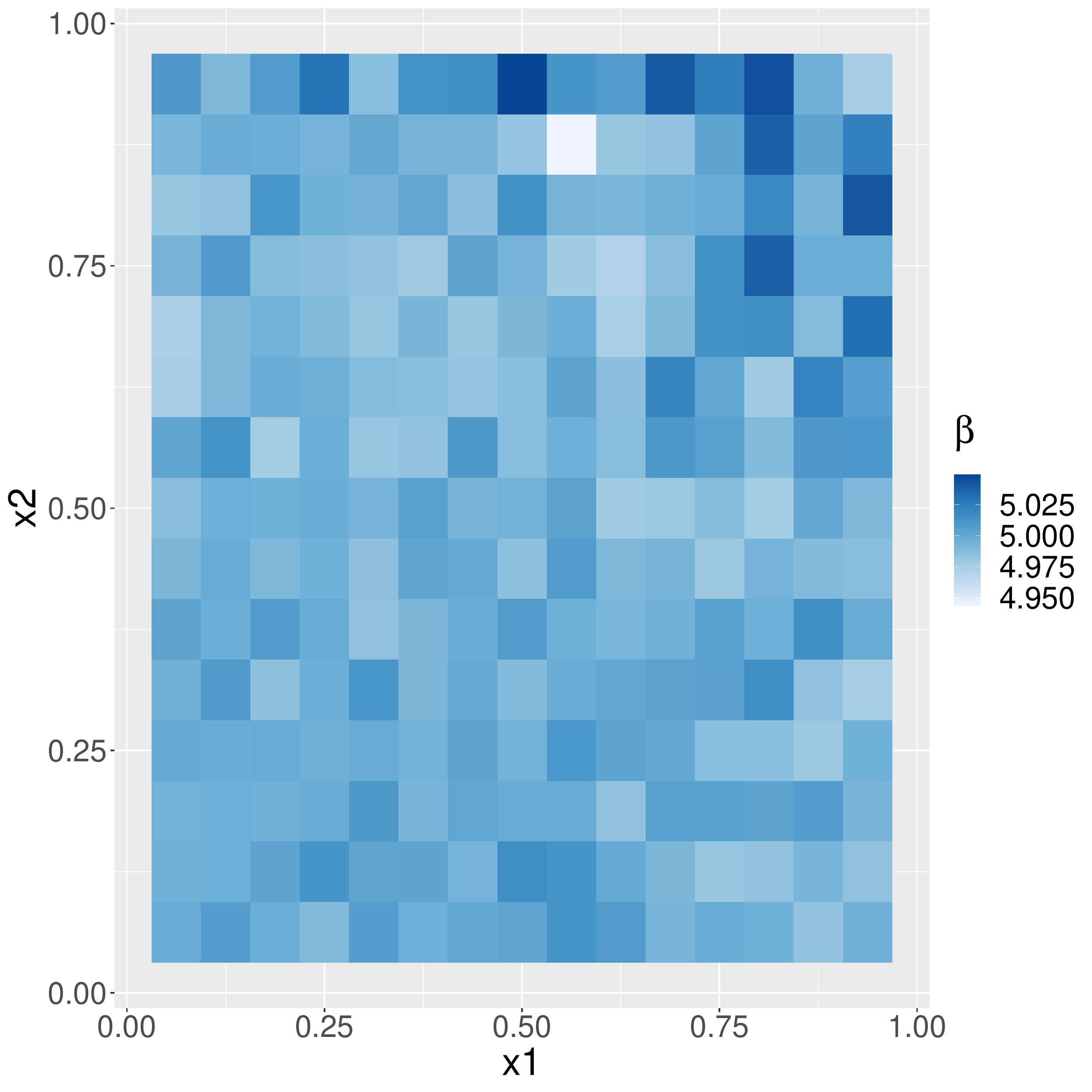}

\end{subfigure}\hfill
\begin{subfigure}[b]{0.25\textwidth}
\centering

\includegraphics[width=\textwidth]{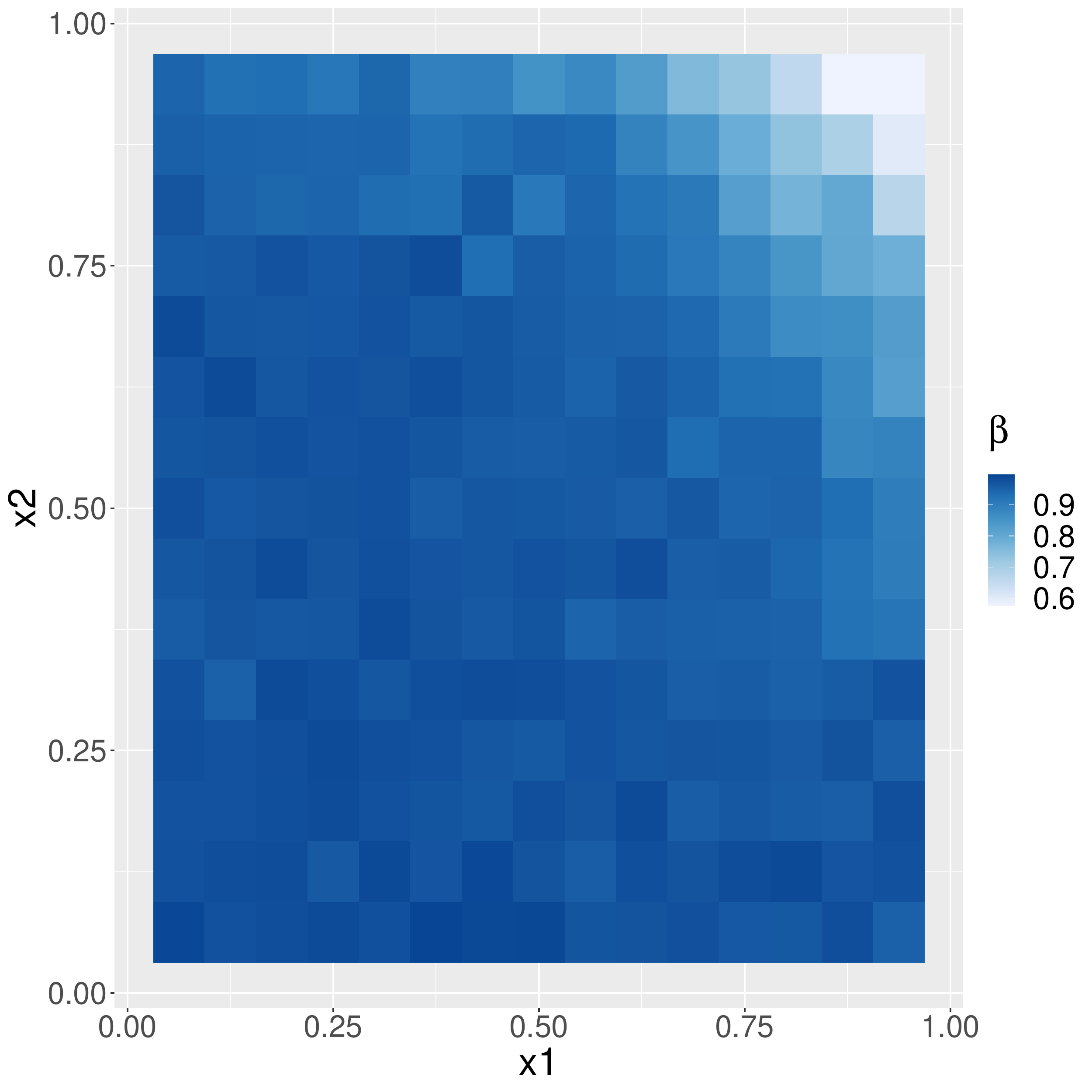}

\end{subfigure}\hfill
\begin{subfigure}[b]{0.25\textwidth}
\centering

\includegraphics[width=\textwidth]{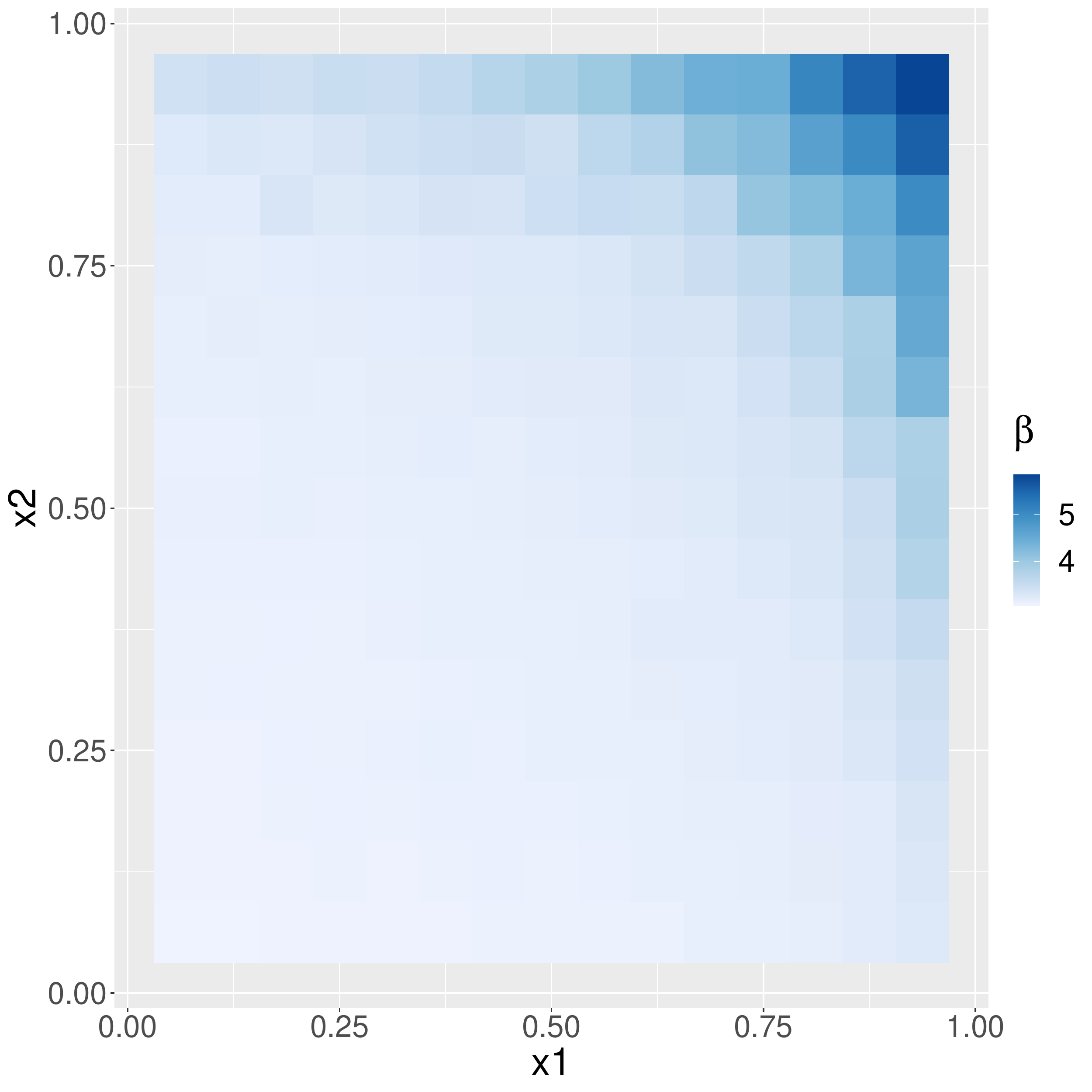}

\end{subfigure}\hfill
\begin{subfigure}[b]{0.25\textwidth}
\centering

\includegraphics[width=\textwidth]{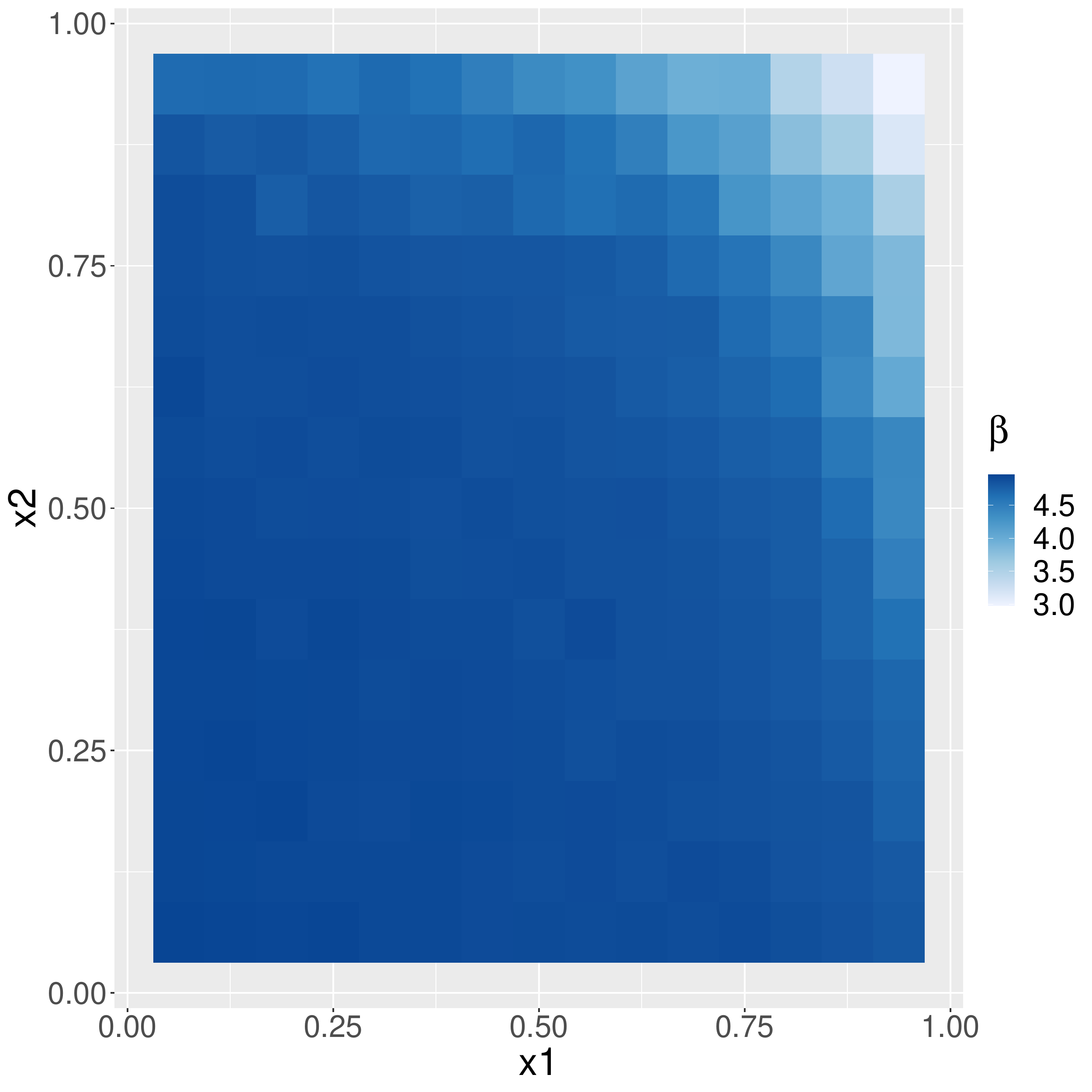}

\end{subfigure}

\end{figure}

\begin{example} \label{ex:boosting} Both $\bZ_i$ and $\bX_i$ were generated randomly from $U(0,1)^{(p+2)}$. The coefficient functions (see Figure \ref{fig:heatmapbeta_boosting}) are
\begin{align*}
	&\beta_0 = 5,\\
	&\bbeta_1(\bX_i) = \begin{pmatrix}
		\beta_1(\bX_i)\\
		\beta_2(\bX_i)\\
		\beta_3(\bX_i)
	\end{pmatrix} = \begin{pmatrix}
		1\\ 
		3\\
		5
	\end{pmatrix}\times\mathbbm 1\big\{X_{i1}^2+X_{i2}^2<1\big\}+\begin{pmatrix}
		0\\
		10\\
		0
	\end{pmatrix}\times\mathbbm 1\big\{X_{i1}^2+X_{i2}^2\ge 1\big\}.
\end{align*} 
The error term $\varepsilon_i = \eta_i-\Phi^{-1}(\tau)$ was generated with $\eta_i\sim N(0,0.25)$. Both the predictive variables $\bZ_i$ and the modifiers $\bX_i$ were uniform random variables generated from $U(0,1)^{(p+2)}$.  The censoring time $C_i$ was generated from $U(0,40)$.
\end{example}

Since $\bZ$ are continuous in \textbf{Scenario \ref{ex:boosting}}, CQRF can no longer be applied to estimate $\bbeta_\tau(\bx)$. We will only compare HCQRF with HCQRF-c and grf-c in this subsection. 
In later Section \ref{sec:condquantile}, we will compare HCQRF with CQRF in estimating the conditional quantile function $Q_\tau(T|\bX = \bx,\bZ = \bz)$ under all the scenarios.  
The MSE and MAE of the estimated $\bbeta_\tau(\bx)$ from HCQRF, HCQRF-c and grf-c are presented in Table \ref{tbl:boosting}. Comparing HCQRF-c and grf-c applied to the complete data, we observe that HCQRF-c have smaller MSE and MAE than grf-c for slope estimations ($\beta_1$, $\beta_2$ and $\beta_3$), while the two methods produce similar results in $\beta_0$. As expected, HCQRF is slightly worse than its benchmark HCQRF-c, and such difference diminishes with the increase of the sample size. In the meantime, HCQRF-c performs better than grf-c in all the slope estimations. Figure \ref{fig:heatmapbeta_boosting} shows the heat maps for both true and estimated quantile coefficients in a grid view of $X_1$ and $X_2$ based on 500 Monte Carlo repetitions. The figure further confirms that the proposed HCQRF approach could effectively capture the complex heterogeneous treatment effect.

\begin{table}
\begin{center}
\begin{threeparttable}[!ht]
\centering
\caption{Estimation performance of the quantile coefficient function at $\tau = 0.5$ for \textbf{Scenario \ref{ex:boosting}} based on  500 simulation runs.}
	\label{tbl:boosting}
\begin{tabular}{cccccccccc}
\hline\hline
 &Method & $\beta_0$ & $\beta_1$&$\beta_2$ &$\beta_3$&$\beta_0$ & $\beta_1$&$\beta_2$&$\beta_3$\\
 \hline
   && \multicolumn{4}{c}{$N_1 = 500$}&\multicolumn{4}{c}{$N_1 = 1000$}\\
   \cline{3-10}
 \multirow{3}{*}{MSE}&HCQRF&0.028&0.123&3.990&2.093&0.017&0.082&2.755&1.443\\
 &HCQRF-c&0.022&0.105&3.602&1.884&0.013&0.069&2.482&1.295\\
 &grf-c&0.011&0.181&8.229&4.239&0.006&0.172&8.101&4.176\\
 \cline{2-10}
  \multirow{3}{*}{ MAE}&HCQRF&0.117&0.204&0.800&0.587&0.093&0.159&0.575&0.427\\
  &HCQRF-c& 0.106&0.183&0.714&0.527&0.084&0.142&0.511&0.381\\
  &grf-c&0.075&0.240&1.400&0.998&0.056&0.225&1.373&0.978\\
%  \cline{2-10}
%   \multirow{3}{*}{Relative MSE}&HCQRF&0.001&-\tnote{a}&0.112&-&0.001&-&0.086&-\\
%  &HCQRF-c&0.001&-&0.104&-&0.001&-&0.078&-\\
%  &grf-c&4E-4&-&0.088&-&2E-4&-&0.086&-\\
%  \cline{2-10}
%   \multirow{3}{*}{Relative MAE}&HCQRF&0.023&-&0.136&-&0.019&-&0.102&-\\
%  &HCQRF-c&0.021&-&0.122&-&0.017&-&0.091&-\\
%  &grf-c&0.015&-&0.180&-&0.011&-&0.174&-\\
\hline\hline

\end{tabular}
\begin{tablenotes}\footnotesize
\item MSE: mean squared error, MAE: mean absolute error
\end{tablenotes}
\end{threeparttable}
\end{center}
\end{table}

\begin{figure}
    \centering
    \caption{Plots of the true and estimated quantile coefficients for \textbf{Scenario \ref{ex:cosine_continuous}} with $N_1 = 500$. The estimated quantile coefficients are averaged based on 500 Monte Carlo repetitions. The ranges of $\bx_1$, $\bx_2$ and $\bx_3$ are divided into 400 equally spaced intervals. The plots for the true quantile coefficients are in the first row. The ones for the estimate quantile coefficients by HCQRF are in the second row. The ones by HCQRF-c are in the third row. The ones by grf-c are in the fourth row.}
    \label{fig:estbeta_cosine_independent}
    \begin{subfigure}[b]{0.25\textwidth}
	\centering
		\caption{$\beta_0(x_3)$}
	\includegraphics[width=\textwidth]{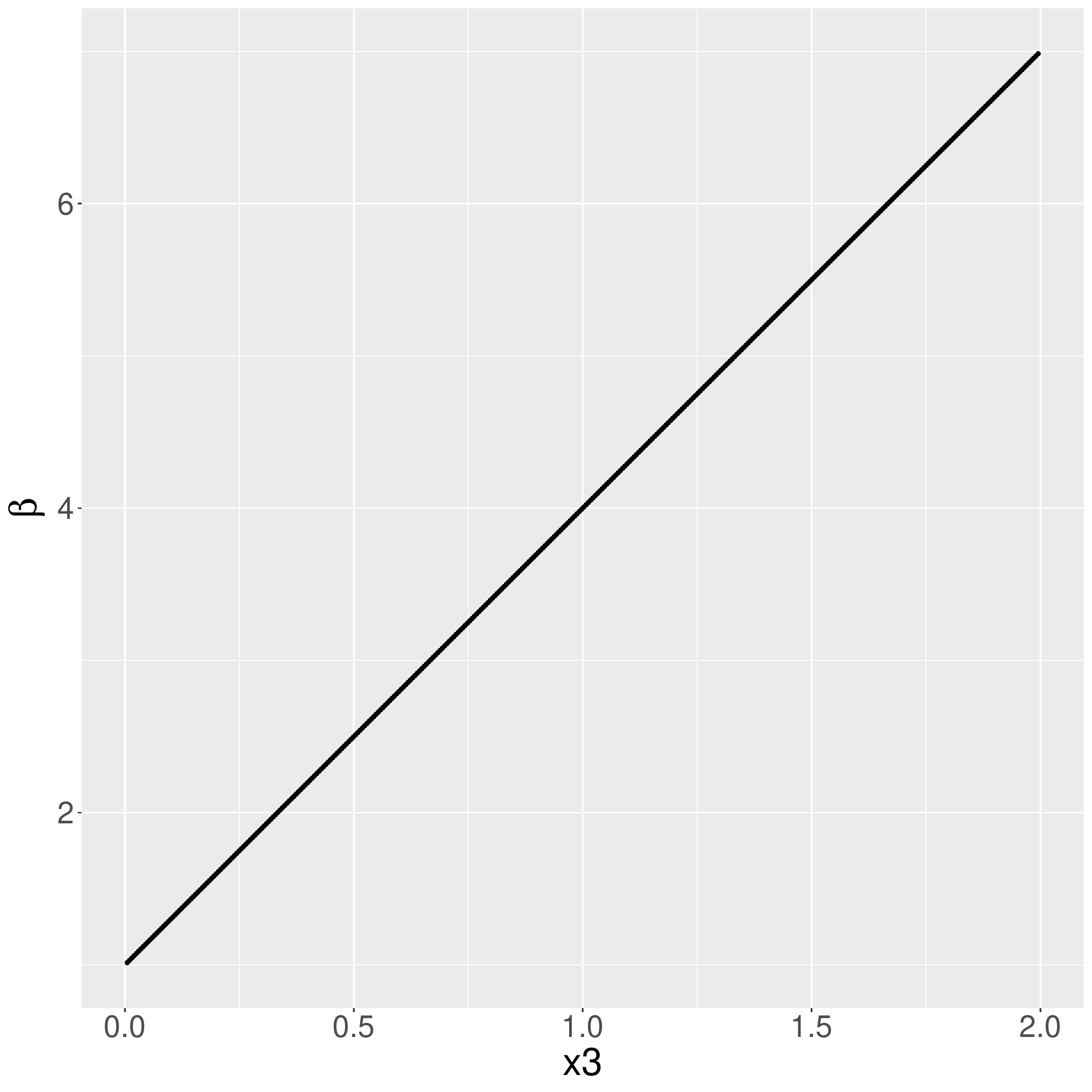}
\end{subfigure}
    \begin{subfigure}[b]{0.25\textwidth}
	\centering
		\caption{$\beta_1(x_1)$}
	\includegraphics[width=\textwidth]{Figures/cosine_continuous_independent/TrueBeta2cosine_continuous_independent.pdf}
\end{subfigure}
\begin{subfigure}[b]{0.25\textwidth}
	\centering
		\caption{$\beta_2(x_2)$}
	\includegraphics[width=\textwidth]{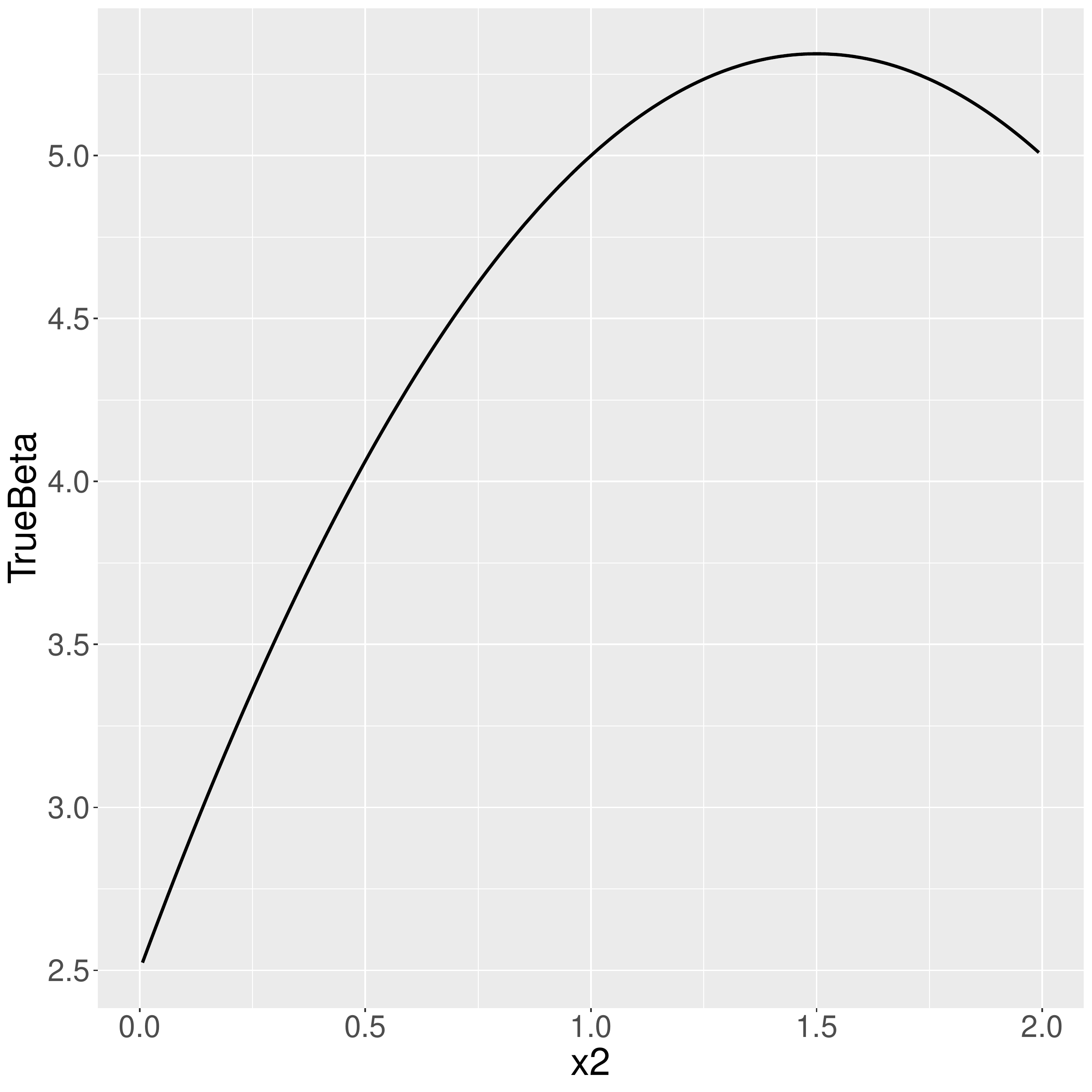}
\end{subfigure}\vfill

\begin{subfigure}[b]{0.25\textwidth}
	\centering
	\includegraphics[width=\textwidth]{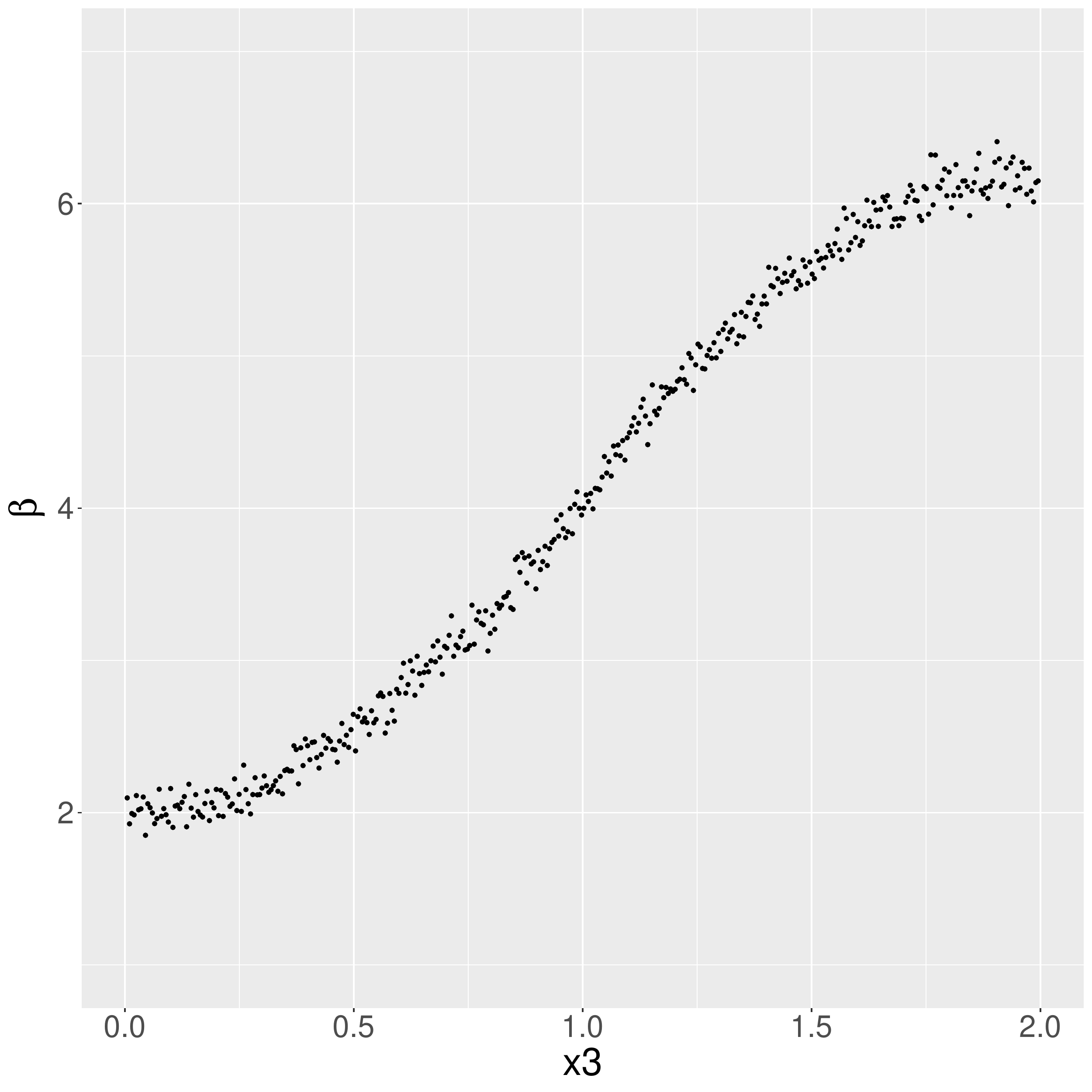}
\end{subfigure}
\begin{subfigure}[b]{0.25\textwidth}
	\centering
	\includegraphics[width=\textwidth]{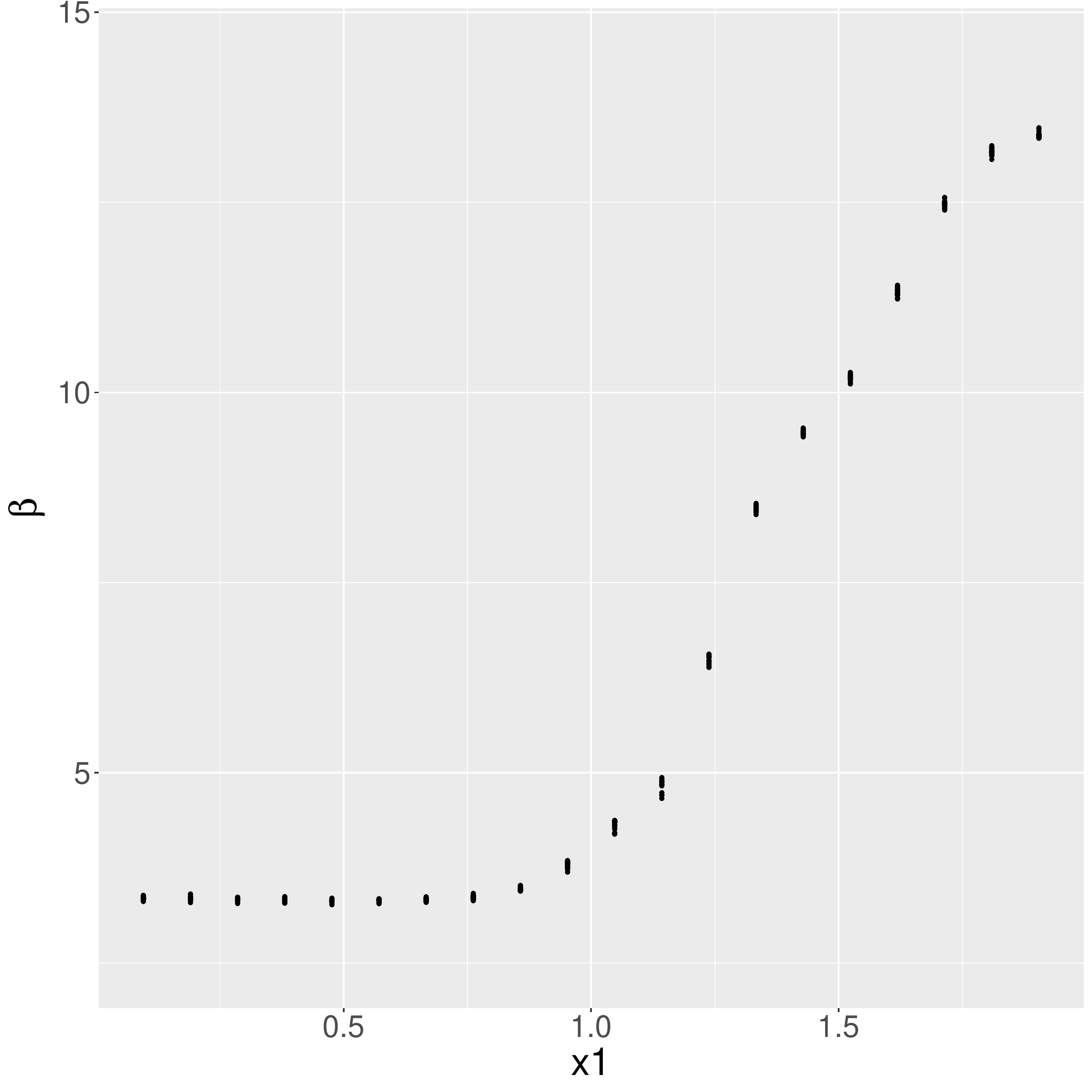}
\end{subfigure}
\begin{subfigure}[b]{0.25\textwidth}
	\centering
	\includegraphics[width=\textwidth]{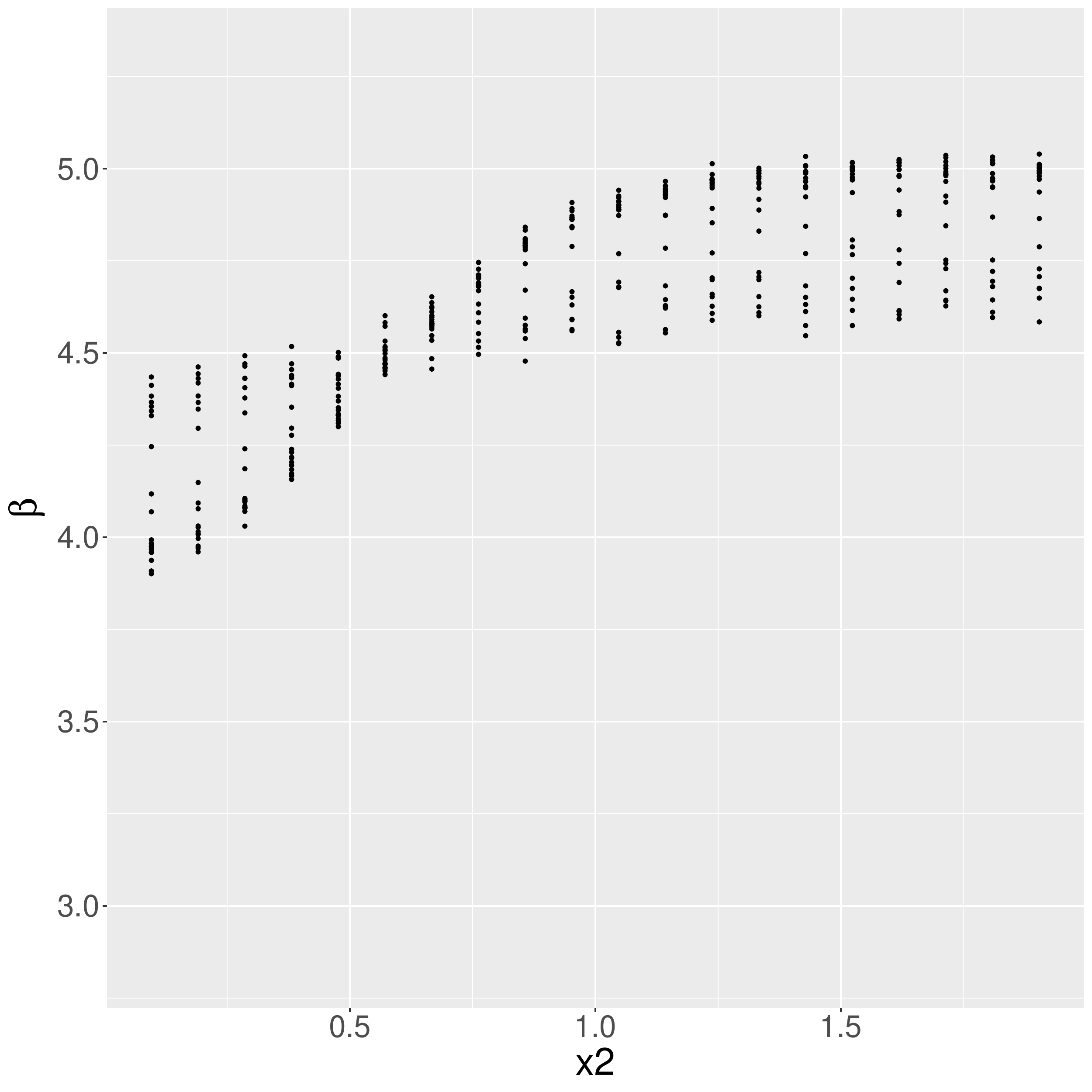}
\end{subfigure}\vfill

\begin{subfigure}[b]{0.25\textwidth}
	\centering
	\includegraphics[width=\textwidth]{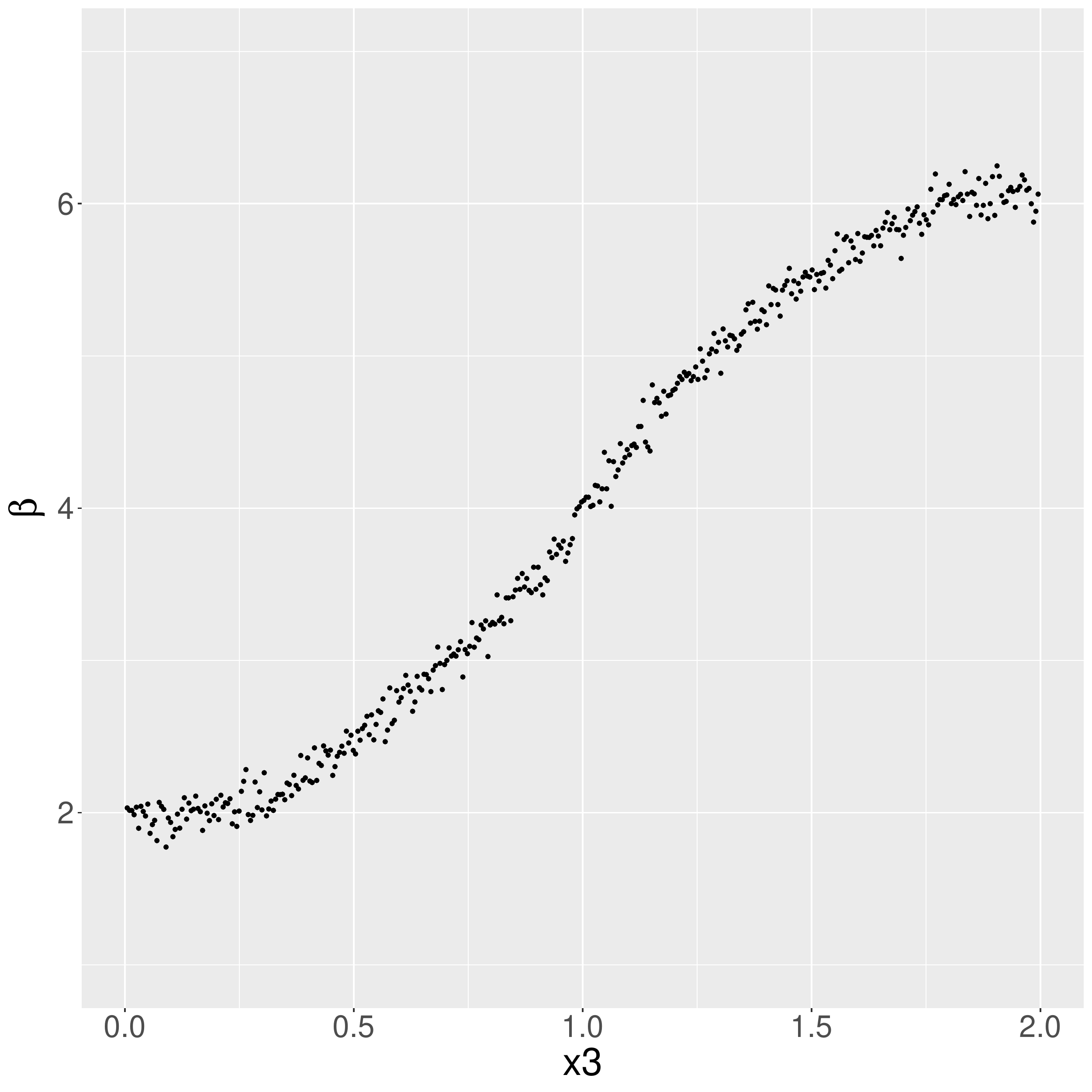}
\end{subfigure}
    \begin{subfigure}[b]{0.25\textwidth}
	\centering
	\includegraphics[width=\textwidth]{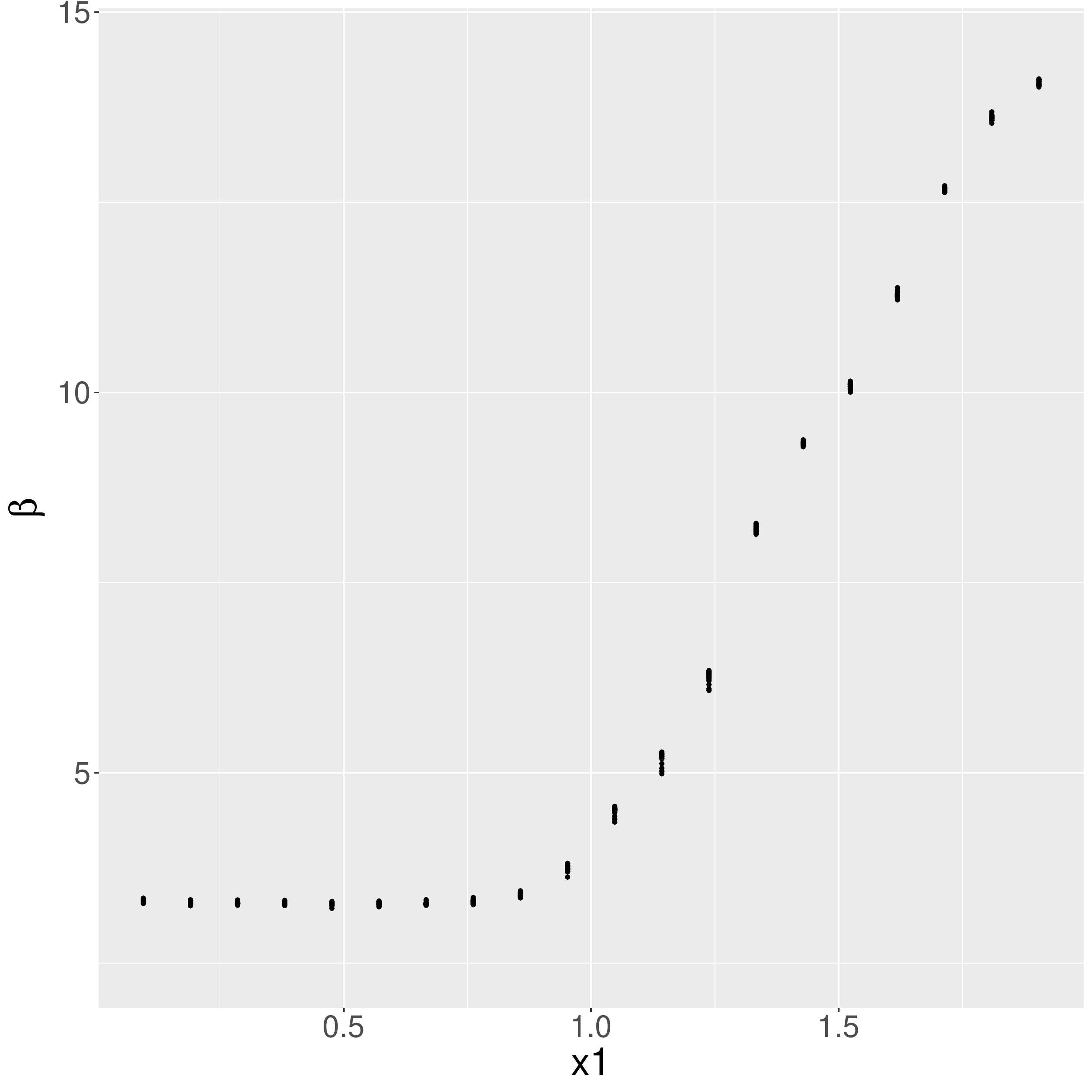}
\end{subfigure}
\begin{subfigure}[b]{0.25\textwidth}
	\centering
	\includegraphics[width=\textwidth]{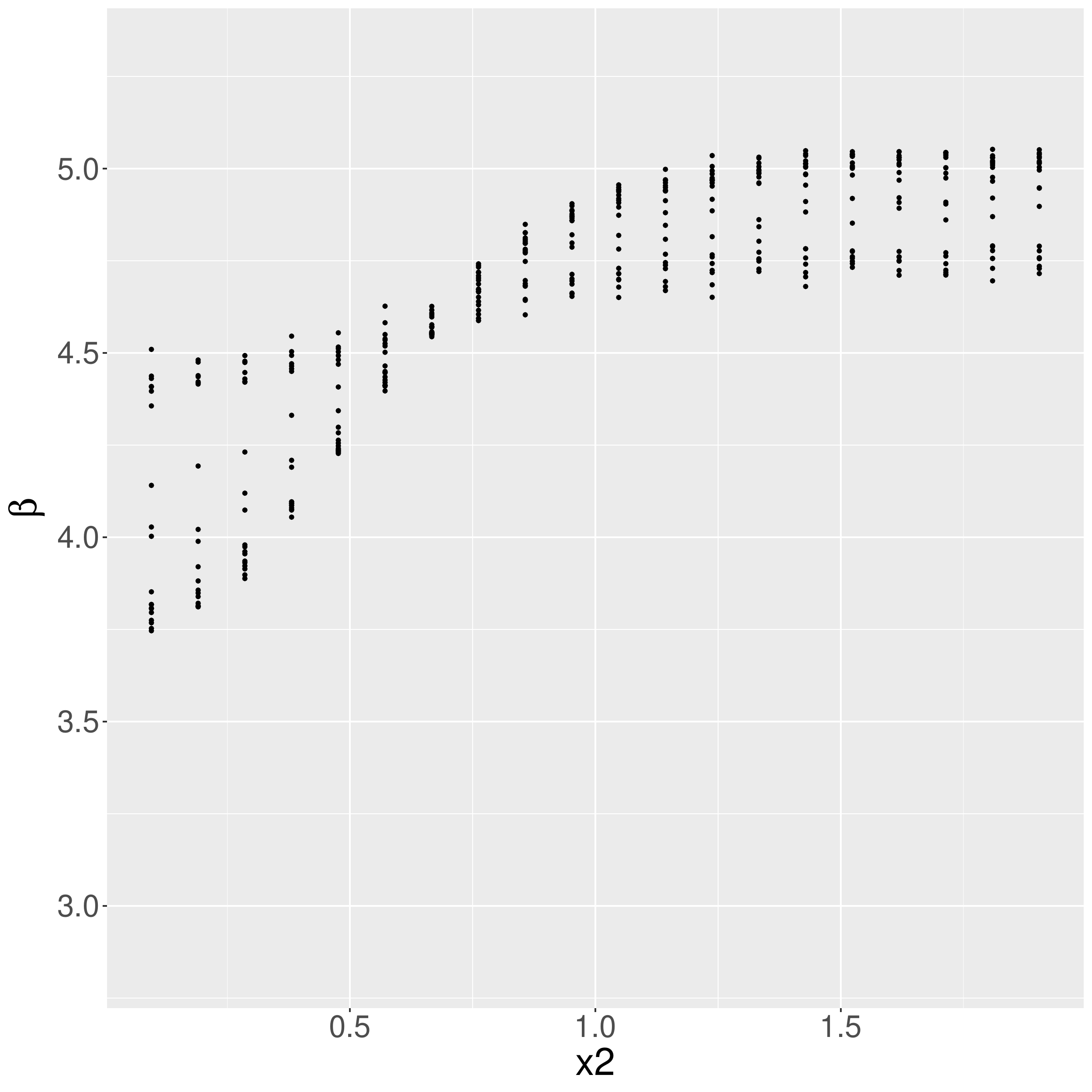}
\end{subfigure}\vfill

\begin{subfigure}[b]{0.25\textwidth}
	\centering
	\includegraphics[width=\textwidth]{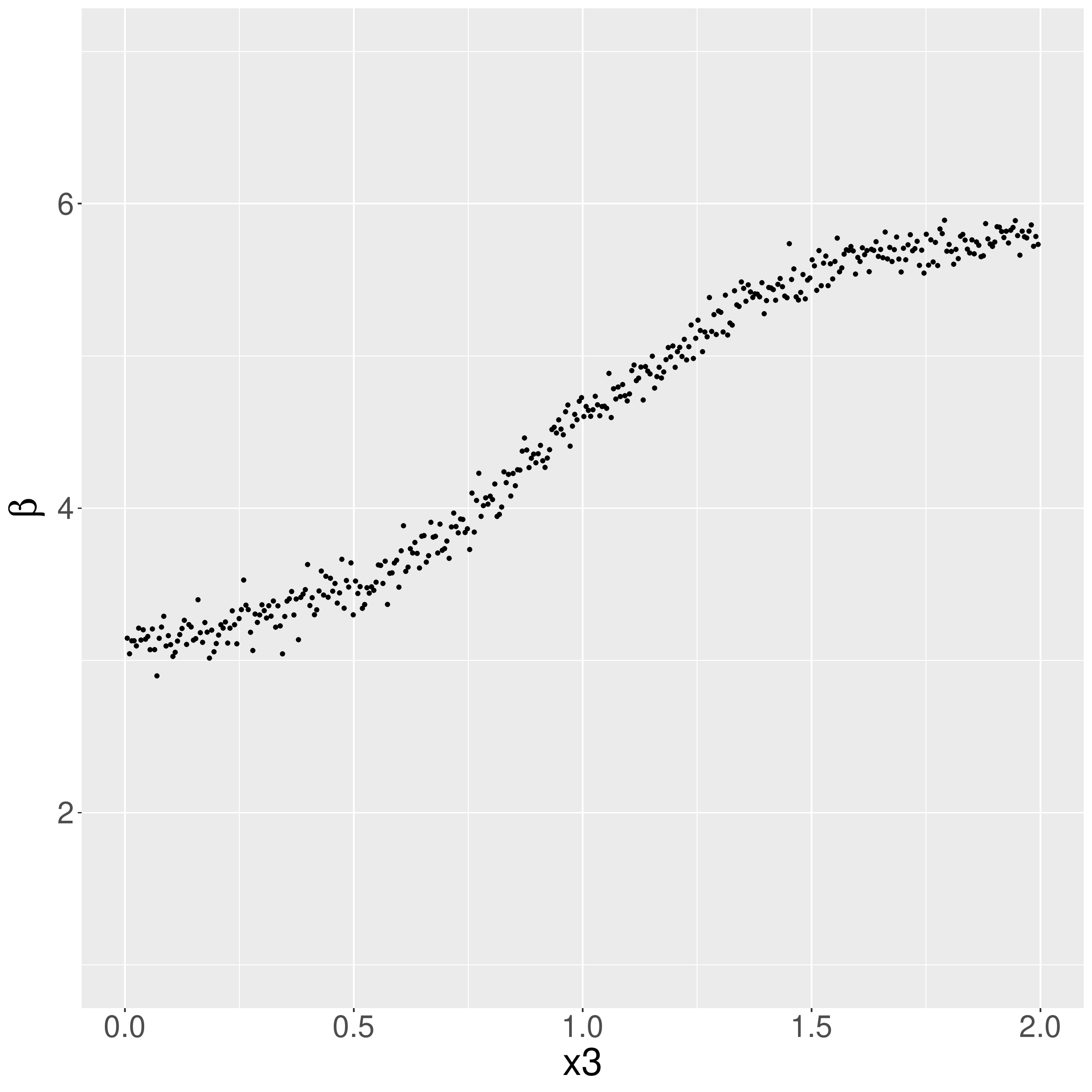}
\end{subfigure}
    \begin{subfigure}[b]{0.25\textwidth}
	\centering
	\includegraphics[width=\textwidth]{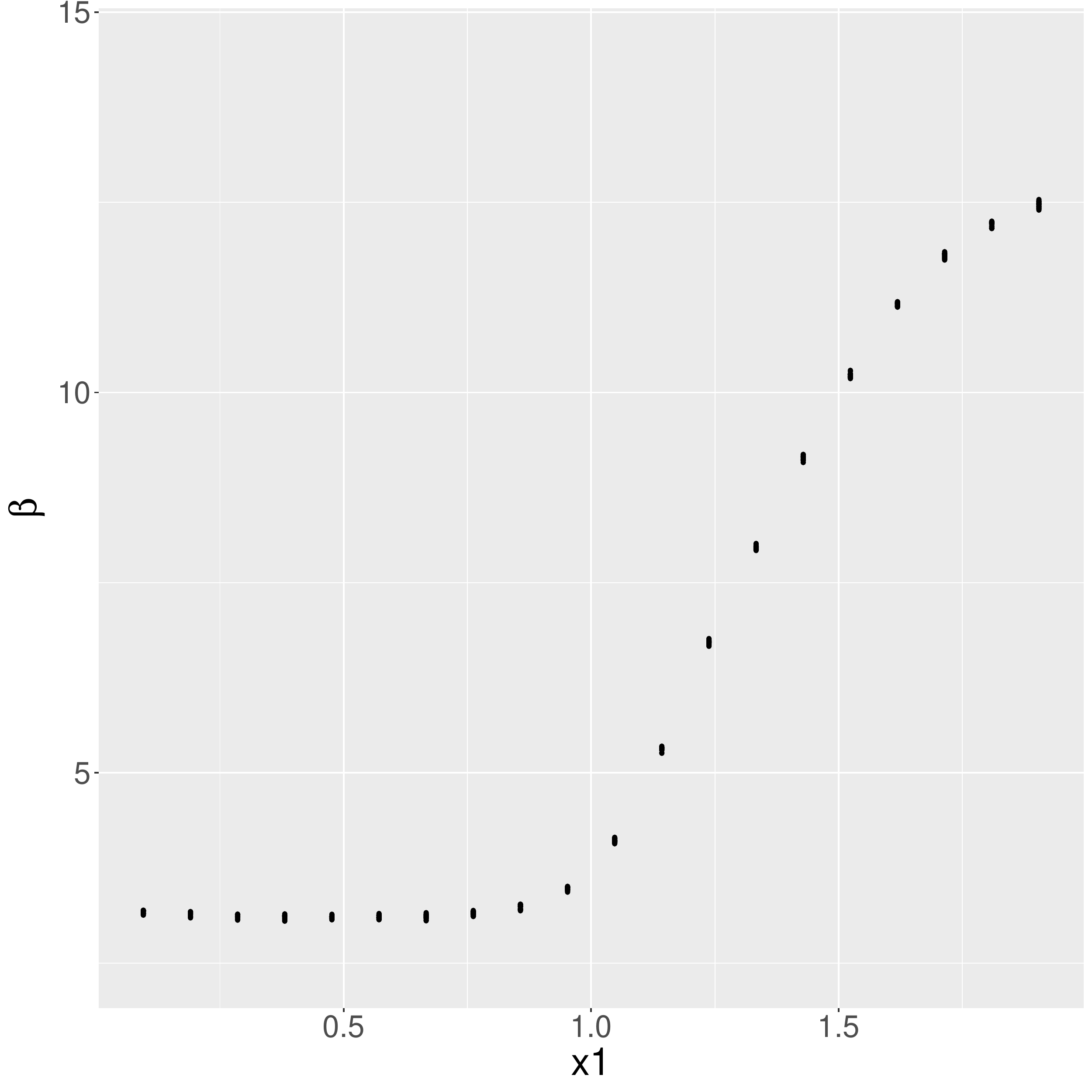}
\end{subfigure}
    \begin{subfigure}[b]{0.25\textwidth}
	\centering
	\includegraphics[width=\textwidth]{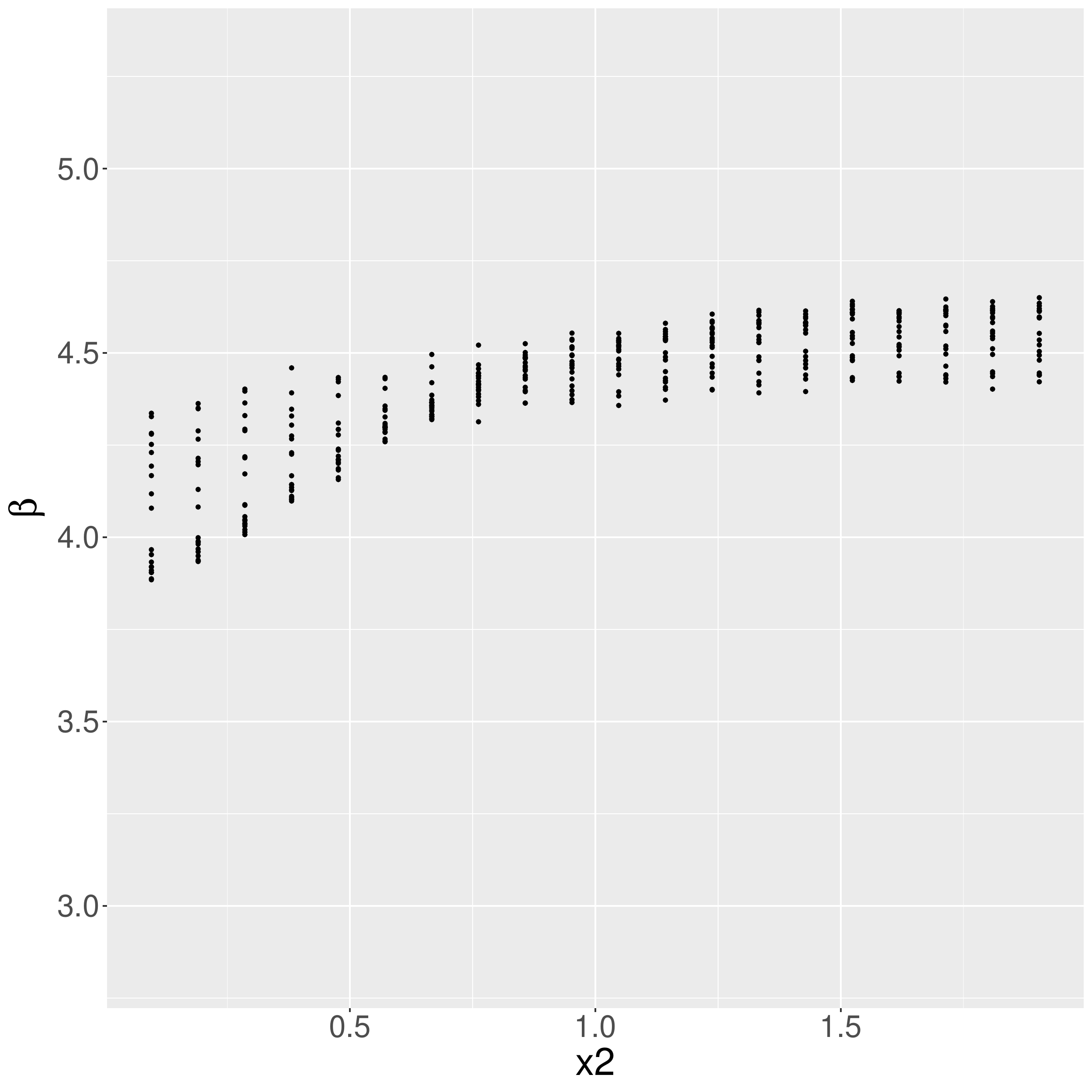}
\end{subfigure}
\end{figure}

\begin{example}\label{ex:cosine_continuous} Both $\bZ_i$ and  $\bX_i$ were generated randomly from $U(0,2)^{(p+2)}$ independently. The  coefficient functions (see the first row in Figure \ref{fig:estbeta_cosine_independent}) are
\begin{align*}
    &\beta_0(\bX_i) = 1+3X_{i3},\\
    &\bbeta_1(\bX_i) = \begin{pmatrix}
		\beta_1(\bX_i)\\
		\beta_2(\bX_i)
	\end{pmatrix} = 
	\begin{pmatrix}
	 10-7.5\cos\left(\frac{\pi}{2}(X_{i1}-0.5)\right)\\
	 0.5X_{i2}(3-X_{i2})+1
	\end{pmatrix}.
\end{align*}
 The error term $\varepsilon_i$ was generated from $N(0,1)$ and the censoring time $C_i$ was generated from $U(0,30)$.  %The set of modifiers $\bx_j^*,~j = 1,2,\ldots,N_2$ were generated randomly from $U(0,2)^p$ and $N_2 = 400$. 
\end{example}

The simulation results are summarized in Table \ref{tbl:cosine_continuous}, where HCQRF-c outperforms grf-c in estimating the quantile coefficient functions, and the estimation performance of HCQRF applied to the censored data is close to the benchmark HCQRF-c. We also demonstrate the average estimation of the coefficient function by different approaches in Figure \ref{fig:estbeta_cosine_independent}.  %and diminishes with the increase of the sample size.
%In this general case, when comparing HCQRF-c and grf-c, Table \ref{tbl:cosine_continuous} shows that the estimation performance of the quantile coefficient function by HCQRF-c is better than the one by grf-c. Table \ref{tbl:cosine_continuous} also shows that 

\begin{table}[!ht]
\centering
\caption{Estimation performance of the quantile coefficient function at $\tau = 0.5$ for \textbf{Scenario \ref{ex:cosine_continuous}} based on 500 simulation runs.}
	\label{tbl:cosine_continuous}
\begin{tabular}{cccccccc}
\hline\hline	
 &Method & $\beta_0$ & $\beta_1$&$\beta_2$ &$\beta_0$ & $\beta_1$&$\beta_2$\\
\hline
  && \multicolumn{3}{c}{$N_1 = 500$}&\multicolumn{3}{c}{$N_1 = 1000$}\\
  \cline{3-8}
\multirow{3}{*}{MSE}&HCQRF&0.977&1.106&0.694&0.666&0.814&0.552\\
& HCQRF-c &0.831&0.778&0.571&0.577&0.585&0.454\\
&grf-c &2.120&1.336&0.725&1.110&1.016&0.565\\
\cline{2-8}
\multirow{3}{*}{MAE}&HCQRF&0.769&0.811&0.628&0.632&0.707&0.563\\
&HCQRF-c&0.718&0.685&0.568&0.592&0.606&0.509\\
&grf-c&1.154&0.866&0.690&0.819&0.772&0.603\\
% \cline{2-8}
% \multirow{3}{*}{Relative MSE}&HCQRF&0.132&0.061&0.037&0.095&0.047&0.030\\
% & HCQRF-c &0.112&0.043&0.030&0.083&0.033&0.025\\
% &grf-c &0.284&0.075&0.038&0.160&0.058&0.031\\
% \cline{2-8}
% \multirow{3}{*}{Relative MAE}&HCQRF&0.244&0.169&0.142&0.203&0.149&0.128\\
% & HCQRF-c &0.227&0.143&0.128&0.191&0.128&0.116\\
% &grf-c &0.365&0.181&0.156&0.264&0.163&0.137\\
\hline\hline
\multicolumn{7}{l}{\footnotesize{MSE: mean squared error, MAE: mean absolute error}}
\end{tabular}
\end{table}

% \begin{table}[!ht]
% 	    \centering
% 	    \caption{Average MSE of the estimate quantile at $\tau = 0.5$ for \textbf{Scenario \ref{ex:cosine_continuous_new}}, based on  500 simulation runs.}
% 	    \label{tbl:cosine_continuousquantile}
% 	    \begin{tabular}{ccc}
% 	    \hline\hline 
% 	    Method & \multicolumn{2}{c}{$Q_\tau(T|\bx,\bz)$}\\
% 	    \hline
% 	 & $N_1 = 500$ & $N_1 = 1000$\\
% 	    \cline{2-3}
% 	      HCQRF &0.356&0.195\\
% 	     cqrf&2.082&1.711\\

% 	     \hline\hline
	     
% 	    \end{tabular}

% 	\end{table}

\subsection{Sensitivity Analysis}\label{sec:complicated}
%In \textbf{Scenario \ref{ex:cosine_continuous}}, we explored a general case where all the coefficient functions are continuous functions with $\bX$, the error term is normally generated and the censoring time is independently generated.
%In this subsection, we present a sensitivity analysis to evaluate the estimation performance when the proposed approach is applied to more complicated scenarios compared with the general case \textbf{Scenario \ref{ex:cosine_continuous}}. 
Under a set of modified \textbf{Scenario \ref{ex:cosine_continuous}} specified as below,  we also investigate the estimation performance under  various error distributions and censoring mechanisms (completely random vs. covariate-dependent), and also investigated estimation consistency across different quantile levels (see Supplementary Materials). %In what follows, we first introduced how we generate dependent and heavy-tailed error term, and covariate-dependent censoring time. 
\begin{itemize}
    \item[]\textbf{Scenario \ref{ex:cosine_continuous}a (Heterogeneous error term)}: The error term $\varepsilon_i$ was generated from $X_{2i}\xi_i/2$, where $\xi_i\sim N(0,1)$. The censoring time $C_i$ was generated from $U(0,60)$ resulting in about $25\%$ censoring.
    \item[] \textbf{Scenario \ref{ex:cosine_continuous}b (Heavy-tailed error term)}: The error term $\varepsilon_i$ was generated from a $t$ distribution with 2 degrees of freedom. The censoring time $C_i$ was generated from $U(0,60)$ resulting in about $25\%$ censoring.
    \item[] \textbf{Scenario \ref{ex:cosine_continuous}c (Covariate-Dependent Censoring)}: The censoring time was generated by \[C_i = -log(\xi_i)/\{0.017\exp(0.1X_{1i})\},\] where $\xi_i\sim U(0,1)$, resulting in about $25\%$ censoring.
%    \item[]\textbf{Heterogeneous quantile coefficients across quantile levels}
\end{itemize}

Table \ref{tbl:cosine_continuous_error} summarizes the resulting estimation accuracy under the modified \textbf{Scenario \ref{ex:cosine_continuous}}, which are consistent with earlier findings.  It suggests that the proposed HCQRF yields robust and stable estimations across various situations. In Supplementary Materials, we also present a modified setting in which quantile coefficient functions vary across quantile levels. Again, the proposed HCQRF delivers accurate estimations, and demonstrates advantages over other approaches.

\begin{table}[!ht]
\centering
\caption{Estimation performance of the quantile coefficient functions at $\tau = 0.5$ for \textbf{Scenario \ref{ex:cosine_continuous}} with heterogeneous error term,  heavy-tailed error term and covariate-dependent censoring time based on 500 simulation runs.}
	\label{tbl:cosine_continuous_error}
\begin{tabular}{ccccccccc}
\hline\hline
\textbf{Scenario} &&Method & $\beta_0$ & $\beta_1$&$\beta_2$ &$\beta_0$ & $\beta_1$&$\beta_2$\\
 \hline
   && &\multicolumn{3}{c}{$N_1 = 500$}&\multicolumn{3}{c}{$N_1 = 1000$}\\
   \cline{4-9}
     \multirow{3}{*}{\ref{ex:cosine_continuous}a}&\multirow{3}{*}{MSE}&HCQRF&0.771&1.052&0.678&0.458&0.658&0.430\\
     && HCQRF-c&0.638&0.752&0.541&0.391&0.471&0.347\\
     &&grf-c&1.588&1.268&0.706&1.009&0.846&0.496\\
     \cline{3-9}
     &\multirow{3}{*}{MAE}&HCQRF&0.663&0.795&0.627&0.518&0.635&0.493\\
     &&HCQRF-c& 0.612&0.677&0.559&0.482&0.541&0.443\\
     &&grf-c& 0.973&0.846&0.684&0.773&0.704&0.570\\
    %  \cline{3-9}
    %  &\multirow{3}{*}{Relative MSE}&HCQRF&0.105&0.044&0.045&0.075&0.033&0.029\\
    %  && HCQRF-c&0.087&0.037&0.037&0.063&0.027&0.024\\
    %  &&grf-c&0.214&0.036&0.042&0.163&0.026&0.031\\
    %  \cline{3-9}
    %  &\multirow{3}{*}{Relative MAE}&HCQRF&0.210&0.158&0.149&0.174&0.133&0.117\\
    %  && HCQRF-c&0.193&0.142&0.134&0.162&0.120&0.105\\
    %  &&grf-c&0.307&0.150&0.157&0.260&0.128&0.132\\
     \hline
 \multirow{3}{*}{\ref{ex:cosine_continuous}b}&\multirow{3}{*}{MSE}&HCQRF&1.370&1.487&1.008&0.906&0.965&0.691\\
     && HCQRF-c&1.162&1.070&0.825&0.779&0.705&0.563\\
     &&grf-c&1.921&1.639&0.863&1.230&1.080&0.622\\
     \cline{3-9}
     &\multirow{3}{*}{MAE}&HCQRF&0.896&0.948&0.772&0.732&0.771&0.630\\
     &&HCQRF-c&0.834&0.811&0.699&0.686&0.666&0.571\\
     &&grf-c& 1.080&0.965&0.749&0.863&0.796&0.634\\
    %  \cline{3-9}
    %  &\multirow{3}{*}{Relative MSE}&HCQRF&0.190&0.090&0.052&0.123&0.060&0.036\\
    %  && HCQRF-c&0.162&0.065&0.043&0.107&0.044&0.029\\
    %  &&grf-c&0.266&0.099&0.045& 0.169&0.067&0.032\\
    %  \cline{3-9}
    %  &\multirow{3}{*}{Relative MAE}&HCQRF&0.288&0.205&0.172&0.234&0.172&0.141\\
    %  && HCQRF-c&0.269&0.175&0.156& 0.219&0.149&0.128\\
    %  &&grf-c&0.348&0.209&0.167&0.276&0.178&0.142\\
     \hline
    \multirow{3}{*}{\ref{ex:cosine_continuous}c}&\multirow{3}{*}{MSE}&HCQRF&0.977&1.095&0.683&0.655&0.686&0.435\\
&&HCQRF-c&0.829&0.804&0.561&0.578&0.509&0.359\\
&&grf-c &2.324&1.435&0.736&1.622&0.971&0.520\\
\cline{3-9}
&\multirow{3}{*}{MAE}&HCQRF&0.773&0.811&0.617&0.653&0.643&0.483\\
&&HCQRF-c&0.724&0.696&0.555&0.622&0.555&0.437\\
&&grf-c&1.222&0.894&0.693&1.038&0.750&0.575\\
% \cline{3-9}
% &\multirow{3}{*}{Relative MSE}&HCQRF&0.134&0.066&0.036&0.088&0.039&0.024\\
% &&HCQRF-c&0.114&0.049&0.029&0.078&0.029&0.020\\
% &&grf-c &0.321&0.086&0.039&0.218&0.055&0.029\\
% \cline{3-9}
% &\multirow{3}{*}{Relative MAE}&HCQRF&0.249&0.174&0.138&0.206&0.135&0.110\\
% &&HCQRF-c&0.233&0.150&0.124&0.196&0.116&0.100\\
% &&grf-c &0.394&0.192&0.155&0.327&0.157&0.131\\
 \hline\hline
 \multicolumn{9}{l}{\footnotesize{MSE: mean squared error, MAE: mean absolute error}}
 \end{tabular}
 \end{table}

\subsection{Comparing HCQRF and CQRF in Estimating the Conditional Quantiles}\label{sec:condquantile}

As mentioned in the earlier section, CQRF cannot be directly applied to estimate the coefficient function $\bbeta_\tau(\bX)$ unless the $\bZ$ is binary. However, we can estimate the conditional quantile function of $T$ given $(\bx_0, \bz)$ from HCQRF by \[\widehat Q_\tau(T | \bX = \bx_0, \bZ = \bz) = \bz^T\widehat\bbeta_{\tau,\mathbb T}(\bx_0),\] where $\widehat \bbeta_{\tau,\mathbb T}(\bx_0)$ is the estimated coefficient at $\bx_0$ by HCQRF. Hence, we compared the  accuracy of the estimated $Q_\tau(T | \bX = \bx_0, \bZ = \bz)$ from the HCQRF and CQRF under the  outlined  scenarios above with continuous $\bZ$. The resulting MSE and MAE are reported in Table \ref{tbl:quantile}. The HCQRF outperforms the CQRF in all the simulation scenarios, and demonstrated robust performances under the complex settings \textbf{Scenario \ref{ex:cosine_continuous}a, b, c}.

%CQRF is an idea estimation approach to capture the heterogeneity of the conditional quantile if it is a tree-structured function of $\bX$ and $\bZ$ like what is in \textbf{Scenario \ref{ex:tree_binary}}. Therefore the estimation performance of both the estimated conditional quantiles and the estimated coefficient functions of CQRF is better than the one of HCQRF. Both MSE and MAE of HCQRF diminishes with the increase of the sample size. By comparing the estimation performance of \textbf{Scenario \ref{ex:cosine_continuous}a, b, c} and \textbf{Scenario \ref{ex:cosine_continuous_scale}} with the one of \textbf{Scenario \ref{ex:cosine_continuous}}, it is obvious that the results are not sensitive to the comlexity of the simulation settings. 

\begin{table}[!ht]
\begin{center}
\begin{threeparttable}
\caption{Estimation performance of the conditional quantile function based on 500 simulation runs.}
	    \label{tbl:quantile}
	    \begin{tabular}{ccccccc}
	    \hline\hline 
	      Scenario &$\tau$&Method & \multicolumn{2}{c}{$N_1 = 500$}&\multicolumn{2}{c}{$N_1 = 1000$}\\
	      \cline{4-7}
	     &&& MSE&MAE&MSE&MAE\\
	    \cline{4-7}
% \ref{ex:tree_binary}&
% 0.5&HCQRF&0.408&0.146&0.112&0.074\\
% && CQRF&0.289&0.128&0.056&0.053\\
% \cline{3-7}
 \ref{ex:boosting}&0.5
&HCQRF&0.531&0.262&0.373&0.194\\
&& CQRF&3.994&1.626&4.001&1.631\\
\cline{3-7}
\ref{ex:cosine_continuous}&0.5&HCQRF &2.076&1.098&1.424&0.899\\
	     &&CQRF&29.25&4.185&29.194&4.172\\
	     \cline{3-7}
	     \ref{ex:cosine_continuous}a&0.5& HCQRF&1.834&1.012&1.160&0.805\\
	     &&CQRF &28.805&4.155&28.262&4.114\\
	     \cline{3-7}
	     \ref{ex:cosine_continuous}b&0.5& HCQRF&2.516&1.196&1.639&0.968\\
	     && CQRF&30.138&4.236&29.226&4.173\\

	     \cline{3-7}
	      \ref{ex:cosine_continuous}c&0.5& HCQRF&2.144&1.121&1.465&0.932\\
	     && CQRF&29.972&4.233&29.598&4.204\\
	    \cline{3-7}
	    \ref{supex:cosine_continuous_scale} &0.25&HCQRF&2.453&1.213&1.713&1.026\\
	    &&CQRF&30.036&4.243&29.629&4.208\\
	    \cline{3-7}
	   \ref{supex:cosine_continuous_scale}&0.5&HCQRF&2.186&1.137&1.520&0.956\\
	    &&CQRF&30.14&4.25&29.779&4.219\\
	     \cline{3-7}
	   \ref{supex:cosine_continuous_scale}&0.75&HCQRF&2.208&1.116&1.475&0.918\\
	   &&CQRF&30.445&4.275&30.058&4.244\\
	     \hline\hline
	    \end{tabular}
\begin{tablenotes}\footnotesize
\item MSE: mean squared error, MAE: mean absolute error
\end{tablenotes}
	\end{threeparttable}
	\end{center}
\end{table}

\subsection{Variable Importance}\label{sec:vi}
\begin{figure}[!ht]
    \centering    \caption{Box plots of the variable importance based on $500$ Monte Carlo repetitions for \textbf{Scenario \ref{ex:tree_binary}} with $N_1 = 1000$ and $p = 30$.}
    \label{fig:VarImp_sim_tree_binary}
    \begin{subfigure}[b]{0.5\textwidth}
    \centering
\caption{Total variable importance}
\label{fig:VarImp_total_sim_tree_binary}
\includegraphics[width = \textwidth]{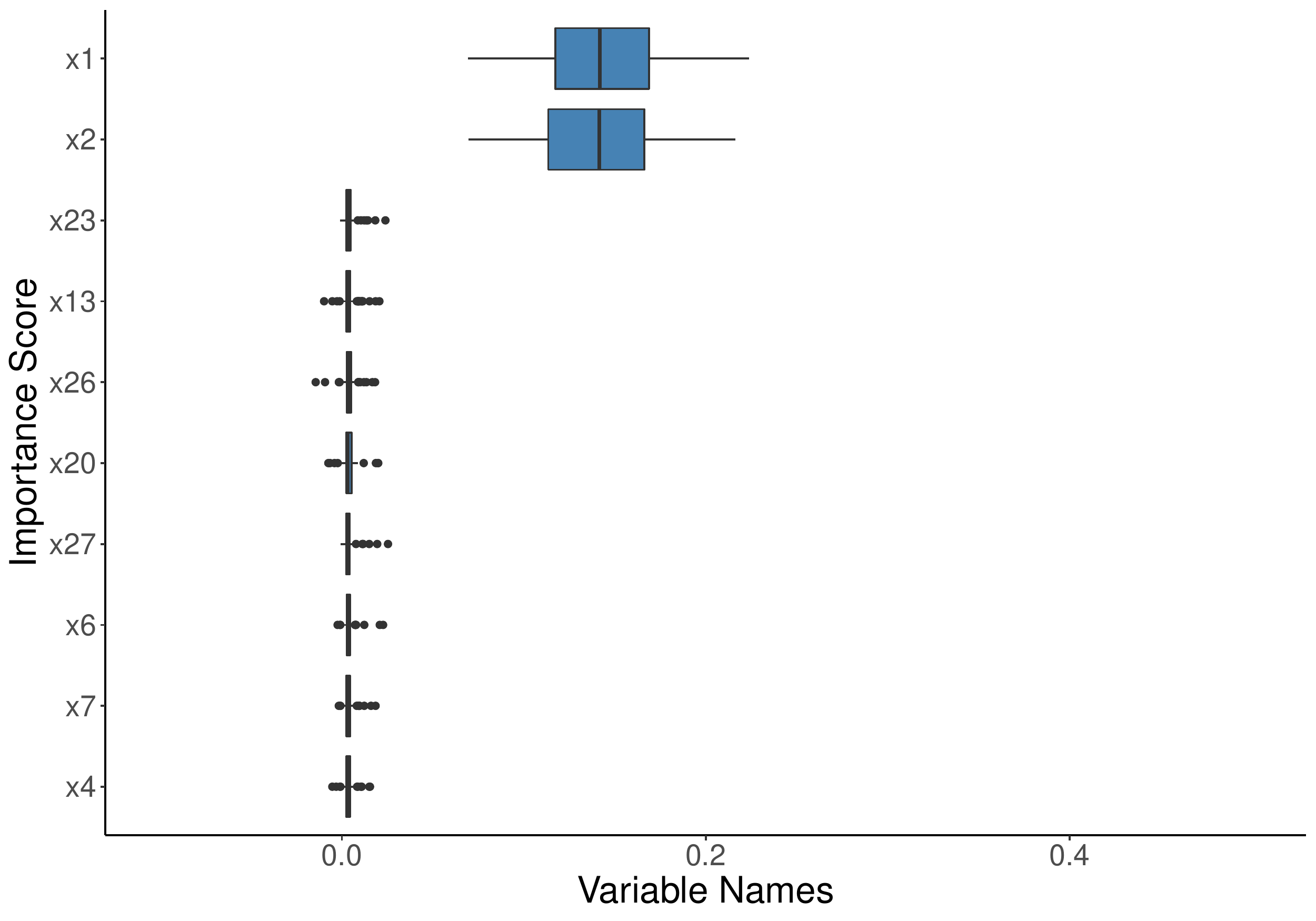}
\end{subfigure}\hfill
\begin{subfigure}[b]{0.5\textwidth}
\caption{Variable importance of $\beta_0(\bX)$}
\label{fig:VarImp_main_sim_tree_binary}
\includegraphics[width = \textwidth]{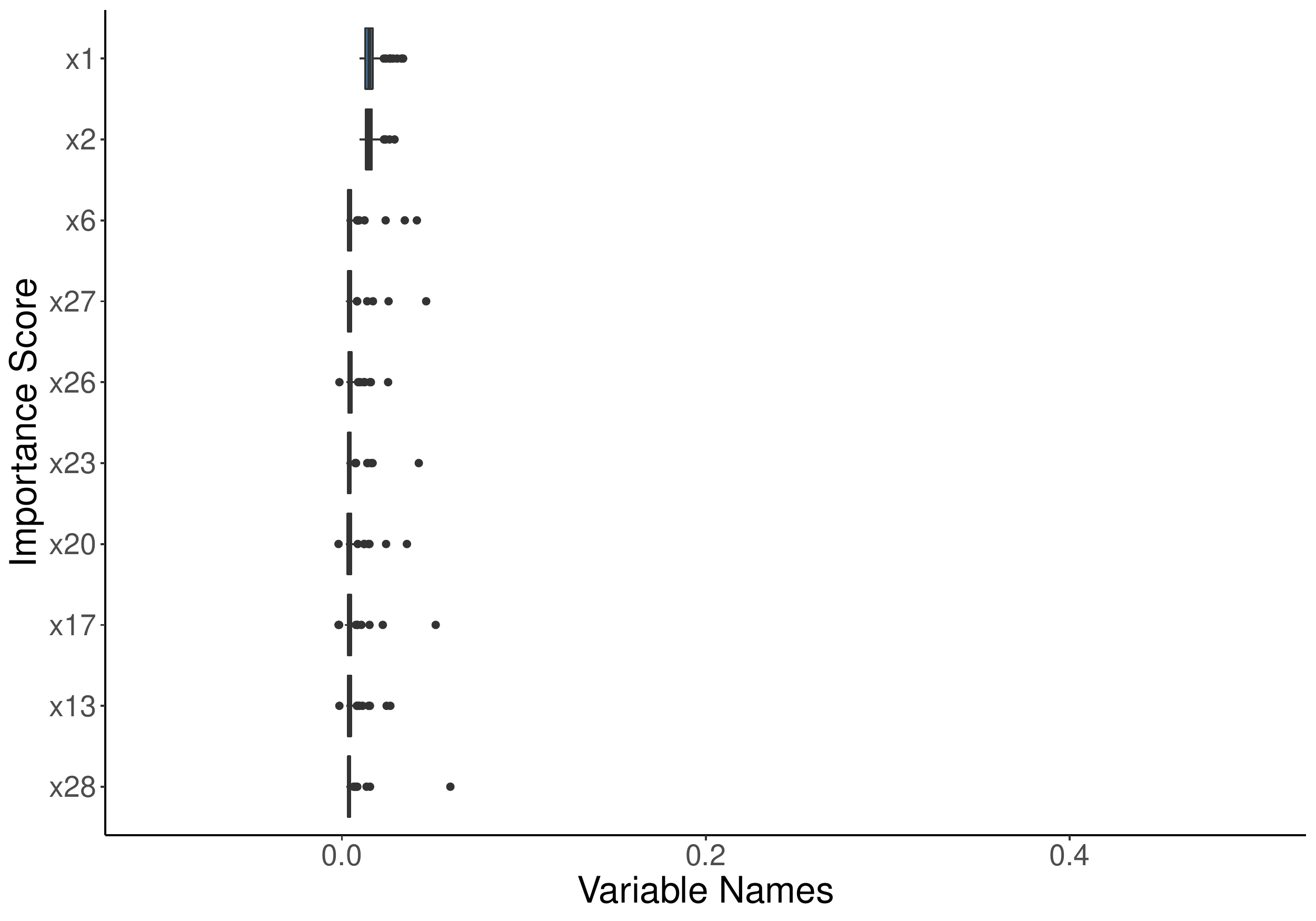}
\end{subfigure}\vfill
\begin{subfigure}[b]{0.5\textwidth}
\caption{Variable importance of $\beta_1(\bX)$}
\label{fig:VarImp_int_sim_tree_binary}
\includegraphics[width = \textwidth]{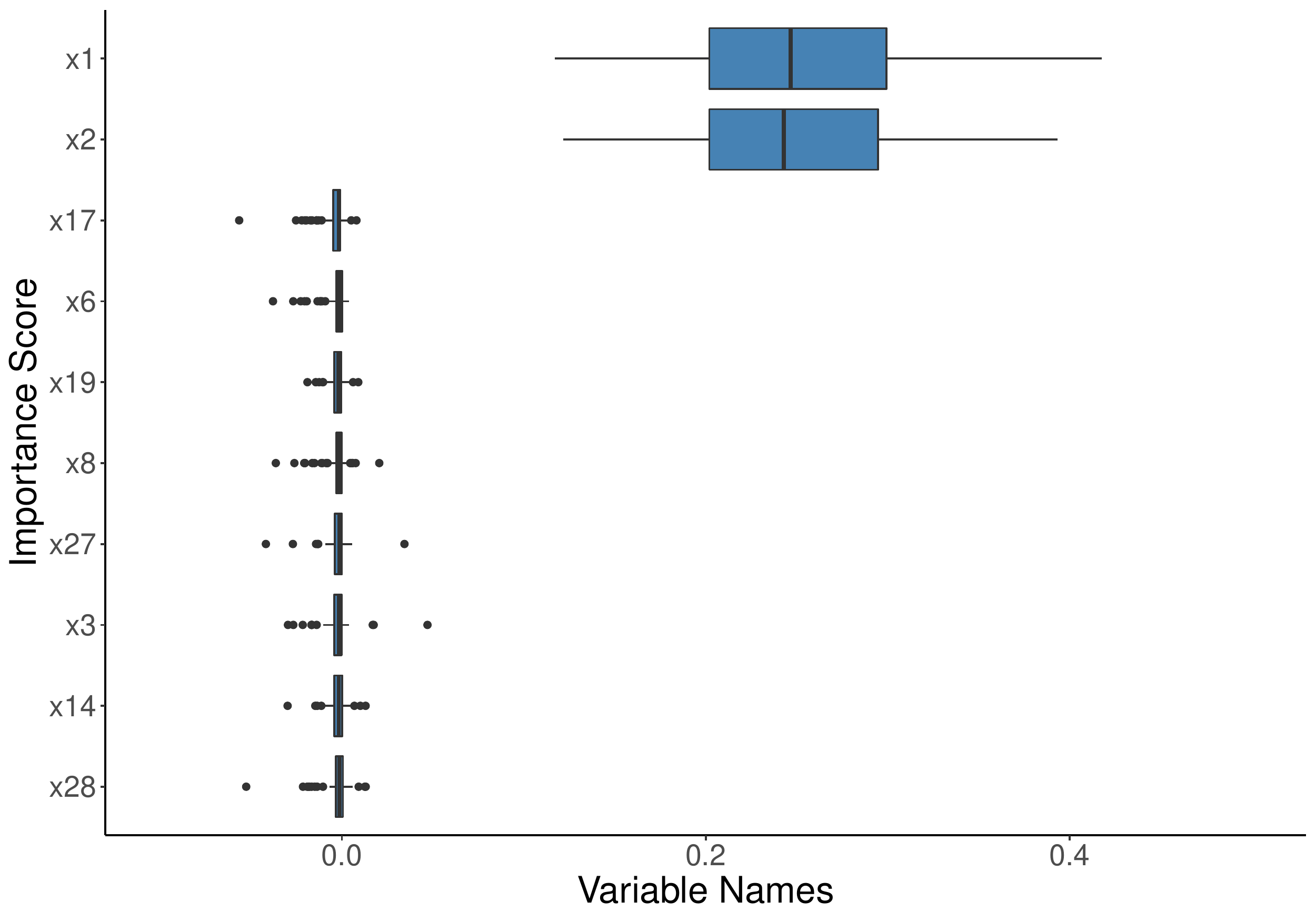}
\end{subfigure}
\end{figure}

In this subsection, we examine the variable importance decomposition proposed in Section \ref{sec:VarImp} using \textbf{Scenario \ref{ex:tree_binary}}, where the predictive variable $\bZ$ is a binary variable. In  \textbf{Scenario \ref{ex:tree_binary}}, $\beta_0(\bX)$ is a constant and $\beta_1(\bX)$ depends on $X_1$ and $X_2$. 
% \begin{example}\label{ex:varimp}
% We generated data $(y_i,\delta_i,\bx_i,\bz_i)_{i = 1}^{N_1}$ from the model \eqref{eq:simmodel} with
% \begin{align}\label{eq:exvarimp}
%     &\beta_0(\bx_i) = 6-x_{i1}+x_{i2},\nonumber\\
%     &\beta_1(\bx_i) = x_{i1}-0.3,
% \end{align} \blue{delete or move to the supplement}
% with $\varepsilon_i\sim \chi^2_2$.  The  predictive variable $\bz_i$ and the modifiers $\bx_i$ were generated the same as in \textbf{Scenario \ref{ex:tree_binary}}. The censoring time $c_i$ was generated from $U(0,38)$, resulting in $25\%$ censoring. The set of modifiers $\bx_j^*,~{j = 1, \ldots, N_2}$ was generated the same as in \textbf{Scenario \ref{ex:tree_binary}}.
% \end{example}
The result of the variable importance decomposition based on \textbf{Scenario \ref{ex:tree_binary}} is reported in Figure \ref{fig:VarImp_sim_tree_binary} at $\tau = 0.5$. In the results, we demonstrate the importance score of $\beta_{\tau,1}(\bX)$ by $|\text{VarImp}_{Z = 1,l,\tau}-\text{VarImp}_{Z = 0,l,\tau}|$.
Figure \ref{fig:VarImp_sim_tree_binary} clearly shows that the importance scores of $X_1$ and $X_2$ are close to 0 for $\beta_{\tau,0}(\bX)$ while they are away from 0 for $\beta_{\tau,1}(\bX)$. The importance plots show that the proposed variable importance decomposition distinguishes the different dependence structures of $\beta_0(\bX)$ and $\beta_1(\bX)$. 
% In \textbf{Scenario \ref{ex:varimp}}, it is clear from \eqref{eq:exvarimp} that when $Z =0$, $X_1$ has a negative effect on the conditional quantile. When $Z = 1$, the effect of $X_1$ is canceled out.
% Figure \ref{fig:VarImp_sim} depicts the results of variable importance based on \textbf{Scenario \ref{ex:varimp}}. It can be seen that both $X_1$ and $X_2$ are important in $\beta_0(\bX)$; moreover, $X_1$ is important in $\beta_1(\bX)$ and has negative variable importance in general. This is as expected, because the situation in \textbf{Scenario \ref{ex:varimp}} is one of the possible explanation of the negative value of $\text{VarImp}_{Z = 1,l,\tau}-\text{VarImp}_{Z = 0,l,\tau}$. 

\section{Real Data}\label{sec:realdata}
We applied both  HCQRF and CQRF to a randomized phase III clinical trial which aims to compare the efficacy of fluorouracil, leucovorin, and irinotecan (FOLFIRI) alone versus panitumumab plus FOLFIRI in patients who were previously treated for metastatic colorectal cancer \citep{peeters2014final}. In this trial, patients were randomly assigned to one of the two treatments, panitumumab plus FOLFIRI or FOLFIRI alone. The primary endpoint is progression-free survival (PFS). Besides estimating their treatment effects on PFS, we are also interested in identifying patient characteristics that maybe modify the treatment effects. Such knowledge help design more targeted treatment plans to improve outcomes. %For example, KRAS is a group of genes involved in the epidermal growth factor receptor pathway. In practice, KRAS mutations were found to be predictive of the efficacy of anti-epidermal growth factor receptor therapies\citep{peeters2014final}, including panitumumab, and hence they may modify the treatment effect.

% The original dataset consists of 946 patients. \blue{I don't think we need to mention the original sample size.}
The study sample includes 830 patients, among those, 783 patients progressed after their treatments (i.e., PFS observed). The main variable of interest $\bZ$ is 
the treatment assignment (FOLFIRI alone = 0, Panitumumab + FOLFIRI = 1). 
The FOLFIRI-alone arm consists of 413 patients %with 229 wild-type KRAS and 184 mutant KRAS, 
while the panitumumab plus FOLFIRI arm has 417 patients. %with 234 wild-type KRAS and 183 mutant KRAS.
We also consider ten covariates as potential effect modifiers, including age (in years) and lactate dehydrogenase value at baseline (LDH, $\times$ upper limit of normal), prior mCRC Bevacizumab use (yes = 1, no = 0), prior mCRC Oxaliplatin Exposure (yes = 1, no = 0), KRAS (wild-type = 1, mutant = 2), prim tumor type (rectum = 0, colon = 1), sex (male = 0, female = 1),  race (white or Caucasian = 1, others = 0), the number of baseline metastatic sites (from 1 to 8), and eastern cooperative oncology group at baseline (ECOG,0, 1, and $\ge2$). We randomly split the data into a training set with 80\% data points(n = 664), and a test set with the remaining 166 observations. We construct both HCQRF and CQRF from the training data, following the outlined algorithms in Section \ref{sec:alg}.
The hyperparameters in the forest models are the same as the specifications in Section \ref{sec:compspec}. 

\subsection{Variable Importance}
We calculate the variable importance from  the constructed HCQRF and CQRF, respectively.   For HCQRF, we follow the algorithm in Section \ref{sec:VarImp} to calculate the variable importance (VI) with $100$ permutations, and future decompose the total VI into VI for $\beta_{\tau,0}(\bX)$  (main effect) and 
VI for $\beta_{\tau,1}(\bX)$ (interactive effects). We note that VI of CQRF was not officially developed in \cite{li2020censored}. As CQRF follows the scheme of generalized random forest \citep{athey2019generalized}, we apply the default variable importance in the \texttt{grf} package \citep{grfPcakge}, which is a weighted sum of how many times a variable was split at each depth in the forest. We set the maximal depth to be 4.

\begin{figure}[!ht]
    \centering
    \caption{Variable importance of $10$ modifiers in the data set at $0.5,~0.25,~0.75$ quantile level by the proposed variable importance decomposition. The first row presents the variable importance for $\tau = 0.5$. The second row presents the variable importance for $\tau = 0.25$. The third row presents the variable importance for $\tau = 0.75$.}
    \label{fig:VarImp}
    \begin{subfigure}[b]{0.33\textwidth}
    \centering
\caption{Total}
\label{fig:VarImp_total0.5}
\includegraphics[width = \textwidth]{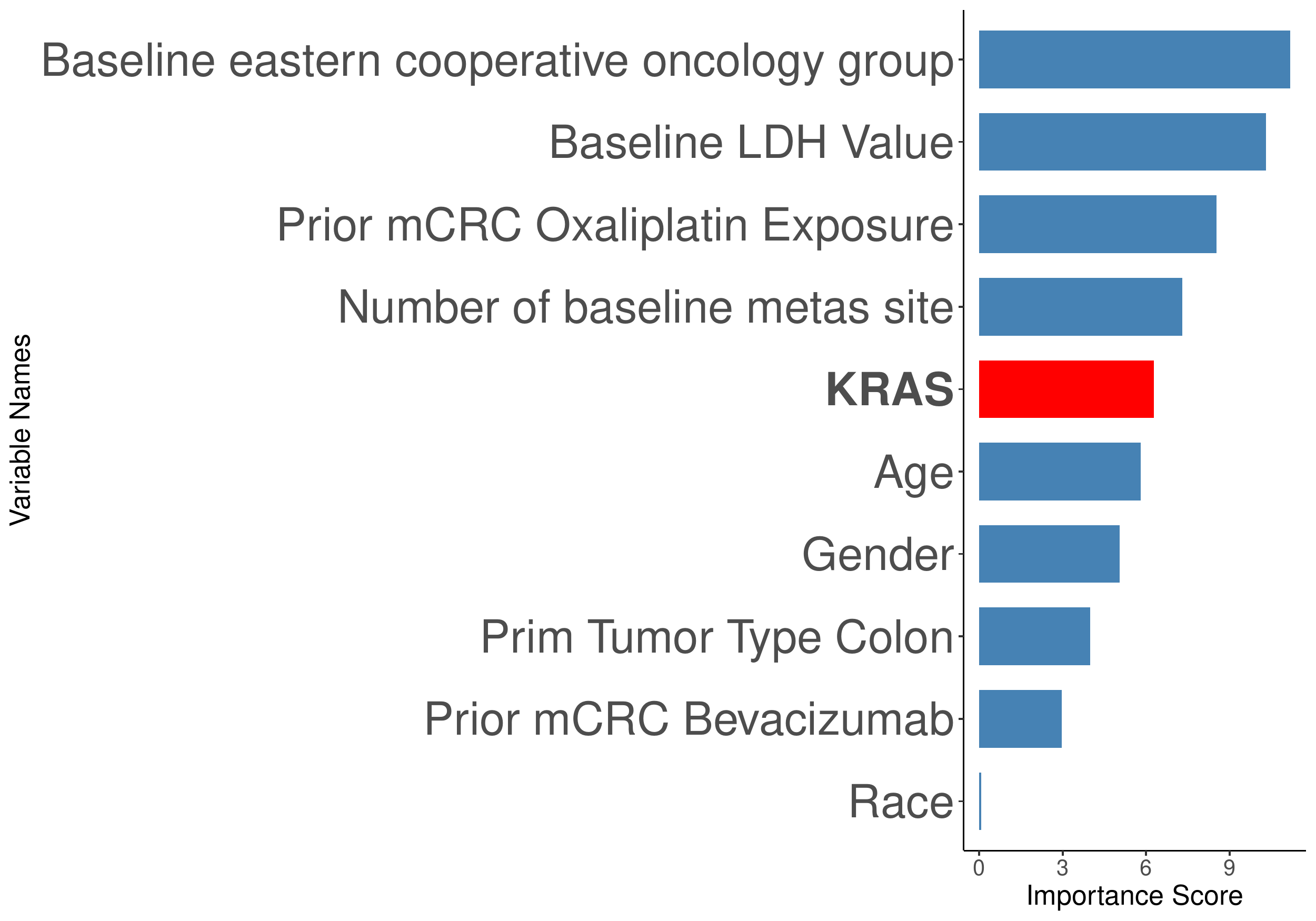}
\end{subfigure}\hfill
\begin{subfigure}[b]{0.33\textwidth}
\caption{The effect of $\beta_{\tau,0}(\bX)$}
\label{fig:VarImp_main0.5}
\includegraphics[width = \textwidth]{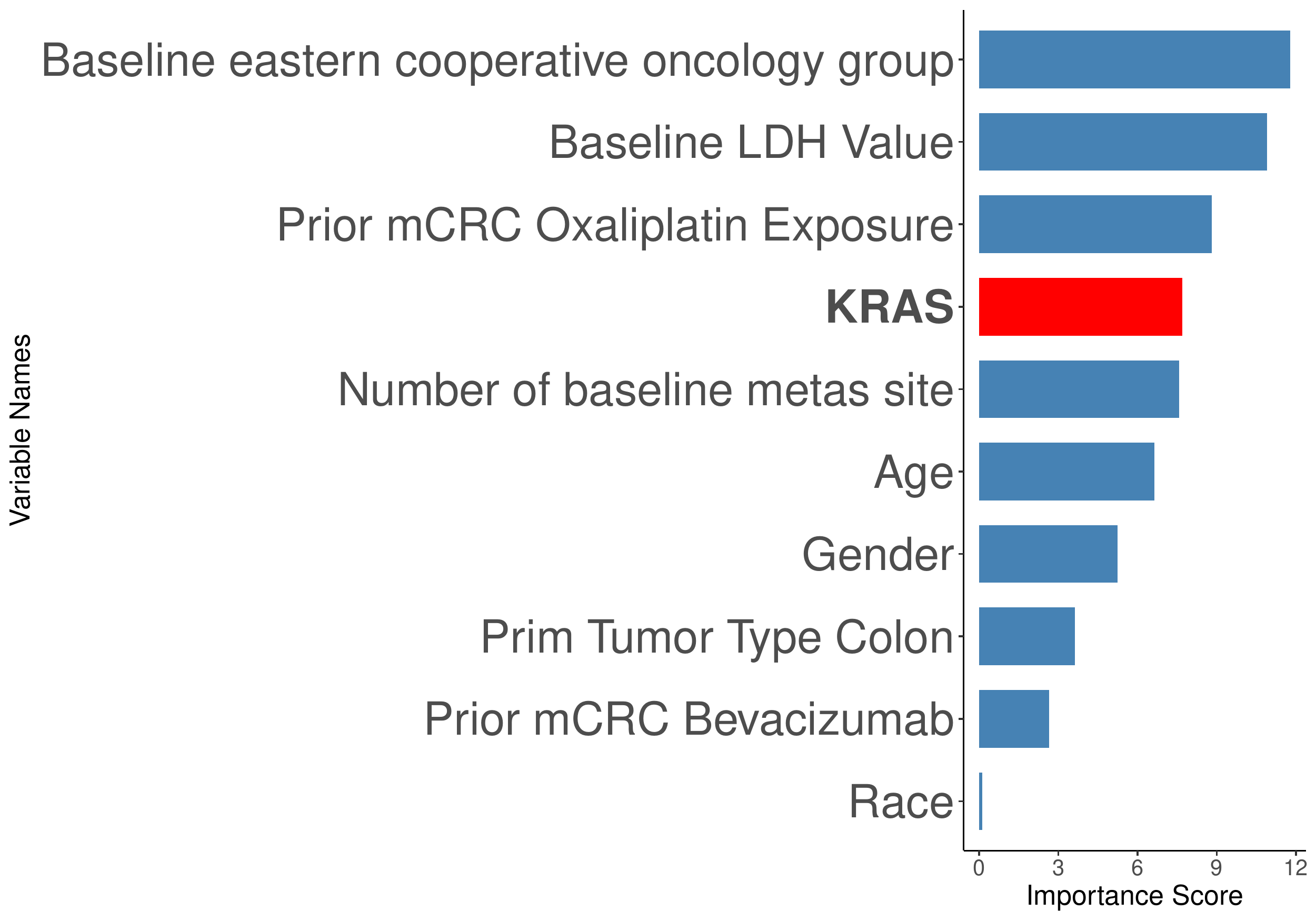}
\end{subfigure}\hfill
\begin{subfigure}[b]{0.33\textwidth}
\caption{The effect of $\beta_{\tau,1}(\bX)$}
\label{fig:VarImp_int0.5}
\includegraphics[width = \textwidth]{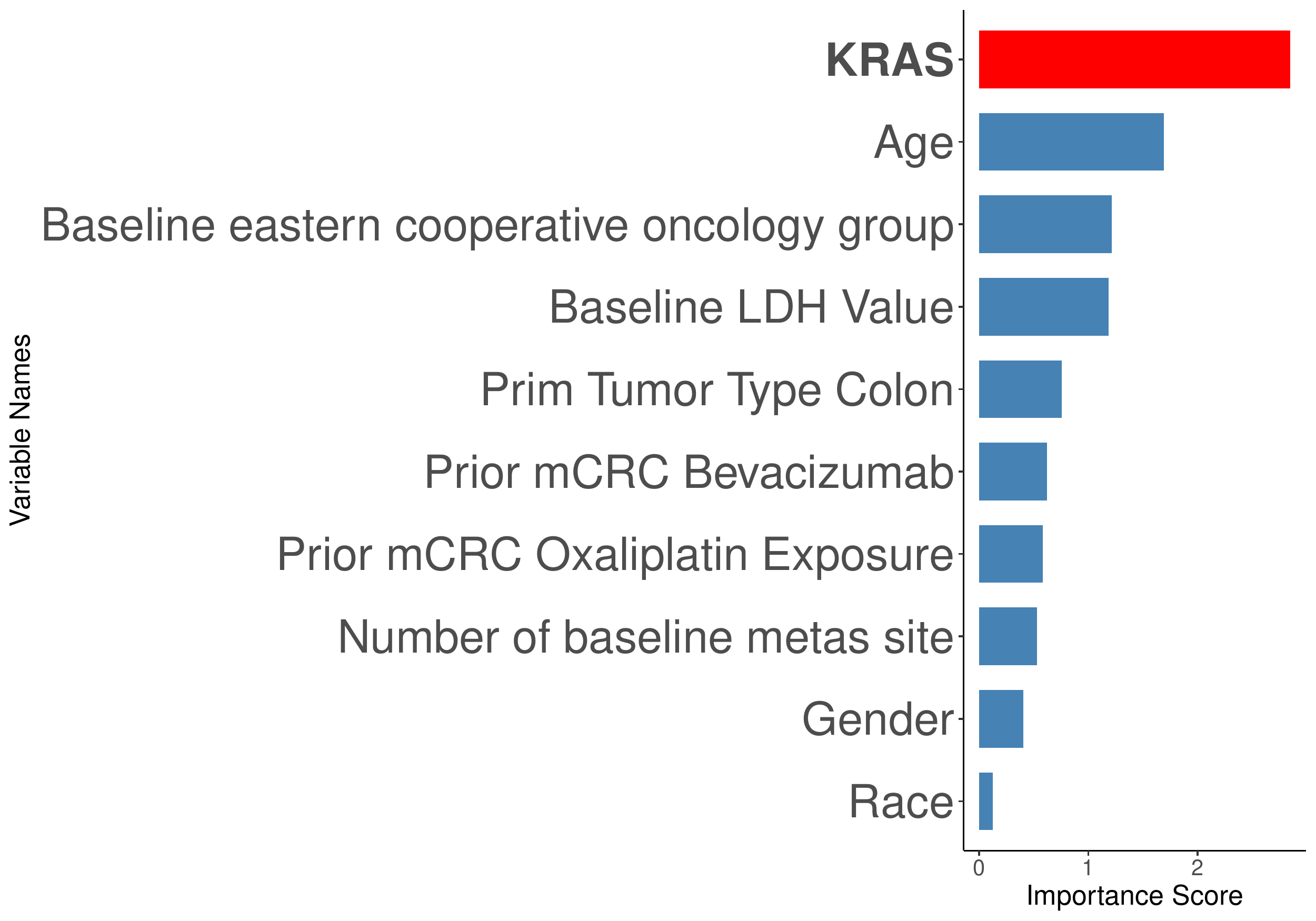}
\end{subfigure}\vfill
\begin{subfigure}[b]{0.33\textwidth}
    \centering
% \caption{Total variable importance}
% \label{fig:VarImp_total0.25}
\includegraphics[width = \textwidth]{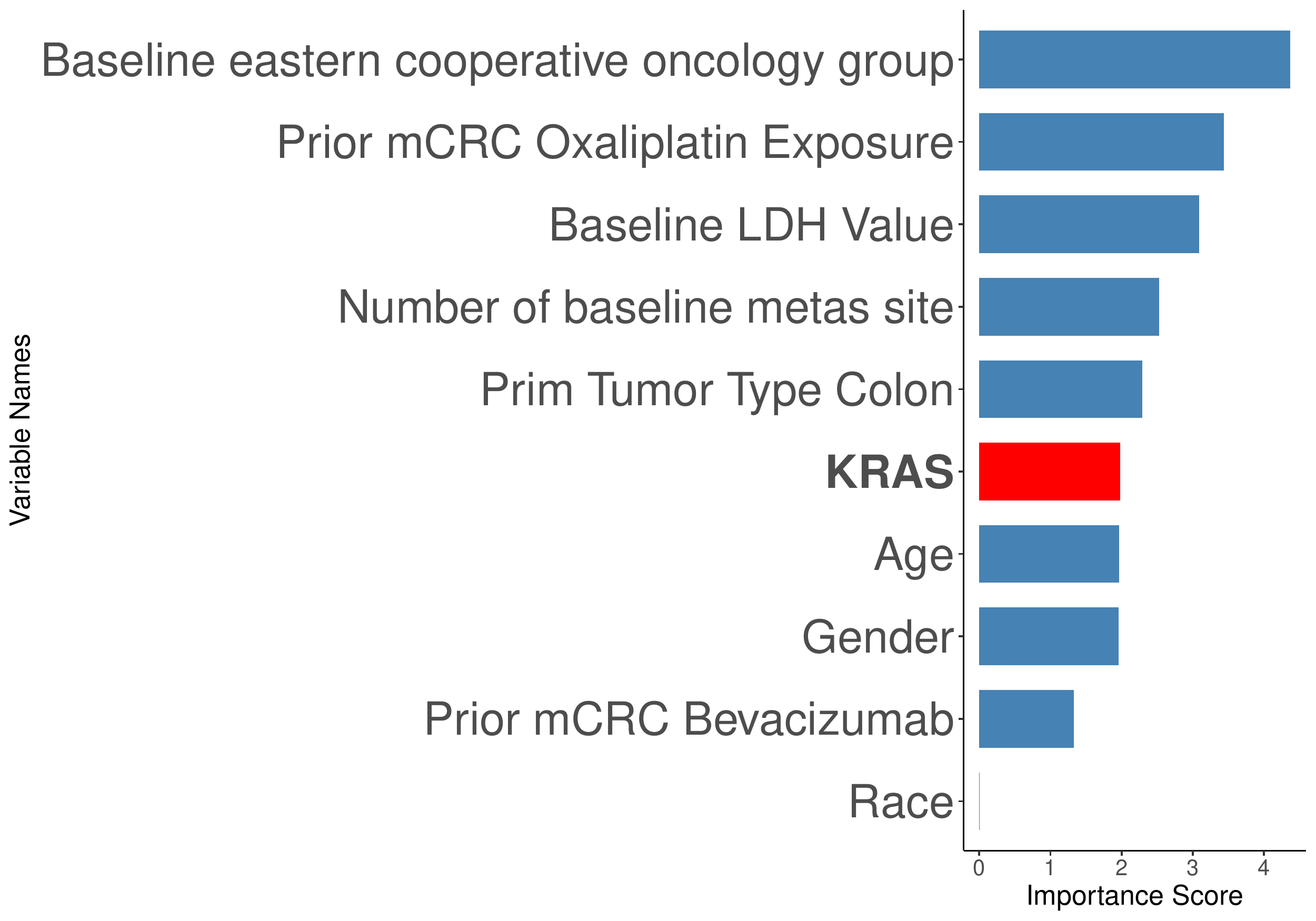}
\end{subfigure}\hfill
\begin{subfigure}[b]{0.33\textwidth}
% \caption{Variable importance of the main effect}
% \label{fig:VarImp_main0.25}
\includegraphics[width = \textwidth]{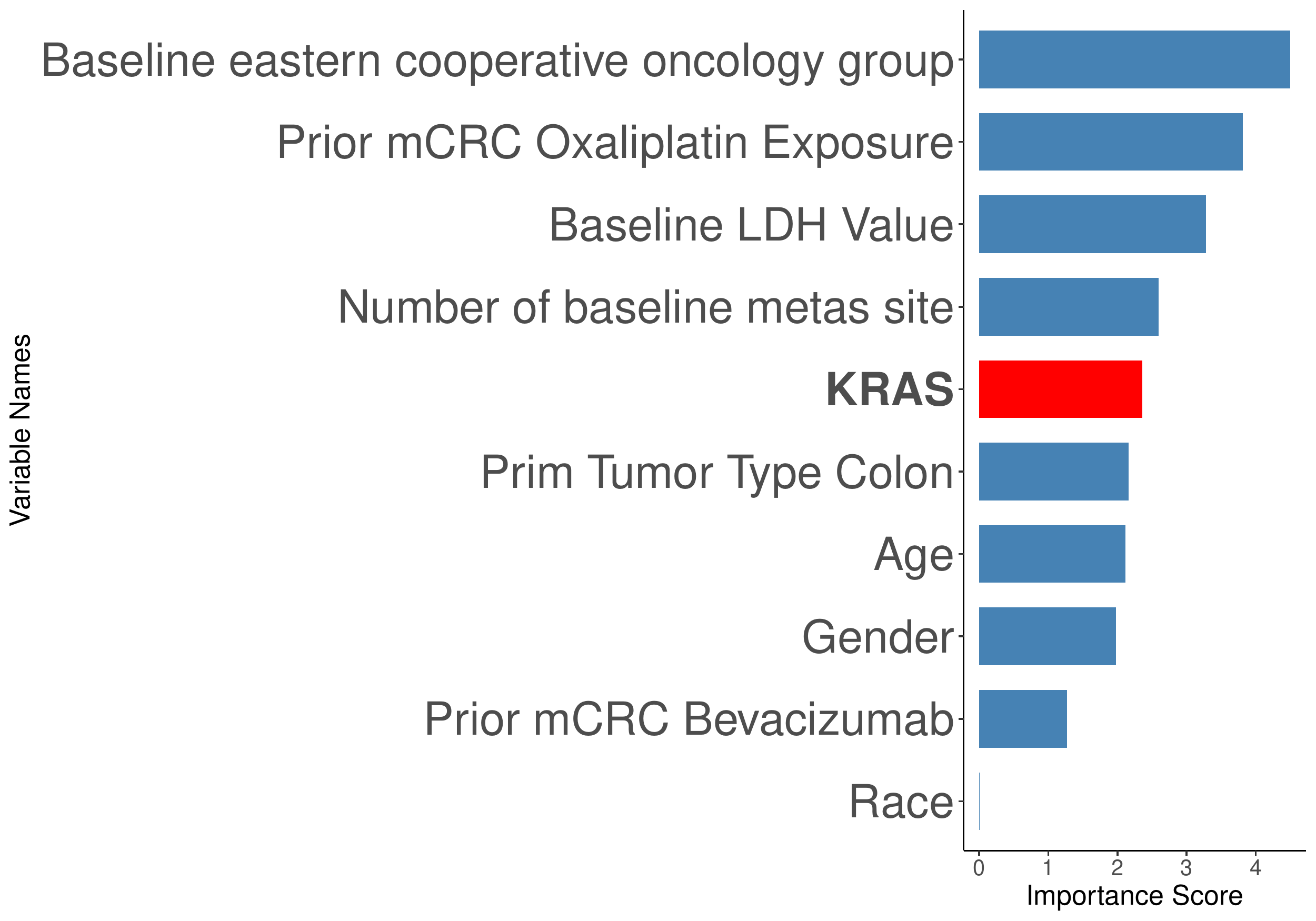}
\end{subfigure}\hfill
\begin{subfigure}[b]{0.33\textwidth}
% \caption{Variable importance of the treatment effect}
% \label{fig:VarImp_int0.25}
\includegraphics[width = \textwidth]{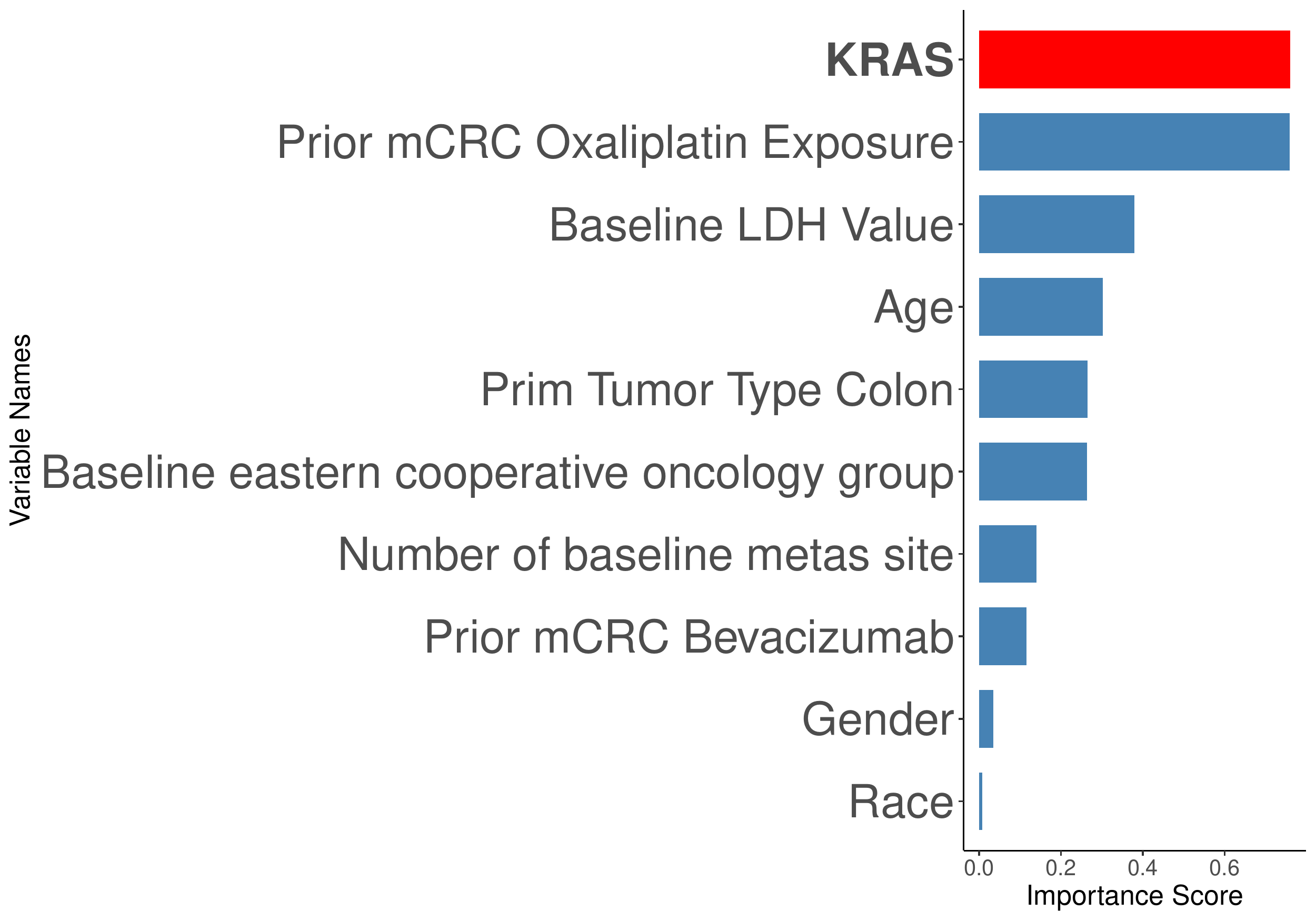}
\end{subfigure}
 \begin{subfigure}[b]{0.33\textwidth}
    \centering
% \caption{Total variable importance}
% \label{fig:VarImp_total0.75}
\includegraphics[width = \textwidth]{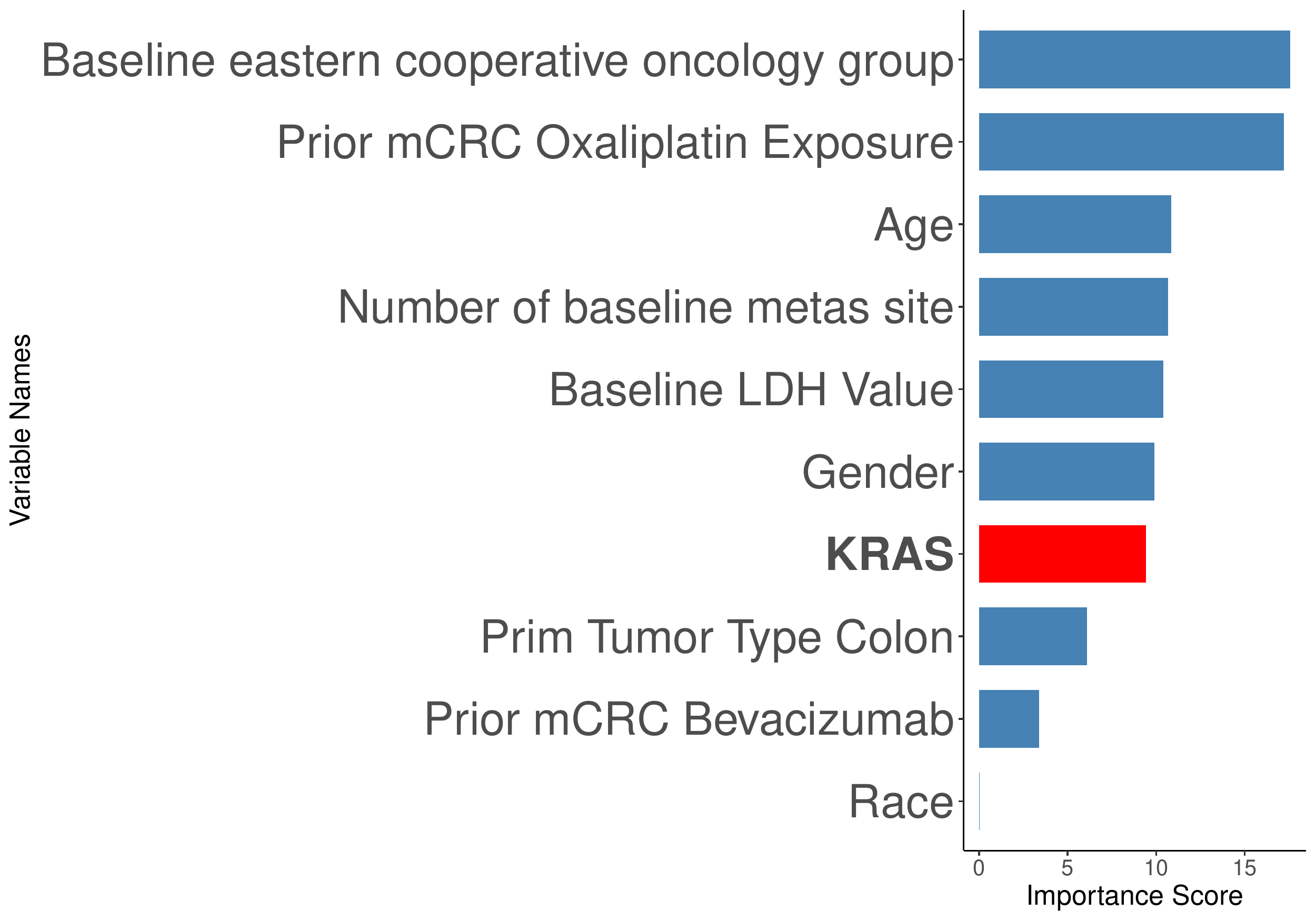}
\end{subfigure}\hfill
\begin{subfigure}[b]{0.33\textwidth}
% \caption{Variable importance of the main effect}
% \label{fig:VarImp_main0.75}
\includegraphics[width = \textwidth]{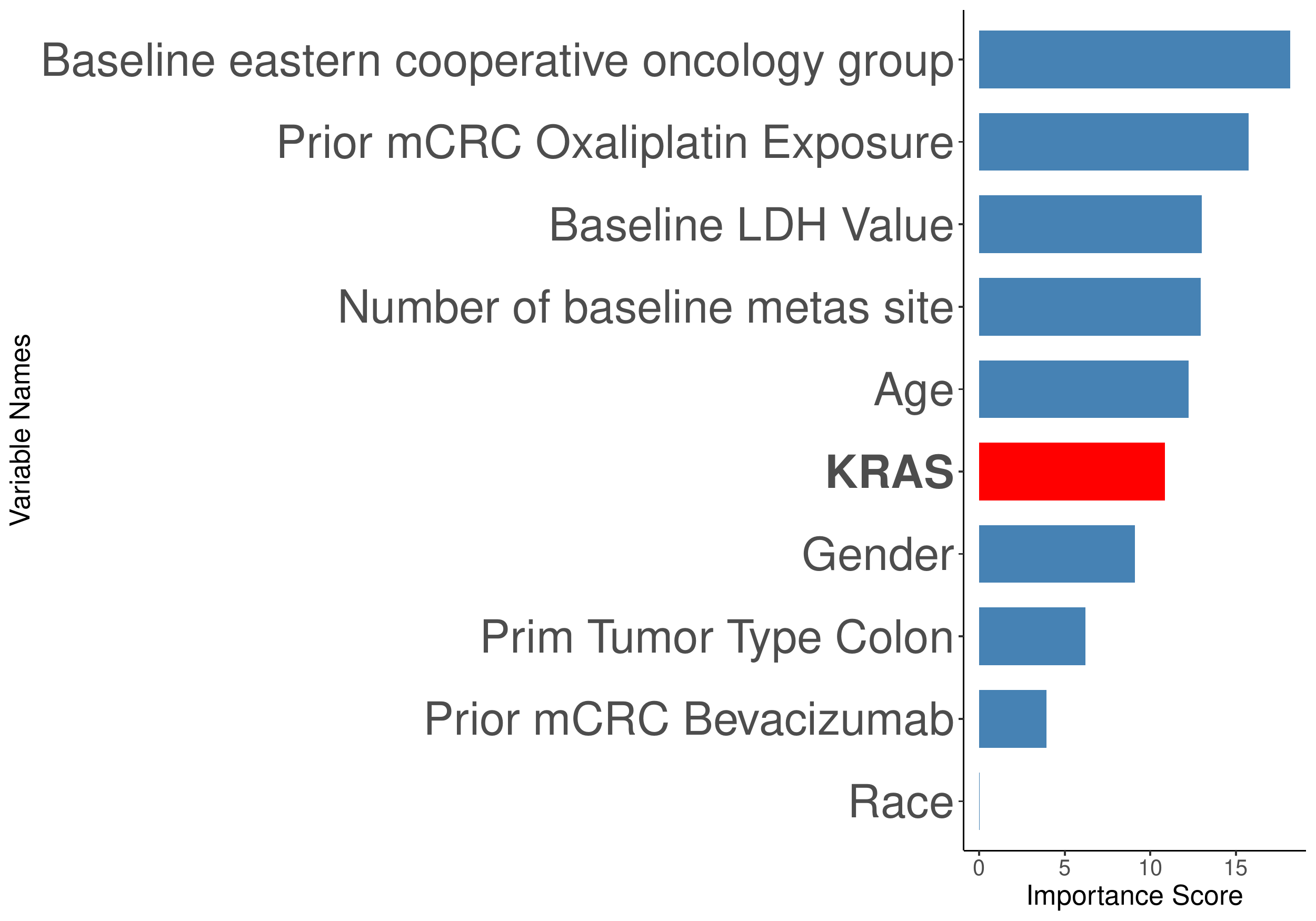}
\end{subfigure}\hfill
\begin{subfigure}[b]{0.33\textwidth}
% \caption{Variable importance of the treatment effect}
% \label{fig:VarImp_int0.75}
\includegraphics[width = \textwidth]{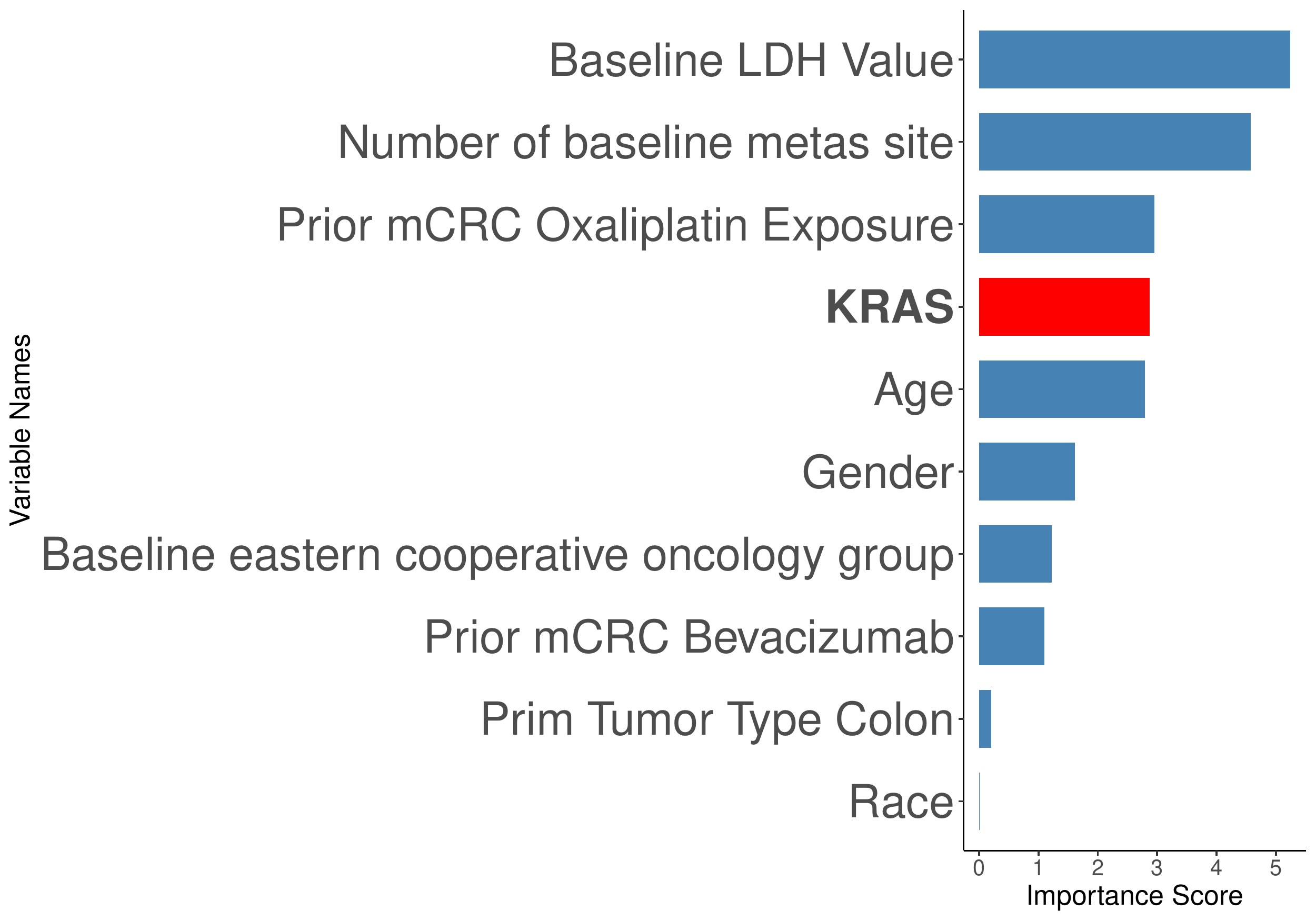}
\end{subfigure}
\end{figure}

\begin{figure}[!ht]
    \centering
    \caption{Variable importance of $10$ modifiers in the data set at $0.5,~0.25,~0.75$ quantile level by CQRF.}
    \label{fig:VarImp_cqrf}
    \begin{subfigure}[b]{0.33\textwidth}
    \caption{$\tau = 0.5$}
    \label{fig:VarImp_cqrf0.5}
    \includegraphics[width = \textwidth]{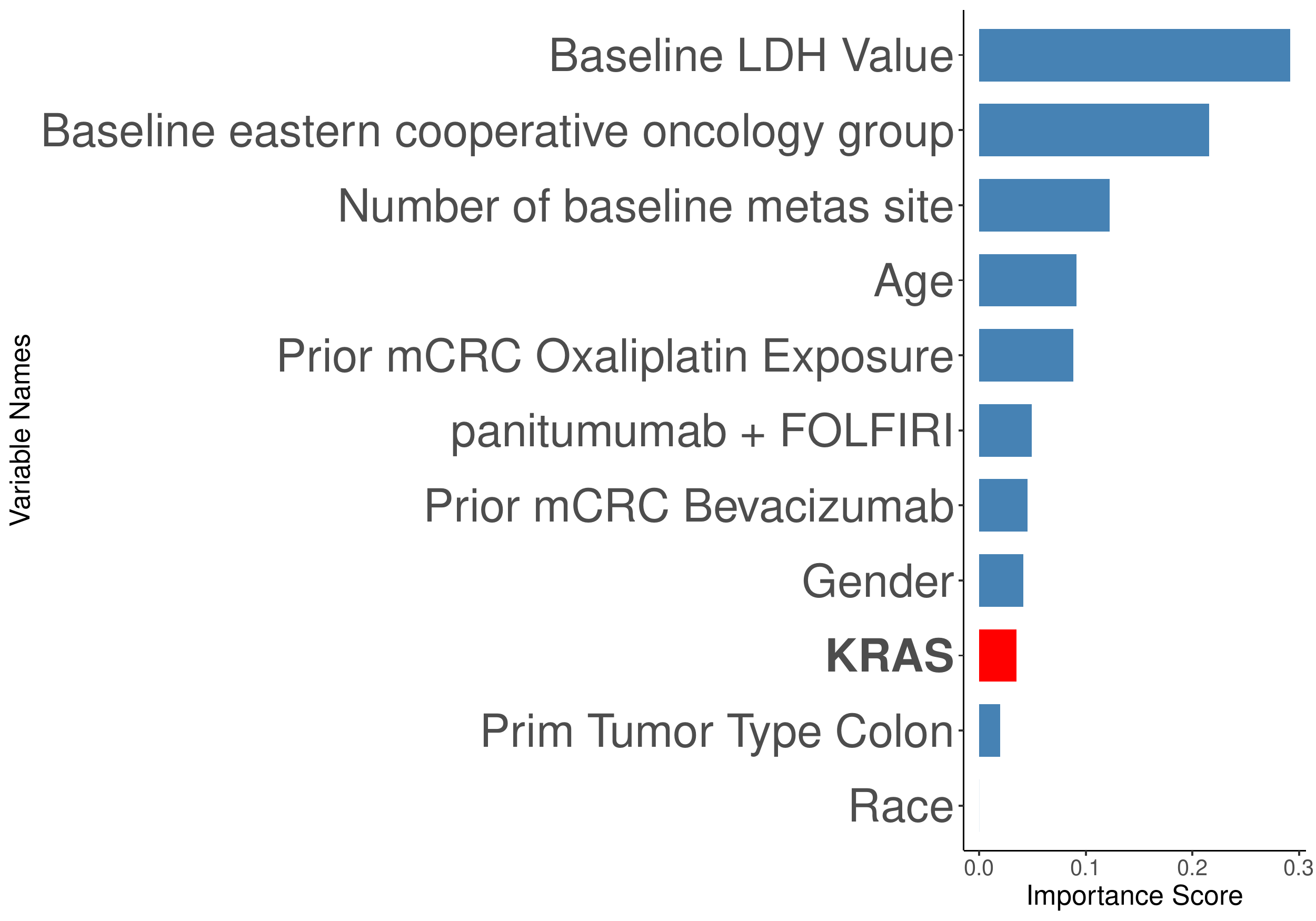}
    \end{subfigure}\hfill
        \begin{subfigure}[b]{0.33\textwidth}
            \caption{$\tau = 0.25$}
    \label{fig:VarImp_cqrf0.25}
    \includegraphics[width = \textwidth]{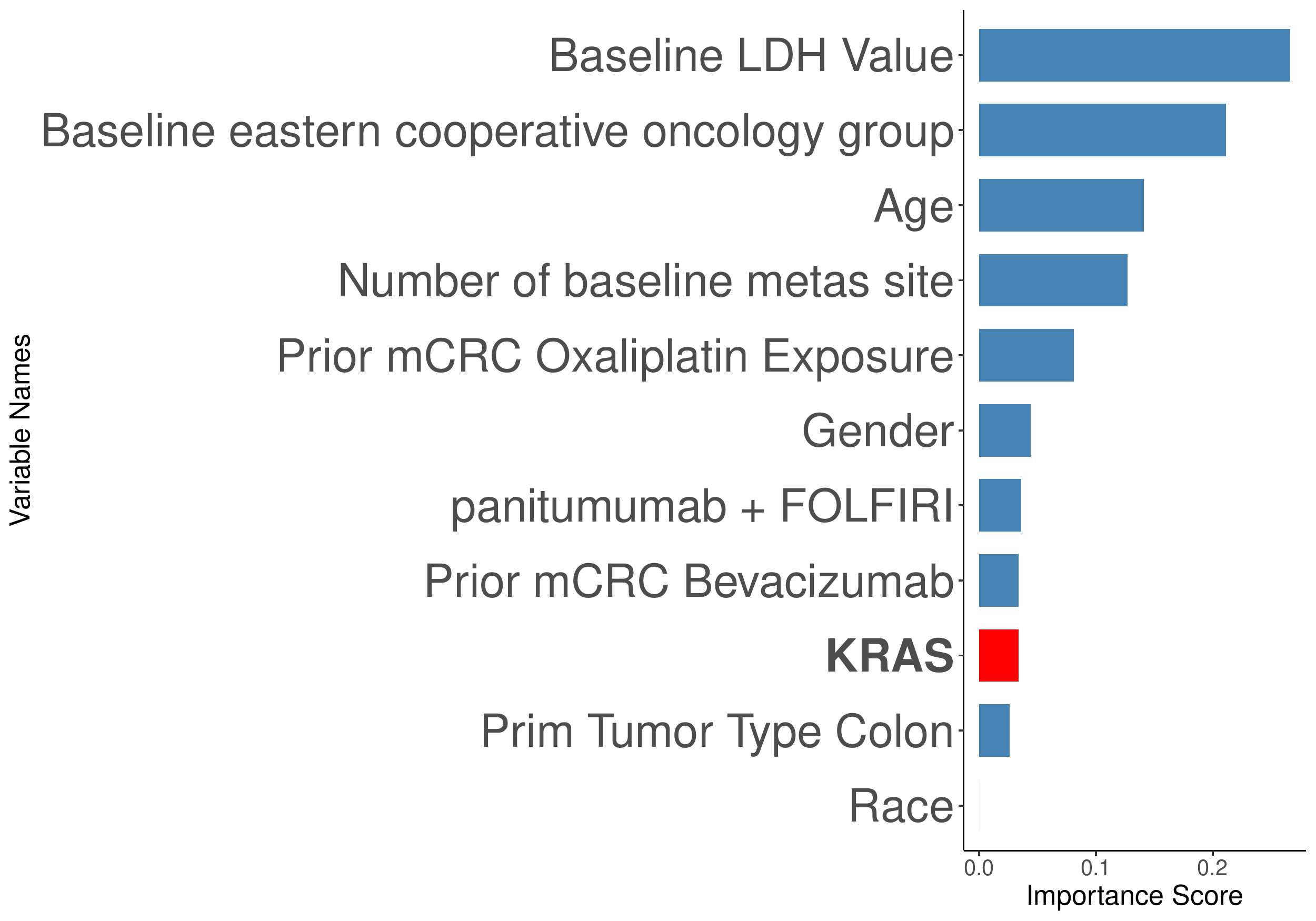}
    \end{subfigure}\hfill
        \begin{subfigure}[b]{0.33\textwidth}
            \caption{$\tau = 0.75$}
    \label{fig:VarImp_cqrf0.75}
    \includegraphics[width = \textwidth]{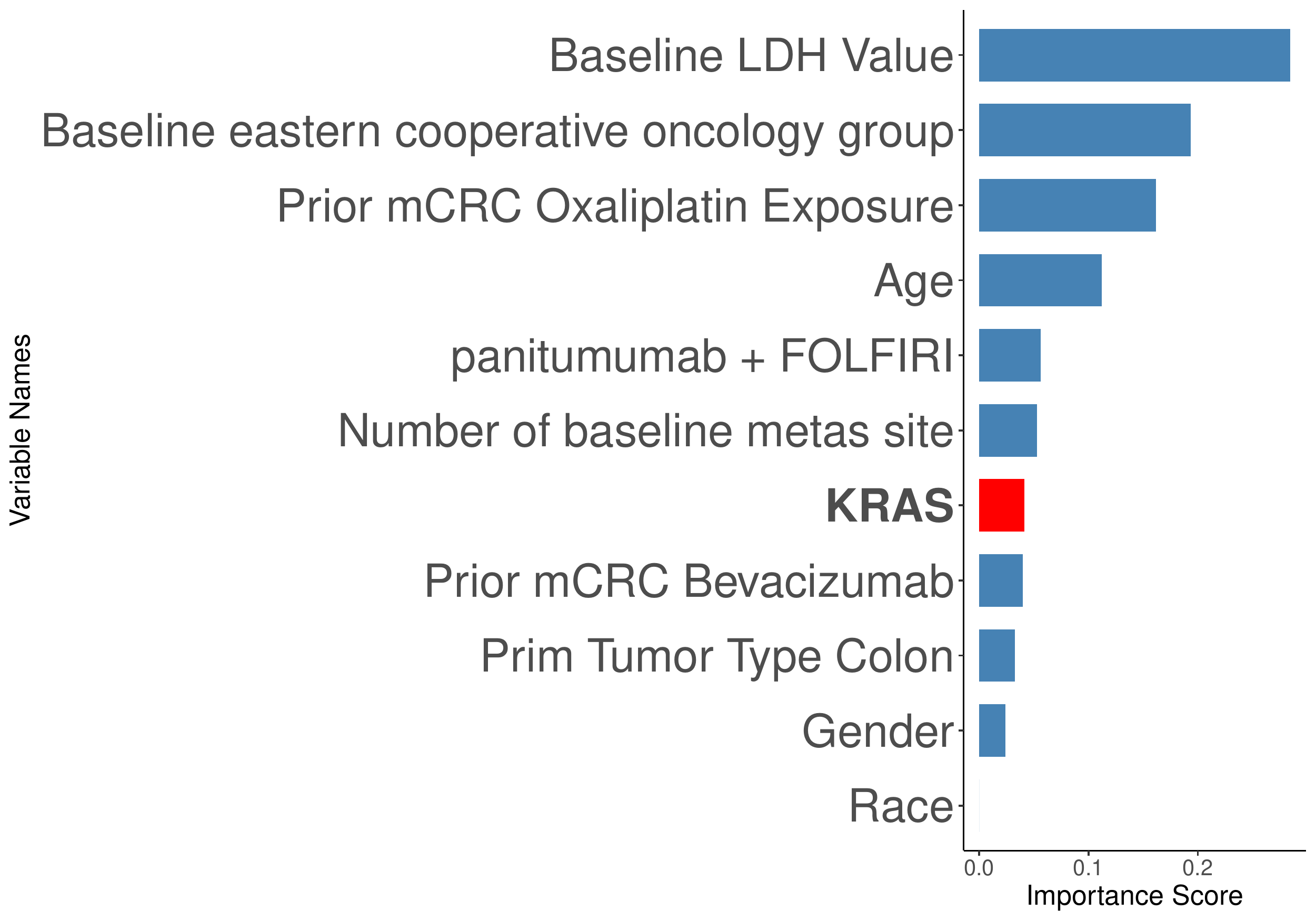}
    \end{subfigure}
\end{figure}

The resulting variable importance from HCQRF and CQRF are presented in Figure \ref{fig:VarImp} and Figure \ref{fig:VarImp_cqrf} respectively.
The first column in Figure \ref{fig:VarImp} presents the total VI at three quantile levels (0.25, 0.5, and 0.75). Top-ranked variables include ECOG, prior Oxalilatin exposure, LDH value, which are consistent with the literature \citep{cohen2009prognostic,li2016prognostic}. The second and the third columns in in Figure \ref{fig:VarImp} present the VI for $\beta_{\tau,0}(\bX)$ and $\beta_{\tau,1}(\bX)$ correspondingly.  
One interesting observation is the importance of KRAS mutation. KRAS is a group of genes involved in the epidermal growth factor receptor pathway. The importance of KRAS is among the middle to lower ranks for the overall importance and effect of $\beta_{\tau,0}$. However, it is ranked at the top when it comes to its importance for 
$\beta_{\tau,1}$. It suggests that KRAS mutation may not affect the disease progress on its own, but could modify the treatment effect on PFS, which is consistent with the literature reports.  For example, \citep{peeters2014final} reported that KRAS mutations predicts the efficacy of anti-epidermal growth factor receptor therapies, including panitumumab, and hence they may modify the treatment effect. The variable importance reported from CQRF (as in Figure \ref{fig:VarImp_cqrf}) are similar to the overall importance of HCQRF and does not recognize the importance of KRAS. 

%KRAS, which is known to be predictive of the efficacy of panitumumab \citep{amado2008wild}, however, is not ranked high in the total variable importance \textcolor{orange}{by CQRF}. The overall variable importance ranking from CQRF is similar, and cannot recognize the importance of KRAS. 
%The results show that the ranking of the variables for $\beta_{\tau,0}(\bX)$ almost remains the same as the one for the total effect. However, when evaluating the importance of the treatment effect, KRAS ascends to one of the top-ranked variables as shown in the third column in Figure \ref{fig:VarImp}.

%For example, KRAS is a group of genes involved in the epidermal growth factor receptor pathway. In practice, 

\subsection{Estimated Quantile Treatment Effect}
\begin{figure}[!ht]
    \centering
    \caption{Box plot of the individualized estimated quantile treatment effect based on HCQRF and CQRF at 0.25, 0.5, and 0.75 quantile levels by different types of KRAS}
    \label{fig:trteff_KRAS}
    \begin{subfigure}[b]{0.5\textwidth}
    \caption{HCQRF: $\tau = 0.25$}
    \label{supfig:trteff_KRAS_hcqrf_0.25}
    \includegraphics[width =  \textwidth]{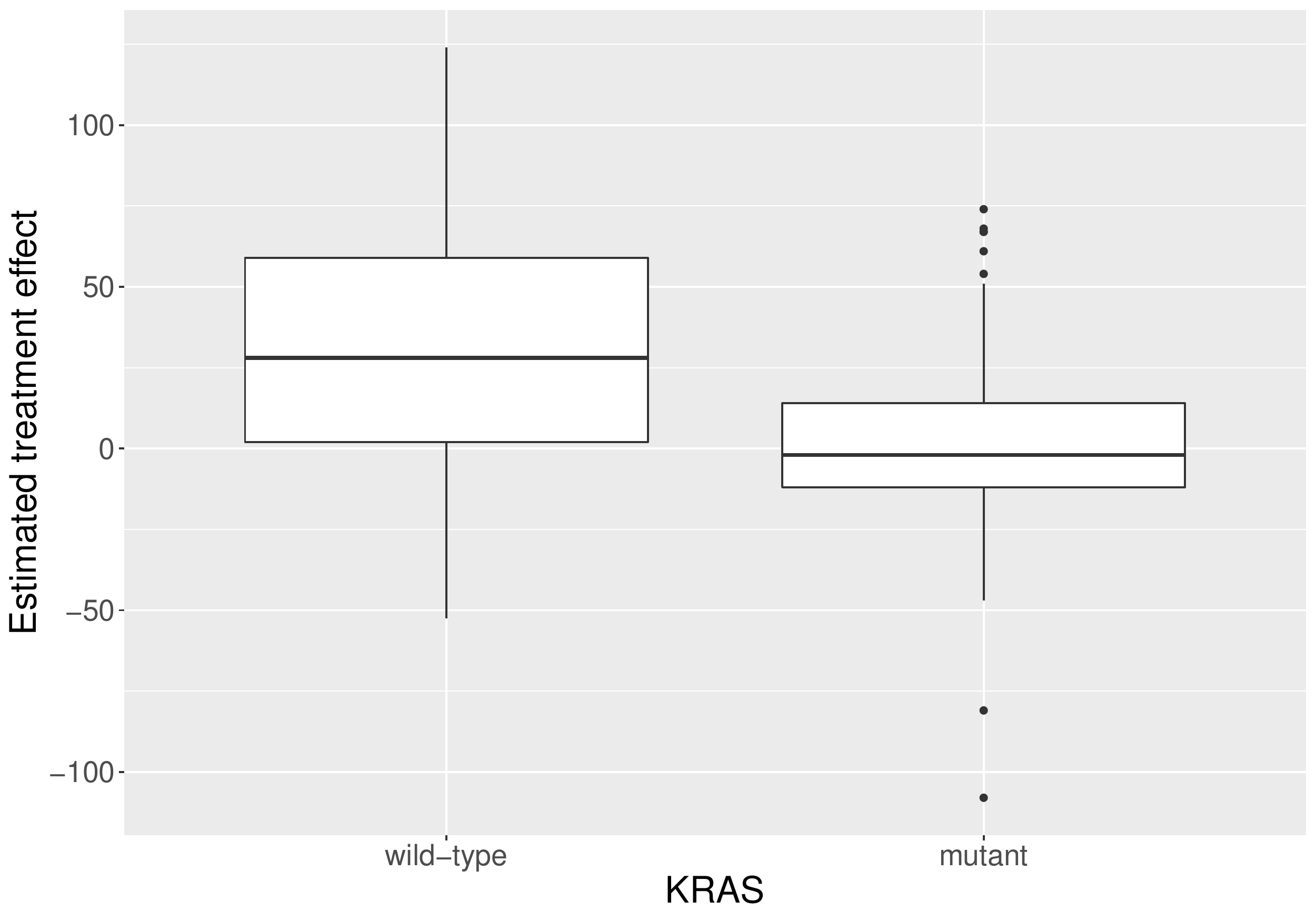}
    \end{subfigure}\hfill
    \begin{subfigure}[b]{0.5\textwidth}
    \caption{CQRF: $\tau = 0.25$}
    \label{supfig:trteff_KRAS_cqrf_0.25}
    \includegraphics[width =  \textwidth]{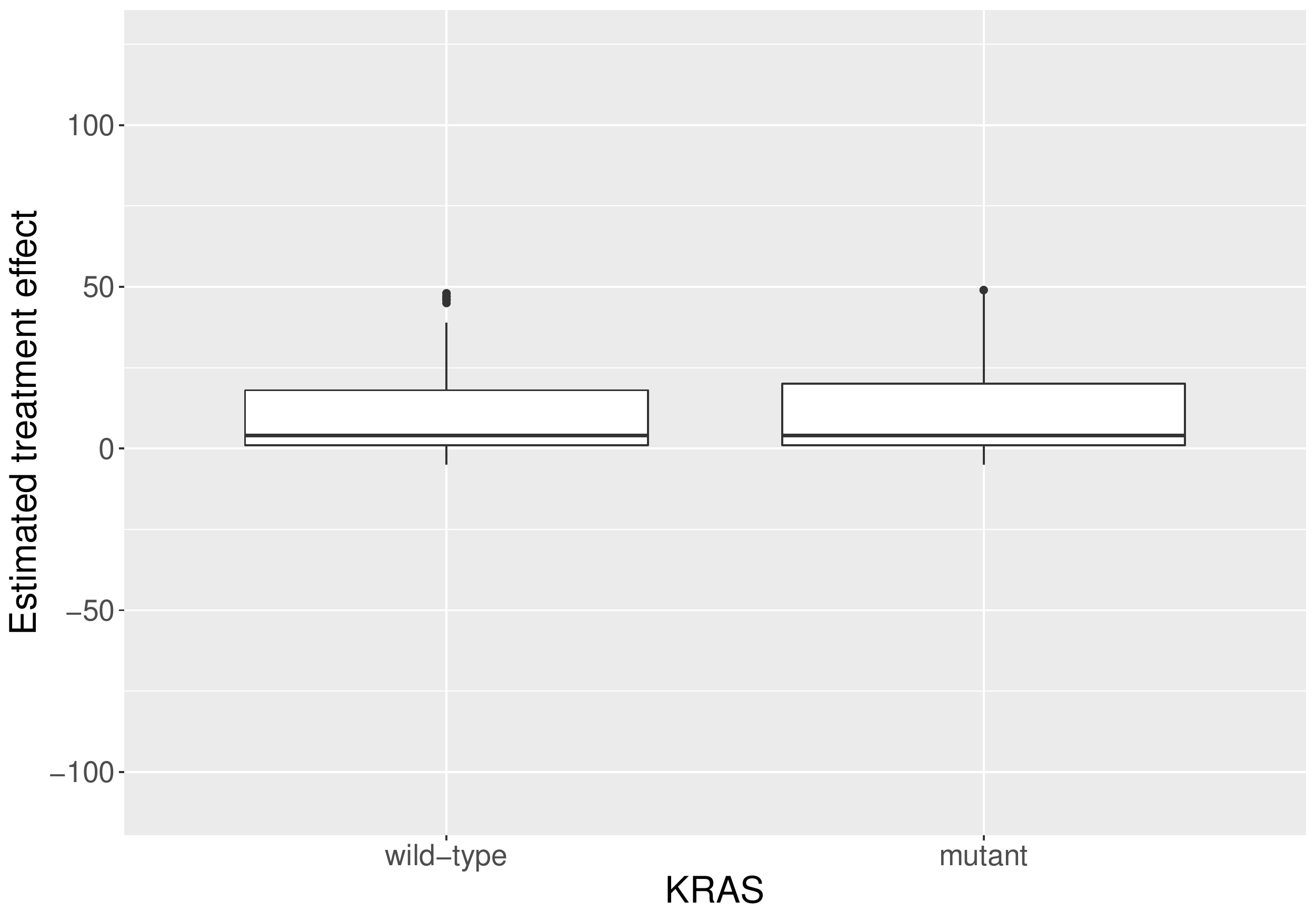}
    \end{subfigure}\vfill
    
    \begin{subfigure}[b]{0.5\textwidth}
    \caption{HCQRF: $\tau = 0.5$}
    \label{supfig:trteff_KRAS_hcqrf_0.5}
    \includegraphics[width =  \textwidth]{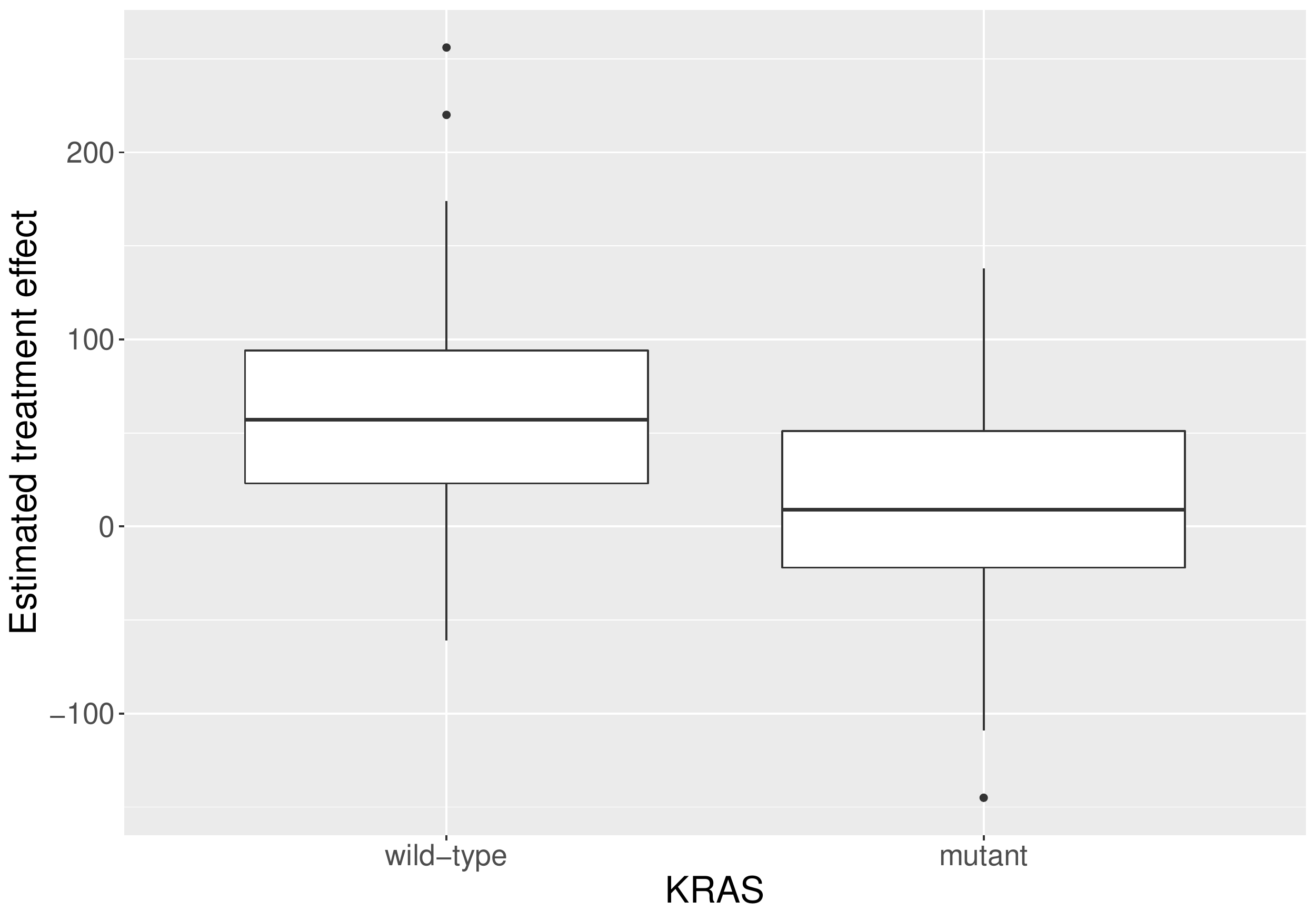}
    \end{subfigure}\hfill
    \begin{subfigure}[b]{0.5\textwidth}
    \caption{CQRF: $\tau = 0.5$}
    \label{supfig:trteff_KRAS_cqrf_0.5}
    \includegraphics[width =  \textwidth]{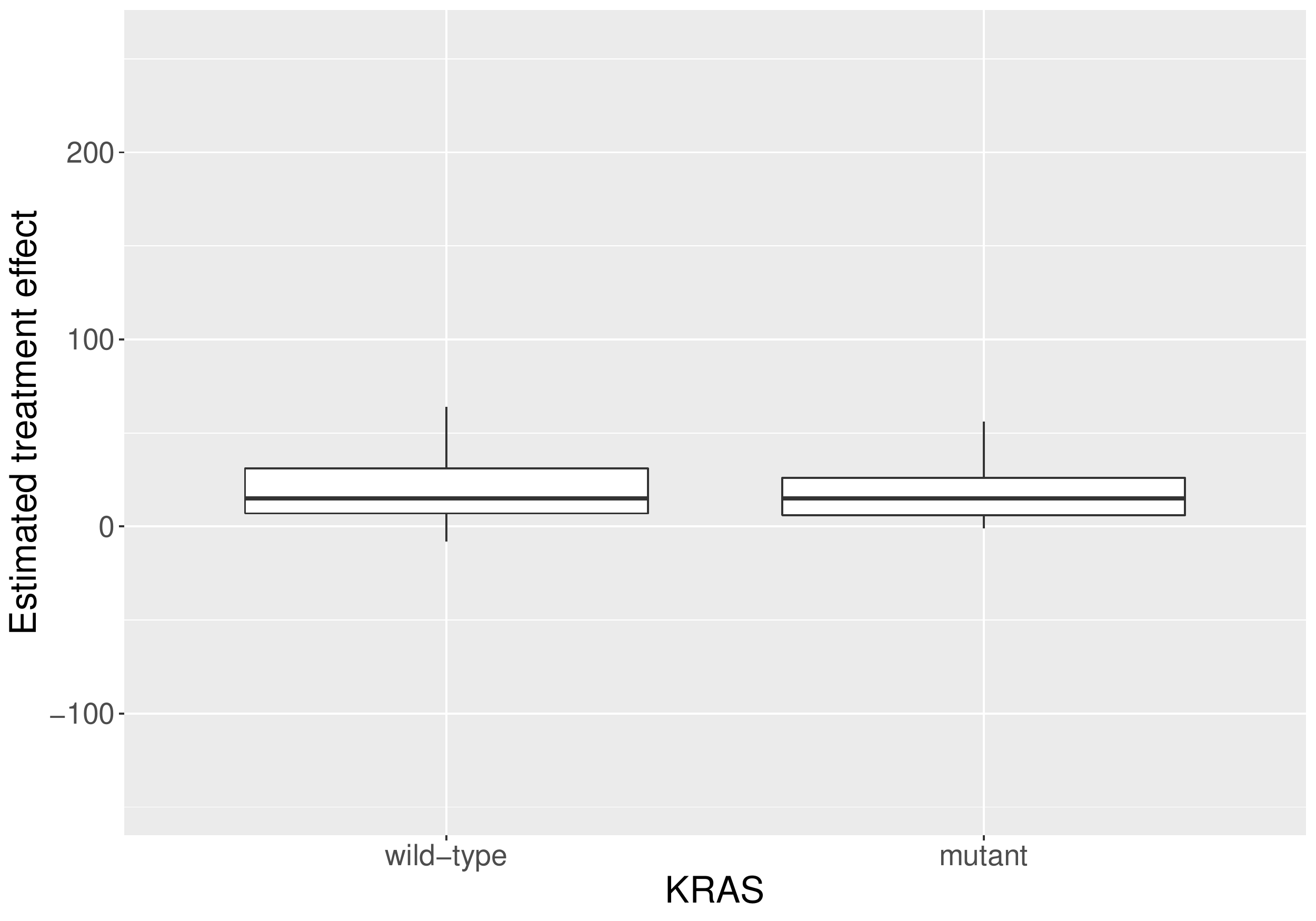}
    \end{subfigure}\vfill
    
    \begin{subfigure}[b]{0.5\textwidth}
    \caption{HCQRF: $\tau = 0.75$}
    \label{supfig:trteff_KRAS_hcqrf_0.75}
    \includegraphics[width =  \textwidth]{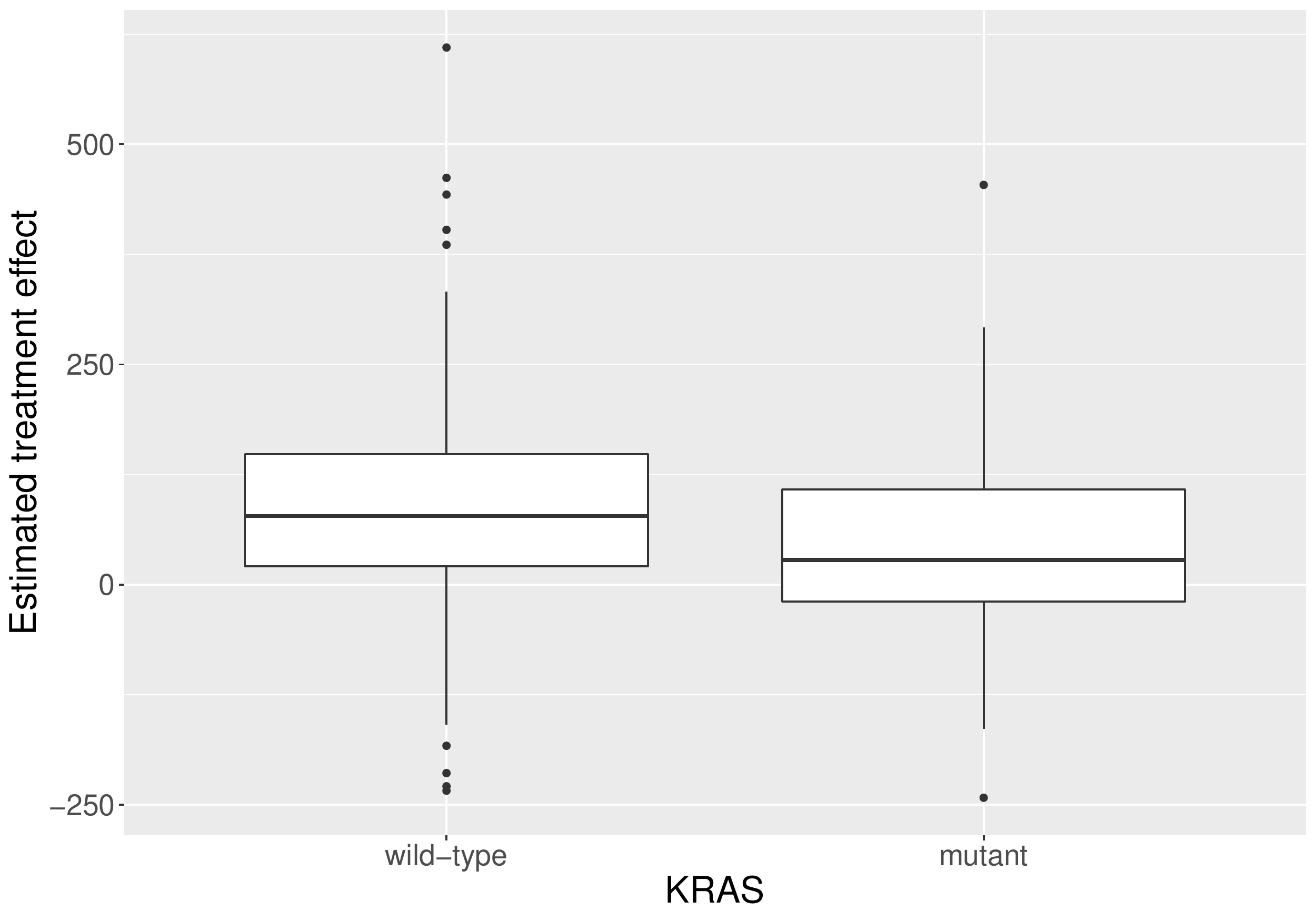}
    \end{subfigure}\hfill
     \begin{subfigure}[b]{0.5\textwidth}
    \caption{CQRF: $\tau = 0.75$}
    \label{supfig:trteff_KRAS_cqrf_0.75}
    \includegraphics[width =  \textwidth]{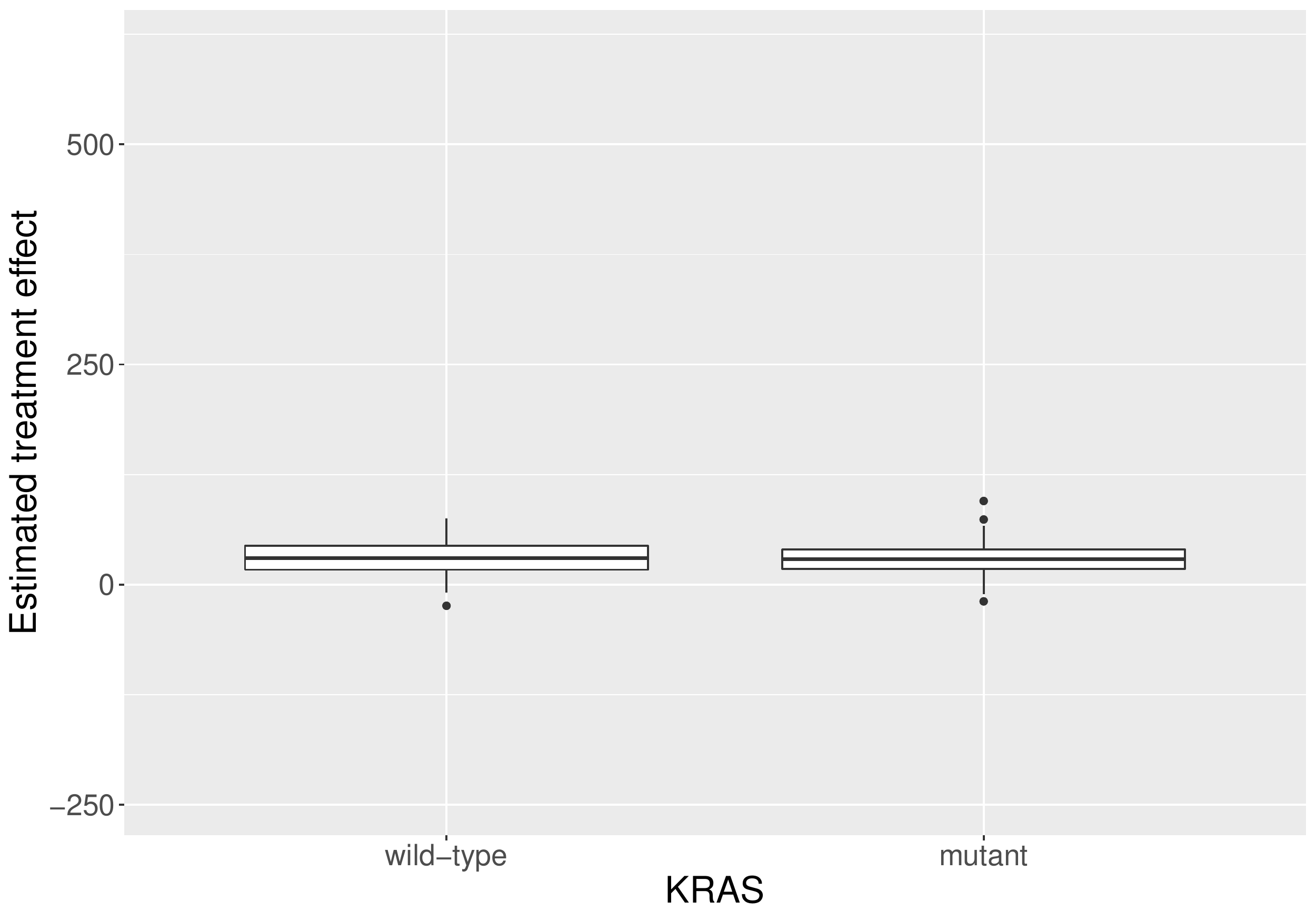}
    \end{subfigure}
\end{figure}

 Since KRAS is identified as a top-ranked effect modifier of the treatment effect, we stratify the estimation of quantile treatment effect (QTE) by KRAS types. In HCQRF,  the estimated $\bbeta_{\tau,1}(\bx_i)$ represents the individualized QTE. In CQRF, we estimate individualized QTE for each observation in the test set by $\widehat{\text{QTE}}_{\tau,i} = \widehat Q_\tau(T|\bX = \bx_i,Z = 1)-\widehat Q_\tau(T|\bX =\bx_i,Z = 0)$. Figure \ref{fig:trteff_KRAS} demonstrates the box plots of the individualized QTE stratified by the two KRAS types at quantile level $0.25,~0.5,~0.75$. The results based on HCQRF clearly manifest that the individualized treatment effects are different between the two KRAS types. However, such differences based on CQRF are less evident. The top panel of Table \ref{tbl:trtKRAS} displays the average quantile treatment effect (AQTE) from HCQRF and CQRF stratified by KRAS types.  CQRF reports comparable treatment effects, while HCQRF suggests very different treatment effects by the KRAS mutation status, across all the quantile levels.

%and The estimated average quantile treatment effect (AQTE) is obtained by $\widehat{\text{AQTE}}_{\tau,\text{KRAS} = j} = \text{Ave}_{\{i:\text{KRAS} = j\}}\widehat{\text{QTE}}_{\tau,i},$ at $\tau = 0.25,0.5,0.75$, where $j = 1,2$.quantile levels.  AQTE results in Table \ref{tbl:trtKRAS} shows similar results.

\subsection{Empirical Validation}

\begin{figure}[!ht]
\caption{Kaplan-Meier curves of progression-free survival stratified by treatment indicator and KRAS.}
\label{fig:kp_KRAS}
% \begin{subfigure}{0.5\textwidth}
% \caption{Esatern cooperative oncology group at baseline}
% \label{fig:kp_ECOG}
% \includegraphics[width = \textwidth]{Figures/Amgen/B_ECOGCD_KP.pdf}
% \end{subfigure}\hfill
\begin{subfigure}{0.5\textwidth}
\caption{KRAS = wild-type}
\label{fig:kp_KRAS_wildtype}
\includegraphics[width = \textwidth]{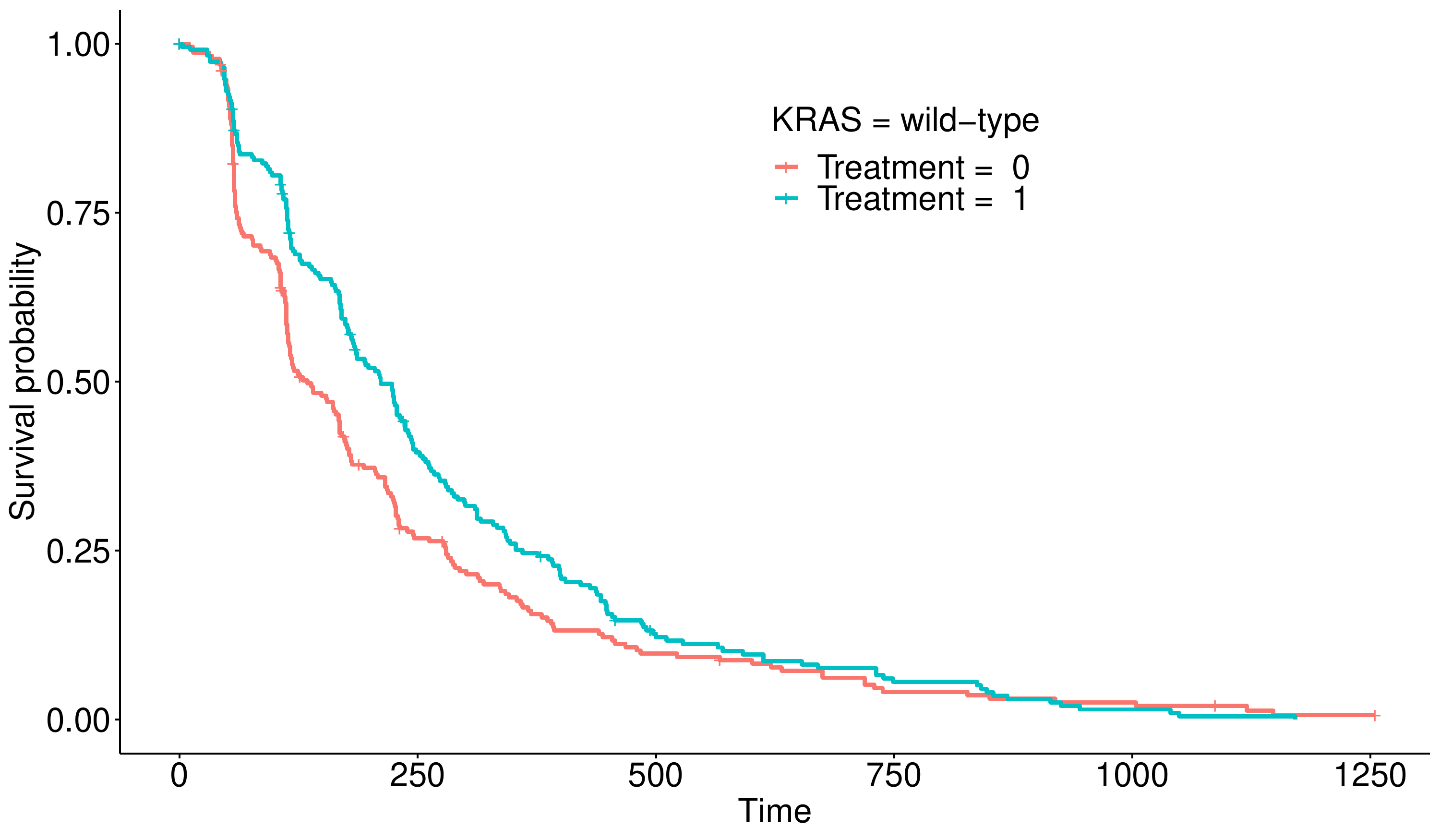}
\end{subfigure}\hfill
% \begin{subfigure}{0.5\textwidth}
% \caption{Prim tumor type}
% \label{fig:kp_tumor}
% \includegraphics[width = \textwidth]{Figures/Amgen/TUMCATCD_KP.pdf}
% \end{subfigure}\hfill
\begin{subfigure}{0.5\textwidth}
\caption{KRAS = mutant}
\label{fig:kp_KRAS_mutant}
\includegraphics[width = \textwidth]{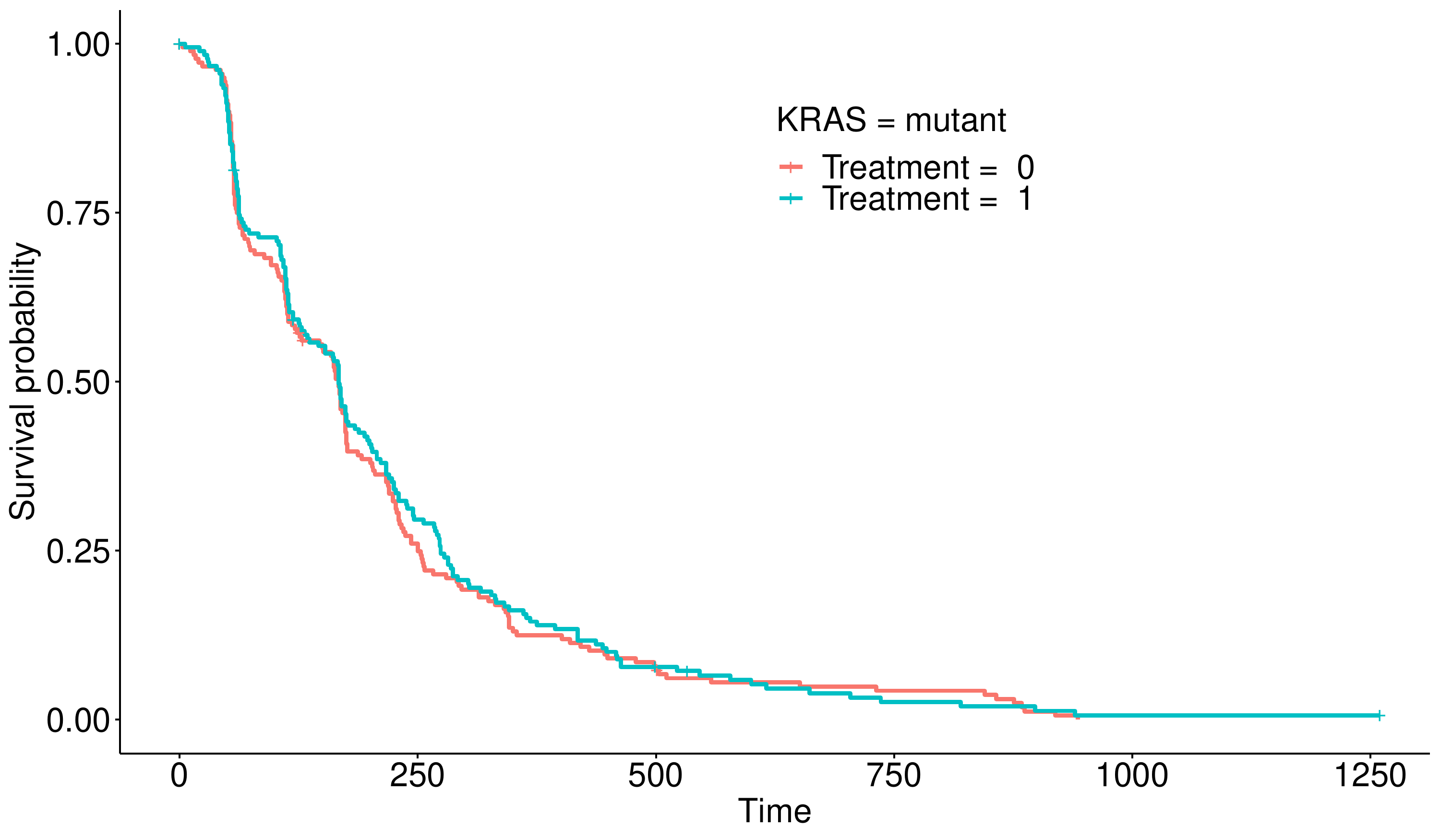}
\end{subfigure}
\end{figure}

The estimates from CQRF and HCQF disagree on the existence of KRAS-treatment interaction. In this subsection, we conduct an empirical validation. We stratify the sample by treatment assignment and KRAS status (229 patients with wild-type KRAS and 184 patients with mutant KRAS taking FOLIFIRI alone; 234 patients with wild-type KRAS and 183 patients with mutant KRAS taking Panitumumab plus FOLFIRI) and construct Kaplan-Meier survival functions of PFS  within each stratum.  The estimated Kaplan-Meier curves in Figure \ref{fig:kp_KRAS} display a noticeably larger difference in PFS between the two treatment arms among the patients with wild-type KRAS. As KS estimation is model-free, we view the agreement with KS estimation as the indication of a good fit to the data. To quantify such agreement, we also measure the distance between the KS-estimated quantile function and the model-derived conditional quantiles from CQRF and HCQRF.  For each observation $\bx_i$, we calculates its conditional quantiles 
$\widehat Q_T(\tau|\bX = \bx_i,Z = z)$ at $\tau = 0.05,0.1,\cdots,0.9,0.95$, and use linear interpolations to construction $\widehat Q_T(\tau|\bX = \bx_i,Z = z)$ for any $\tau\in (0,1)$. Since KS-estimates are marginal quantile functions in each stratum, while $\widehat Q_T(\tau|\bX = \bx_i,Z = z)$ are conditional quantiles, they are not directly comparable. To assess their agreement, we calculate 
$$\widehat \tau_{j,k} = \text{Ave}_{\{i:\text{KRAS}_i = j,Z_i = k\}}\left[\widehat F\left\{\widehat Q_T^{KM}(\tau|\text{KRAS} = j,Z = k)\Big|\bx_i, Z_i = k\right\}\right],$$
where $\widehat F(t|\bx_i,z_i)  = \sup\{\tau:\widehat  Q_T(\tau|\bX = \bx_i,Z = z_i)\le t\}$ is the distribution function induced from the conditional quantile function $Q_T(\tau|\bX = \bx_i,Z = z_i)$, and $Q_T^{KM}(\tau|\text{KRAS} = j,Z = k)$ is KS-estimated stratum-specific $\tau$th quantile. If the model fits the data well, we expect the resulting $\widehat \tau_{j,k}$ to be equal to its nominal level $\tau$. The bottom half of Table \ref{tbl:trtKRAS} lists the $\widehat \tau_{j,k}$ at quantile levels $\tau = 0.25,0.5,0.75$ in each stratum (i.e., $j = 1,2$ and  $k = 0,1$). Clearly, the ones derived from HCQRF are consistently closer to their nominal level $\tau$. We conclude that there is empirical evidence supporting the KRAS-treatment interaction.

%The estimated quantiles in different subgroups are obtained by $\widehat Q_\tau^{KM}(T|\text{KRAS} = j,Z = z),~j = 1,2$, at $\tau = 0.25,0.5,0.75$. We access the estimated conditional cumulative distribution function of the survival time at $t$ by the inverse of the estimated conditional quantile function, which is $\widehat F(t|\bx,z)  = \sup\{\tau:\widehat  Q_\tau(T|\bX = \bx,Z = z)\le t\}$. We obtain $\widehat Q_\tau(T|\bX = \bx,Z = z)$ at $\tau = 0.05,0.1,\cdots,0.9,0.95$ for each observation in the test set based on HCQRF or CQRF with $z\in\{0,1\}$ and process the interpolation to get $\widehat Q_\tau(T|\bX = \bx,Z = z)$ as a function of $\tau$. The estimated quantile levels by HCQRF or CQRF of the estimated quantiles by Kaplan-Meier curve is 

%at $\tau = 0.25,0.5,0.75$, where $j = 1,2$ and $k = 0,1$. The closer the estimated quantile level is to its corresponding nominal level, \textcolor{orange}{the more consistent the estimation is with the Kaplan-Meier estimate}. 
% \red{- the statement is too strong, not sure what does ``better'' mean - lower bias?. KM also has uncertainty, so it is hard to say better here}. 
%The result in demonstrates the estimated quantile levels $\widehat \tau_{j,k}$ and it shows that the ones derived from HCQRF are closer to $\tau$.}

\begin{table}[!ht]
    \centering
        \caption{The estimated AQTE stratified by different KRAS types and the estimated quantile levels at $\widehat \tau_{j,k}$ by HCQRF and CQRF. }
    \label{tbl:trtKRAS}
    \begin{tabular}{ccccccccc}
    \hline\hline
        &&& \multicolumn{3}{c}{KRAS = wild-type}&\multicolumn{3}{c}{KRAS = mutant}\\
       
        &&&$\tau = 0.25$&$0.5$&$0.75$&$0.25$&$0.5$&$ 0.75$\\
         \cline{2-9}
         \multirow{2}{*}{AQTE}&HCQRF &&30.788&61.584&82.529&1.681&13.469&44.617\\
        &CQRF&&8.502&12.918&27.088&5.973&7.716&18.070\\
       \cline{2-9}
        % Kaplan-Meier&53&77&80&1&0&24\\ 
        \multirow{4}{*}{$\widehat \tau_{j,k}$}
        % &\multirow{4}{*}{Whole}&HCQRF &treatment & {0.278}& {0.518}& {0.751}& {0.222}& {0.483}& {0.731}\\
        % &&&control & {0.249}& {0.497}&0.764& {0.249}&0.530& {0.741}\\
        % &&CQRF&treatment &0.341&0.581&0.806&0.202&0.462&0.709\\
        % &&& control&0.226&0.441& {0.751}&0.230& {0.506}&0.708\\
        % \cline{2-10}
        &HCQRF&treatment& {0.278}& {0.503}& {0.729}& {0.228}& {0.503}&0.716\\
        &&control& {0.251}& {0.484}& {0.765}& {0.237}&0.538& {0.748}\\
        &CQRF&treatment&0.325&0.558&0.779&0.224&0.497& {0.728}\\
        && control&0.211&0.421&0.725& 0.233& {0.517}&0.719\\
        % \cline{2-10}
        % &\multirow{4}{*}{Training}&HCQRF& treatment &\tbf{0.285}&\tbf{0.517}&\tbf{0.739}&\tbf{0.225}&\tbf{0.505}&\tbf{0.734}\\
        % &&& control & \tbf{0.238}&\tbf{0.491}&0.765&\tbf{0.228}&0.529&\tbf{0.735}\\
        % &&CQRF & treatment &0.358&0.598&0.806&0.195&0.462&0.696\\
        % &&& control &0.225&0.451&\tbf{0.750}&0.215&\tbf{0.493}&0.695\\
        
        % \multirow{6}{*}{$\widehat q$}&KM&treatment &113&211&360&62&167&274\\
        % &&control&60&134&280&61&167&250\\
        % &HCQRF&treatment &113.38&212.770&378.402&88.152&167.086&302.281\\
        % && control &82.750&150.965&298.68&85.787&152.895&257.910\\
        % & CQRF&treatment &90.588&182.035&329.235&82.173&163.617&293.556\\
        % &&control&82.541&169.118&302.200&76.16&156.593&275.481\\
        % \hline
        %  & KM &&53&77&80&1&0&24\\

        \hline\hline
    \end{tabular}

\end{table}

In the analysis of the randomized clinical trial data, KRAS is identified as a top-ranked important variable of the treatment effect through the proposed variable importance decomposition. In contrast, the ranking of the importance in CQRF does not demonstrate a similar result. Additional comparisons of the individualized QTE and AQTE between HCQRF and CQRF confirm our conclusion from the variable importance. Moreover, the result of the empirical validation also shows that the estimation obtained by HCQRF is more congruent with the data.

% \vskip 10pt
% \textbf{\textcolor{red}{\large{Please re-write/re-organize this section with the following order of paragraphs}}}

% \paragraph{Paragraph 1----The modeling fitting procedure, including choices of parameters, training-testing set participation in comparison methods}

% \paragraph{Paragraph 2----Comparing the variable importance from he two methods (HCQRF vs CQRF) as a way to identify treatment effect modifier.  }

% \paragraph{Paragraph 3----Since KRAS is identified as top effect modifier, we will estimate and compare the individualized treatment effects as well as average treatment effect between the two methods and stratified by the KRAS status.Box-plots go here; Also the AQTE part of Table 7 also goes here; Focus on focus on test data for this part;}

% \paragraph{Paragraph 4---Validation with KM estimate --- Plot the stratified KM curves (based on whole data);Figure 8 goes here; Introduce the $\tau$-hat (based on the testing data) as a way assess the fitness or estimation accuracy. need to explain the rational of $\tau$-hat; Table 7 tau-hat part goes here}

% \paragraph{Paragraph 5----Summary and Conclusion of the real data example. }

\section{Discussion}
In this paper, we develop a hybrid censored quantile regression forest to assess the heterogeneous quantile effects of censored data. The proposed estimation procedure takes advantage of both random forest and censored quantile regression. We also develop a variable importance decomposition to measure the impact of a modifier only on the treatment effect if the predictive variable is binary. Both the simulation studies and the real data analysis demonstrate that the proposed approach can achieve a better estimation performance and insightful variable importance results.

In Section \ref{sec:VarImp}, we assume that the treatments are randomly assigned, and the method can be easily extended to handle the case where the treatments assignments satisfy the ignorability assumptions \citep{rosenbaum1983central}. Specifically, we replace the term $\mathbbm 1\{Z_i=j\}$ in \eqref{eq:varimp_z0} by a weight $\nu_{i,1} = \frac{\mathbbm 1\{Z_i=1\}}{N\widehat \pi(\bX_i)}$ if $j = 1$ or $\nu_{i,0} = \frac{\mathbbm 1\{Z_i = 0\}}{N\left( 1-\widehat\pi(\bX_i)\right)}$ if $j = 0$, where $\widehat\pi(\bX_i)$ is an estimated probability of $Z_i = 1$ given $\bX_i$ and can be estimated by a logistic regression model.

Based on the simulation studies and real data analysis, the proposed HCQRF provides a good estimation of the coefficient function as well as an insightful variable ranking. 
In this paper, we focus on the case of a finite number of predictive variables. One promising future work is to extend the proposed approach to high-dimensional predictors. Constructing the confidence interval of the estimated coefficient is also one of the promising future research. The bootstrap of little bags technique used in grf \citep{athey2019generalized} may help us to quantify the uncertainty of the estimate by our proposed approach.

% Acknowledgements should go at the end, before appendices and references

\acks{This is partially supported by NSF/NIH (DMS-1953527), NIH (R21HL156288), and Direct Grants for the Chinese University of Hong Kong (171428926).}


\vskip 0.2in
\bibliography{ref}
\clearpage
\newpage
% \widetext
\begin{center}
\huge Supplementary Materials of "Hybrid Censored Quantile Regression Forest to Assess the Heterogeneous Effects"
\end{center}

\setcounter{equation}{0}
\setcounter{figure}{0}
\setcounter{table}{0}
\setcounter{page}{1}
\setcounter{section}{0}
\makeatletter
\renewcommand\thesection{\Alph{section}}
\renewcommand{\thefigure}{S\arabic{figure}}
\renewcommand{\thetable}{S\arabic{table}}
\renewcommand{\bibnumfmt}[1]{[S#1]}
\renewcommand{\citenumfont}[1]{S#1}

\section{Extra Simulation Results}
\begin{table}[ht]
\begin{center}
\begin{threeparttable}
% \centering
\caption{Estimation performance of the quantile coefficient function at $\tau = 0.5$ for \textbf{Scenario \ref{ex:tree_binary}}, based on 500 simulation runs.}
	\label{suptbl:tree_binary}
\begin{tabular}{cccccc}
\hline\hline	
&Method & $\beta_0$ & $\beta_1$ &$\beta_0$ & $\beta_1$\\
\hline
  && \multicolumn{2}{c}{$N_1 = 500$}&\multicolumn{2}{c}{$N_1 = 1000$}\\
\cline{3-6}

% \multirow{4}{*}{MSE}&HCQRF &0.004&0.815&0.003&0.225\\
% &CQRF&0.002&0.578&0.002&0.112\\
%  &HCQRF-c  &0.004&0.386&0.003&0.100\\
% &grf-c & 0.001&3.636&0.001&2.765\\
% \cline{2-6} 
% \multirow{4}{*}{MAE}&HCQRF&0.052&0.256&0.046&0.119\\
% &CQRF&0.039&0.225&0.034&0.084\\
% &HCQRF-c&0.050&0.156&0.044&0.084\\
% &grf-c&0.028&0.869&0.023&0.686\\
% % \cline{3-7}
\multirow{3}{*}{Relative MSE}&HCQRF &1.9e-04&0.001&1.5e-04&6.1e-05\\
&CQRF&9.6e-05&0.018&7.1e-05&0.008\\
&HCQRF-c  &1.7e-04&0.000&1.4e-04&4.9e-05\\
&grf-c &5.0e-05&0.028&3.4e-05&0.025\\
\cline{2-6}
\multirow{3}{*}{Relative MAE}&HCQRF &0.011&0.011&0.010&0.006\\
&CQRF&0.008&0.067&0.007&0.033\\
&HCQRF-c  &0.011&0.007&0.009&0.006\\
&grf-c & 0.006&0.097&0.005&0.089\\

\hline\hline
\end{tabular}
\begin{tablenotes}\footnotesize
\item MSE: mean squared error, MAE: mean absolute error.
\end{tablenotes}
\end{threeparttable}
\end{center}
\end{table}

\begin{table}[!ht]
\begin{center}
\begin{threeparttable}
\centering
\caption{Estimation performance of the quantile coefficient function at $\tau = 0.5$ for \textbf{Scenario \ref{ex:boosting}} based on 500 simulation runs.}
	\label{suptbl:boosting}
\begin{tabular}{cccccccccc}
\hline\hline
 &Method & $\beta_0$ & $\beta_1$&$\beta_2$ &$\beta_3$&$\beta_0$ & $\beta_1$&$\beta_2$&$\beta_3$\\
 \hline
   && \multicolumn{4}{c}{$N_1 = 500$}&\multicolumn{4}{c}{$N_1 = 1000$}\\
   \cline{3-10}
%  \multirow{3}{*}{MSE}&HCQRF&0.028&0.123&3.990&2.093&0.017&0.082&2.755&1.443\\
%  &HCQRF-c&0.022&0.105&3.602&1.884&0.013&0.069&2.482&1.295\\
%  &grf-c&0.011&0.181&8.229&4.239&0.006&0.172&8.101&4.176\\
%  \cline{2-10}
%   \multirow{3}{*}{ MAE}&HCQRF&0.117&0.204&0.800&0.587&0.093&0.159&0.575&0.427\\
%   &HCQRF-c& 0.106&0.183&0.714&0.527&0.084&0.142&0.511&0.381\\
%   &grc-c&0.075&0.240&1.400&0.998&0.056&0.225&1.373&0.978\\
%  \cline{2-10}
  \multirow{3}{*}{Relative MSE}&HCQRF&0.001&-\tnote{a}&0.112&-&0.001&-&0.086&-\\
 &HCQRF-c&0.001&-&0.104&-&0.001&-&0.078&-\\
 &grf-c&4E-4&-&0.088&-&2E-4&-&0.086&-\\
 \cline{2-10}
  \multirow{3}{*}{Relative MAE}&HCQRF&0.023&-&0.136&-&0.019&-&0.102&-\\
 &HCQRF-c&0.021&-&0.122&-&0.017&-&0.091&-\\
 &grf-c&0.015&-&0.180&-&0.011&-&0.174&-\\
\hline\hline

\end{tabular}
\begin{tablenotes}\footnotesize
\item MSE: mean squared error, MAE: mean absolute error
\item a: relative evaluation measurements are unavailable for $\beta$ because $\beta_1 = 0$.
\end{tablenotes}
\end{threeparttable}
\end{center}
\end{table}

\begin{table}[!ht]
\centering
\caption{Estimation performance of the quantile coefficient function at $\tau = 0.5$ for \textbf{Scenario \ref{ex:cosine_continuous}} based on 500 simulation runs.}
	\label{suptbl:cosine_continuous}
\begin{tabular}{cccccccc}
\hline\hline	
 &Method & $\beta_0$ & $\beta_1$&$\beta_2$ &$\beta_0$ & $\beta_1$&$\beta_2$\\
\hline
  && \multicolumn{3}{c}{$N_1 = 500$}&\multicolumn{3}{c}{$N_1 = 1000$}\\
  \cline{3-8}
% \multirow{3}{*}{MSE}&HCQRF&0.977&1.106&0.694&0.666&0.814&0.552\\
% & HCQRF-c &0.831&0.778&0.571&0.577&0.585&0.454\\
% &grf-c &2.120&1.336&0.725&1.110&1.016&0.565\\
% \cline{2-8}
% \multirow{3}{*}{MAE}&HCQRF&0.769&0.811&0.628&0.632&0.707&0.563\\
% &HCQRF-c&0.718&0.685&0.568&0.592&0.606&0.509\\
% &grf-c&1.154&0.866&0.690&0.819&0.772&0.603\\
% \cline{2-8}
\multirow{3}{*}{Relative MSE}&HCQRF&0.132&0.061&0.037&0.095&0.047&0.030\\
& HCQRF-c &0.112&0.043&0.030&0.083&0.033&0.025\\
&grf-c &0.284&0.075&0.038&0.160&0.058&0.031\\
\cline{2-8}
\multirow{3}{*}{Relative MAE}&HCQRF&0.244&0.169&0.142&0.203&0.149&0.128\\
& HCQRF-c &0.227&0.143&0.128&0.191&0.128&0.116\\
&grf-c &0.365&0.181&0.156&0.264&0.163&0.137\\
\hline\hline
\multicolumn{7}{l}{\footnotesize{MSE: mean squared error, MAE: mean absolute error}}
\end{tabular}
\end{table}

\begin{table}[!ht]
\centering
\caption{Estimation performance of the quantile coefficient function at $\tau = 0.5$ for \textbf{Scenario \ref{ex:cosine_continuous}} with heterogeneous error term,  heavy-tailed error term and covariate-dependent censoring time based on 500 simulation runs.}
	\label{suptbl:cosine_continuous_error}
\begin{tabular}{ccccccccc}
\hline\hline
\textbf{Scenario} &&Method & $\beta_0$ & $\beta_1$&$\beta_2$ &$\beta_0$ & $\beta_1$&$\beta_2$\\
 \hline
   && &\multicolumn{3}{c}{$N_1 = 500$}&\multicolumn{3}{c}{$N_1 = 1000$}\\
   \cline{4-9}
     \multirow{3}{*}{\ref{ex:cosine_continuous}a}
    %  &\multirow{3}{*}{MSE}&HCQRF&0.771&1.052&0.678&0.458&0.658&0.430\\
    %  && HCQRF-c&0.638&0.752&0.541&0.391&0.471&0.347\\
    %  &&grf-c&1.588&1.268&0.706&1.009&0.846&0.496\\
    %  \cline{3-9}
    %  &\multirow{3}{*}{MAE}&HCQRF&0.663&0.795&0.627&0.518&0.635&0.493\\
    %  &&HCQRF-c& 0.612&0.677&0.559&0.482&0.541&0.443\\
    %  &&grf-c& 0.973&0.846&0.684&0.773&0.704&0.570\\
    %  \cline{3-9}
     &\multirow{3}{*}{Relative MSE}&HCQRF&0.105&0.044&0.045&0.075&0.033&0.029\\
     && HCQRF-c&0.087&0.037&0.037&0.063&0.027&0.024\\
     &&grf-c&0.214&0.036&0.042&0.163&0.026&0.031\\
     \cline{3-9}
     &\multirow{3}{*}{Relative MAE}&HCQRF&0.210&0.158&0.149&0.174&0.133&0.117\\
     && HCQRF-c&0.193&0.142&0.134&0.162&0.120&0.105\\
     &&grf-c&0.307&0.150&0.157&0.260&0.128&0.132\\
     \hline
 \multirow{3}{*}{\ref{ex:cosine_continuous}b}
%  &\multirow{3}{*}{MSE}&HCQRF&1.370&1.487&1.008&0.906&0.965&0.691\\
%      && HCQRF-c&1.162&1.070&0.825&0.779&0.705&0.563\\
%      &&grf-c&1.921&1.639&0.863&1.230&1.080&0.622\\
%      \cline{3-9}
%      &\multirow{3}{*}{MAE}&HCQRF&0.896&0.948&0.772&0.732&0.771&0.630\\
%      &&HCQRF-c&0.834&0.811&0.699&0.686&0.666&0.571\\
%      &&grf-c& 1.080&0.965&0.749&0.863&0.796&0.634\\
    %  \cline{3-9}
     &\multirow{3}{*}{Relative MSE}&HCQRF&0.190&0.090&0.052&0.123&0.060&0.036\\
     && HCQRF-c&0.162&0.065&0.043&0.107&0.044&0.029\\
     &&grf-c&0.266&0.099&0.045& 0.169&0.067&0.032\\
     \cline{3-9}
     &\multirow{3}{*}{Relative MAE}&HCQRF&0.288&0.205&0.172&0.234&0.172&0.141\\
     && HCQRF-c&0.269&0.175&0.156& 0.219&0.149&0.128\\
     &&grf-c&0.348&0.209&0.167&0.276&0.178&0.142\\
     \hline
    \multirow{3}{*}{\ref{ex:cosine_continuous}c}
%     &\multirow{3}{*}{MSE}&HCQRF&0.977&1.095&0.683&0.655&0.686&0.435\\
% &&HCQRF-c&0.829&0.804&0.561&0.578&0.509&0.359\\
% &&grf-c &2.324&1.435&0.736&1.622&0.971&0.520\\
% \cline{3-9}
% &\multirow{3}{*}{MAE}&HCQRF&0.773&0.811&0.617&0.653&0.643&0.483\\
% &&HCQRF-c&0.724&0.696&0.555&0.622&0.555&0.437\\
% &&grf-c&1.222&0.894&0.693&1.038&0.750&0.575\\
% \cline{3-9}
&\multirow{3}{*}{Relative MSE}&HCQRF&0.134&0.066&0.036&0.088&0.039&0.024\\
&&HCQRF-c&0.114&0.049&0.029&0.078&0.029&0.020\\
&&grf-c &0.321&0.086&0.039&0.218&0.055&0.029\\
\cline{3-9}
&\multirow{3}{*}{Relative MAE}&HCQRF&0.249&0.174&0.138&0.206&0.135&0.110\\
&&HCQRF-c&0.233&0.150&0.124&0.196&0.116&0.100\\
&&grf-c &0.394&0.192&0.155&0.327&0.157&0.131\\
 \hline\hline
 \multicolumn{9}{l}{\footnotesize{MSE: mean squared error, MAE: mean absolute error}}
 \end{tabular}
 \end{table}

\begin{table}[!ht]
\begin{threeparttable}
\centering
\caption{Estimation performance of the conditional quantile based on 500 simulation runs.}
	    \label{suptbl:quantile}
	    \begin{tabular}{ccccccccc}
	    \hline\hline 
	      Scenario &$\tau$&Method & \multicolumn{3}{c}{$N_1 = 500$}&\multicolumn{3}{c}{$N_1 = 1000$}\\
	      \cline{4-9}
	     &&& MSE&RMSE&RMAE&MSE&RMSE&RMAE\\
	    \cline{4-9}
\ref{ex:tree_binary}&
0.5&HCQRF&0.428&0.003&0.013&0.112&0.001&0.007\\
&& CQRF&0.287&0.002&0.011&0.056&3E-4&0.005\\
\cline{3-9}
 \ref{ex:boosting}&0.5
&HCQRF&0.531&0.004&0.024&0.373&0.003&0.018\\
&& CQRF&3.994&0.042&0.168&4.001&0.040&0.165\\
\cline{3-9}
\ref{ex:cosine_continuous}&0.5&HCQRF &2.076&0.015&0.081&1.424&0.015&0.073\\
	     &&CQRF&29.250&0.224&0.310&29.194&0.298&0.339\\
	     \cline{3-9}
	     \ref{ex:cosine_continuous}a&0.5& HCQRF&1.834&0.013&0.074&1.160&0.009&0.061\\
	     &&CQRF &28.805&0.112&0.263&28.262&0.112&0.262\\
	     \cline{3-9}
	     \ref{ex:cosine_continuous}b&0.5& HCQRF&2.516&0.021&0.091&1.639&0.016&0.078\\
	     && CQRF&30.138&0.246&0.323&29.226&0.286&0.335\\

	     \cline{3-9}
	      \ref{ex:cosine_continuous}c&0.5& HCQRF&2.144&0.019&0.086&1.465&0.014&0.073\\
	     && CQRF&29.972&0.268&0.327&29.598&0.278&0.327\\
	    \cline{3-9}
	    \ref{supex:cosine_continuous_scale} &0.25&HCQRF&2.453&0.022&0.094&1.713&0.016&0.080\\
	    &&CQRF&30.036&0.266&0.328&29.629&0.277&0.327\\
	    \cline{3-9}
	   \ref{supex:cosine_continuous_scale}&0.5&HCQRF&2.186&0.019&0.087&1.520&0.014&0.074\\
	    &&CQRF&30.14&0.266&0.327&29.779&0.278&0.327\\
	     \cline{3-9}
	   \ref{supex:cosine_continuous_scale}&0.75&HCQRF&2.208&0.019&0.085&1.475&0.014&0.071\\
	   &&CQRF&30.445&0.266&0.327&30.058&0.277&0.327\\
	     \hline\hline
	    \end{tabular}
\begin{tablenotes}\footnotesize
\item MSE: mean squared error, RMSE: relative mean squared error RMAE: relative mean absolute error
\end{tablenotes}
	\end{threeparttable}
\end{table}

\newpage

\section{Extra Simulations}

\begin{example}\label{supex:nonVCM} 
We generated data $(Y_i,\Delta_i,\bX_i,\bZ_i)_{i = 1}^{N_1}$ from the model \eqref{eq:simmodel} with $\beta_0(\bX_i) = 5$,  $\beta_1(\bX_i) = 10$ and $\varepsilon_i\sim \chi^2_2$. The predictive variable $\bZ_i$ and the modifiers $\bX_i$ were generated from $U(0,1)^{(p+1)}$. The censoring time $C_i$ was generated from $U(0,48)$, resulting in $25\%$ censoring. The set of modifiers $\bX_j^*,~j = 1,2,\ldots,N_2$ were generated randomly from $U(0,1)^p$ and $N_2 = 200$.
\end{example}

In \textbf{Scenario \ref{supex:nonVCM}}, the coefficient functions are constants. Censored quantile regression (cqr, \cite{wang2009locally}) can also be applied to this scenario. Thus, we compare the estimation performance of HCQRF and cqr in Scenario \ref{supex:nonVCM}. When the quatile coefficients are constants, censored quantile regression (cqr) achieves the optimal estimation performance. It is inefficient to apply a complex estimation framework such as random forest on the data with simple true underlying model. The result in Table \ref{suptbl:nonVCM} shows that the proposed \textbf{HCQRF} can still achieve a good estimation performance.

\begin{table}[!ht]
\begin{center}
\caption{Estimation performance of the quantile coefficient function at $\tau = 0.5$ for \textbf{Scenario \ref{supex:nonVCM}} based on 500 simulation runs.}
\label{suptbl:nonVCM}
   \begin{threeparttable}
% \centering
\begin{tabular}{ccccccc}
\hline\hline	
 Scenario&&Method & $\beta_0$ & $\beta_1$ &$\beta_0$ & $\beta_1$\\
\hline
  &&& \multicolumn{2}{c}{$N_1 = 500$}&\multicolumn{2}{c}{$N_1 = 1000$}\\
\cline{4-7} 
\multirow{2}{*}{\textbf{ \ref{supex:nonVCM}}}&\multirow{2}{*}{MSE}&HCQRF &0.302&0.890&0.114&0.664\\
%&grf &0.015&0.176&7.805&4.058&0.008&0.160&7.471&3.878\\
&&cqr  &0.036 &0.111&0.007&0.069\\
\cline{3-7}
&\multirow{2}{*}{Relative MSE}&HCQRF &0.007&0.009&0.003&0.007\\
&&cqr&0.002&0.001&0.001&0.000\\
\cline{3-7}
&\multirow{2}{*}{Relative MAE}&HCQRF &0.068&0.074&0.042&0.065\\
&&cqr&0.041&0.026&0.027&0.017\\

\hline\hline
\end{tabular}
\begin{tablenotes}\footnotesize
\item MSE: mean squared error, MAE: mean absolute error, cqr: censored quantile regression
\end{tablenotes}
\end{threeparttable}
\end{center}
\end{table}

% \begin{example}\label{supex:VCM}
% We generated data $(t_i,\bX_i,\bZ_i,\delta_i)_{i = 1}^{N_1}$ from the model \eqref{eq:simmodel} with
% \begin{align*}
% &\beta_0(\bX_i) = 2+2X_{i1},\\
% &\beta_1(\bX_i) = 2-3/2\cos\left((X_{i1}-0.5)\pi/2\right).
% \end{align*}
% The error term $\varepsilon_i$ was generated from a $t$ distribution with a degree freedom of $10$. The  predictive variable $\bZ_i$ and the modifiers $\bX_i$ were generated from $U(0,1)^2$. The censoring time $C_i$ was generated from $U(0,30)$, resulting in $15\%$ censoring when $\tau = 0.5$. The set of modifiers $\bX_j^*,~j = 1,2,\cdots,N_2$ were generated as follows: $x^*_{j1}$ takes the values of $(1/201,2/201\cdots,200/201)$ and the other values of $\bX_j^*$ were generated from $U(0,2)^{(p-1)}$. Thus, $N_2 = 200$.
% \end{example}

% Table \ref{tbl:ipw} summarizes the simulation results of \textbf{Example \ref{supex:VCM}}, which is a classical example with varying-coefficient model for censored data. The results demonstrate that the estimation performance of our proposed approach is always better than \textbf{CQRF-IPW}.

% \todo[color = green,inline]{the coefficients settings are different.}
In \textbf{Scenario \ref{supex:cosine_continuous_scale}}, we consider an extended scenario where $\beta_2(\bX)$ is a coefficient function varying across different quantile levels 
\begin{example}\label{supex:cosine_continuous_scale}
We generated data $(Y_i,\delta_i,\bX_i,\bZ_i)_{i = 1}^{N_1}$ from the model \eqref{eq:simmodel} with
\begin{align}\label{supeq:varyquantile}
    &\beta_0(\bX_i) = 1+3X_{i3},\nonumber\\
    &\bbeta_1(\bX_i) = \begin{pmatrix}
		\beta_1(\bX_i)\\
		\beta_2(\bX_i)
	\end{pmatrix} = 
	\begin{pmatrix}
	 10-7.5\cos\left(\frac{\pi}{2}(X_{i1}-0.5)\right)\\
	 0.5X_{i2}(3-X_{i2})+1+X_{i2}\xi_i/10,
	\end{pmatrix}
\end{align}
where $\xi_i$ is generated from a Chi-sqaured distribution with  degree freedom 1. It indicates from \eqref{supeq:varyquantile} that $\beta_2(\bX_i)$ can be rewrited as a function depending on both $\bX$ and quantile level $\tau$, $\beta_2(\bX_i,\tau) = 0.5X_{i2}(3-X_{i2})+1+X_{i2}Q_{\xi_i}(\tau)/10$, where $Q_{\xi_i}(\tau)$ is the $\tau$th quantile of $
\xi_i$.
The predictive variables $\bZ_i$ and the modifiers $\bX_i$ were randomly generated from $U(0,2)^{(p+2)}$ independently. The error term $\varepsilon_i$ was generated from $U(0,1)$.
The censoring time $C_i$ was generated from $U(0,67)$. The set of modifiers $\bX_j^*,~j = 1,2,\ldots,N_2$ were generated randomly from $U(0,2)^p$ and $N_2 = 400$. 
\end{example}

Table \ref{tbl:cosine_continuous_scale} summarizes the simulation results of \textbf{Scenario \ref{supex:cosine_continuous_scale}} at quantile levels $\tau = 0.25,0.5,0.75$. Table \ref{tbl:quantile} summarizes the estimation performance of the conditional quantile for the four different extended scenarios. 

\begin{table}[!ht]
\centering
\caption{Estimation performance of the quantile coefficient function at $\tau = 0.25,~0.5,~0.75$ for \textbf{Scenario \ref{supex:cosine_continuous_scale}} based on 500 simulation runs.}
	\label{suptbl:cosine_continuous_scale}
\begin{tabular}{ccccccccc}
\hline\hline	
$\tau$&&Method & $\beta_0$ & $\beta_1$ &$\beta_2$ &$\beta_0$ & $\beta_1$&$\beta_2$\\
\hline
  &&& \multicolumn{3}{c}{$N_1 = 500$}&\multicolumn{3}{c}{$N_1 = 1000$}\\
\cline{4-9} 
\multirow{3}{*}{$0.25$}
% &\multirow{3}{*}{MSE}&HCQRF&1.190&1.073&0.709&0.819&0.713&0.498\\
% &&HCQRF-c &0.823&0.796&0.523&0.593&0.527&0.359\\
% &&grf-c&2.364&1.283&0.662&1.648&0.981&0.488\\
% \cline{3-9} 
% &\multirow{3}{*}{MAE}&HCQRF&0.860&0.837&0.649&0.726&0.660&0.516\\
% &&HCQRF-c& 0.734&0.707&0.556&0.629&0.565&0.435\\
% &&grf-c&1.230&0.901&0.662&1.046&0.755&0.541"\\

% \cline{3-9} 
&\multirow{3}{*}{Relative MSE}&HCQRF&0.162&0.067&0.037&0.110&0.041&0.027\\
&&HCQRF-c &0.112&0.050&0.027&0.080&0.030&0.020\\
&&grf-c&0.327&0.082&0.035&0.222&0.056&0.027\\
\cline{3-9} 
&\multirow{3}{*}{Relative MAE}&HCQRF&0.274&0.175&0.140&0.228&0.138&0.118\\
&&HCQRF-c &0.233&0.149&0.120&0.198&0.118&0.099\\
&&grf-c&0.398&0.186&0.145&0.329&0.158&0.123\\
\hline
\multirow{3}{*}{$0.5$}
% &\multirow{3}{*}{MSE}&HCQRF&0.987&1.097&0.691&0.679&0.701&0.435\\
% &&HCQRF-c &0.856&0.833&0.575&0.593&0.527&0.359\\
% &&grf-c&2.356&1.452&0.732&1.648&0.981&0.510\\
% \cline{3-9} 
% &\multirow{3}{*}{MAE}&HCQRF&0.779&0.810&0.617&0.663&0.651&0.483\\
% &&HCQRF-c&0.734&0.707&0.560&0.629&0.565&0.435\\
% &&grf-c&1.230&0.901&0.682&1.046&0.755&0.560\\
% \cline{3-9} 
&\multirow{3}{*}{Relative MSE}&HCQRF&0.136&0.066&0.036&0.091&0.040&0.024\\
&&HCQRF-c &0.118&0.050&0.030&0.080&0.030&0.020\\
&&grf-c&0.325&0.087&0.038&0.222&0.056&0.028\\
\cline{3-9} 
&\multirow{3}{*}{Relative MAE}&HCQRF&0.251&0.175&0.137&0.209&0.136&0.109\\
&&HCQRF-c &0.236&0.152&0.124&0.198&0.118&0.099\\
&&grf-c&0.397&0.194&0.151&0.329&0.158&0.127\\
\hline
\multirow{3}{*}{$0.75$}
% &\multirow{3}{*}{MSE}&HCQRF&1.021&1.414&0.697&0.715&0.955&0.439\\
% &&HCQRF-c &0.856&0.833&0.601&0.593&0.527&0.375\\
% &&grf-c&2.356&1.452&0.818&1.648&0.981&0.580\\
% \cline{3-9} 
% &\multirow{3}{*}{MAE}&HCQRF&0.798&0.897&0.634&0.689&0.740&0.499\\
% &&HCQRF-c&0.734&0.707&0.580&0.629&0.565&0.452\\
% &&grf-c&1.230&0.901&0.736&1.046&0.755&0.614\\
% \cline{3-9} 
&\multirow{3}{*}{Relative MSE}&HCQRF&0.141&0.085&0.035&0.097&0.055&0.023\\
&&HCQRF-c &0.118&0.050&0.030&0.080&0.030&0.020\\
&&grf-c&0.325&0.087&0.041&0.222&0.056&0.031\\
\cline{3-9} 
&\multirow{3}{*}{Relative MAE}&HCQRF&0.258&0.193&0.138&0.217&0.155&0.111\\
&&HCQRF-c &0.236&0.152&0.126&0.198&0.118&0.101\\
&&grf-c&0.397&0.194&0.160&0.329&0.158&0.137\\
\hline\hline
\multicolumn{9}{l}{\footnotesize{MSE: mean squared error, MAE: mean absolute error}}
\end{tabular}
\end{table}

\begin{table}[!ht]
\centering
\caption{Estimation performance of the quantile coefficient function at $\tau = 0.25,~0.5,~0.75$ for \textbf{Scenario \ref{supex:cosine_continuous_scale}} based on 500 simulation runs.}
	\label{tbl:cosine_continuous_scale}
\begin{tabular}{ccccccccc}
\hline\hline	
$\tau$&&Method & $\beta_0$ & $\beta_1$ &$\beta_2$ &$\beta_0$ & $\beta_1$&$\beta_2$\\
\hline
  &&& \multicolumn{3}{c}{$N_1 = 500$}&\multicolumn{3}{c}{$N_1 = 1000$}\\
\cline{4-9} 
\multirow{3}{*}{$0.25$}&\multirow{3}{*}{MSE}&HCQRF&1.190&1.073&0.709&0.819&0.713&0.498\\
&&HCQRF-c &0.823&0.796&0.523&0.593&0.527&0.359\\
&&grf-c&2.364&1.283&0.662&1.648&0.981&0.488\\
\cline{3-9} 
&\multirow{3}{*}{MAE}&HCQRF&0.860&0.837&0.649&0.726&0.660&0.516\\
&&HCQRF-c& 0.734&0.707&0.556&0.629&0.565&0.435\\
&&grf-c&1.230&0.901&0.662&1.046&0.755&0.541"\\

% \cline{3-9} 
% &\multirow{3}{*}{Relative MSE}&HCQRF&0.162&0.067&0.037&0.110&0.041&0.027\\
% &&HCQRF-c &0.112&0.050&0.027&0.080&0.030&0.020\\
% &&grf-c&0.327&0.082&0.035&0.222&0.056&0.027\\
% \cline{3-9} 
% &\multirow{3}{*}{Relative MAE}&HCQRF&0.274&0.175&0.140&0.228&0.138&0.118\\
% &&HCQRF-c &0.233&0.149&0.120&0.198&0.118&0.099\\
% &&grf-c&0.398&0.186&0.145&0.329&0.158&0.123\\
\hline
\multirow{3}{*}{$0.5$}&\multirow{3}{*}{MSE}&HCQRF&0.987&1.097&0.691&0.679&0.701&0.435\\
&&HCQRF-c &0.856&0.833&0.575&0.593&0.527&0.359\\
&&grf-c&2.356&1.452&0.732&1.648&0.981&0.510\\
\cline{3-9} 
&\multirow{3}{*}{MAE}&HCQRF&0.779&0.810&0.617&0.663&0.651&0.483\\
&&HCQRF-c&0.734&0.707&0.560&0.629&0.565&0.435\\
&&grf-c&1.230&0.901&0.682&1.046&0.755&0.560\\
% \cline{3-9} 
% &\multirow{3}{*}{Relative MSE}&HCQRF&0.136&0.066&0.036&0.091&0.040&0.024\\
% &&HCQRF-c &0.118&0.050&0.030&0.080&0.030&0.020\\
% &&grf-c&0.325&0.087&0.038&0.222&0.056&0.028\\
% \cline{3-9} 
% &\multirow{3}{*}{Relative MAE}&HCQRF&0.251&0.175&0.137&0.209&0.136&0.109\\
% &&HCQRF-c &0.236&0.152&0.124&0.198&0.118&0.099\\
% &&grf-c&0.397&0.194&0.151&0.329&0.158&0.127\\
\hline
\multirow{3}{*}{$0.75$}&\multirow{3}{*}{MSE}&HCQRF&1.021&1.414&0.697&0.715&0.955&0.439\\
&&HCQRF-c &0.856&0.833&0.601&0.593&0.527&0.375\\
&&grf-c&2.356&1.452&0.818&1.648&0.981&0.580\\
\cline{3-9} 
&\multirow{3}{*}{MAE}&HCQRF&0.798&0.897&0.634&0.689&0.740&0.499\\
&&HCQRF-c&0.734&0.707&0.580&0.629&0.565&0.452\\
&&grf-c&1.230&0.901&0.736&1.046&0.755&0.614\\
% \cline{3-9} 
% &\multirow{3}{*}{Relative MSE}&HCQRF&0.141&0.085&0.035&0.097&0.055&0.023\\
% &&HCQRF-c &0.118&0.050&0.030&0.080&0.030&0.020\\
% &&grf-c&0.325&0.087&0.041&0.222&0.056&0.031\\
% \cline{3-9} 
% &\multirow{3}{*}{Relative MAE}&HCQRF&0.258&0.193&0.138&0.217&0.155&0.111\\
% &&HCQRF-c &0.236&0.152&0.126&0.198&0.118&0.101\\
% &&grf-c&0.397&0.194&0.160&0.329&0.158&0.137\\
\hline\hline
\multicolumn{9}{l}{\footnotesize{MSE: mean squared error, MAE: mean absolute error}}
\end{tabular}
\end{table}

\section{Extra Figures in Real Data Analysis}
\begin{figure}[!ht]
\caption{Kaplan-Meier curves of progression-free survival stratified by treatment indicator and KRAS.}
\label{supfig:kp_KRAS}
% \begin{subfigure}{0.5\textwidth}
% \caption{Esatern cooperative oncology group at baseline}
% \label{fig:kp_ECOG}
% \includegraphics[width = \textwidth]{Figures/Amgen/B_ECOGCD_KP.pdf}
% \end{subfigure}\hfill
\begin{subfigure}{0.5\textwidth}
\caption{KRAS = wild-type}
\label{supfig:kp_KRAS_wildtype}
\includegraphics[width = \textwidth]{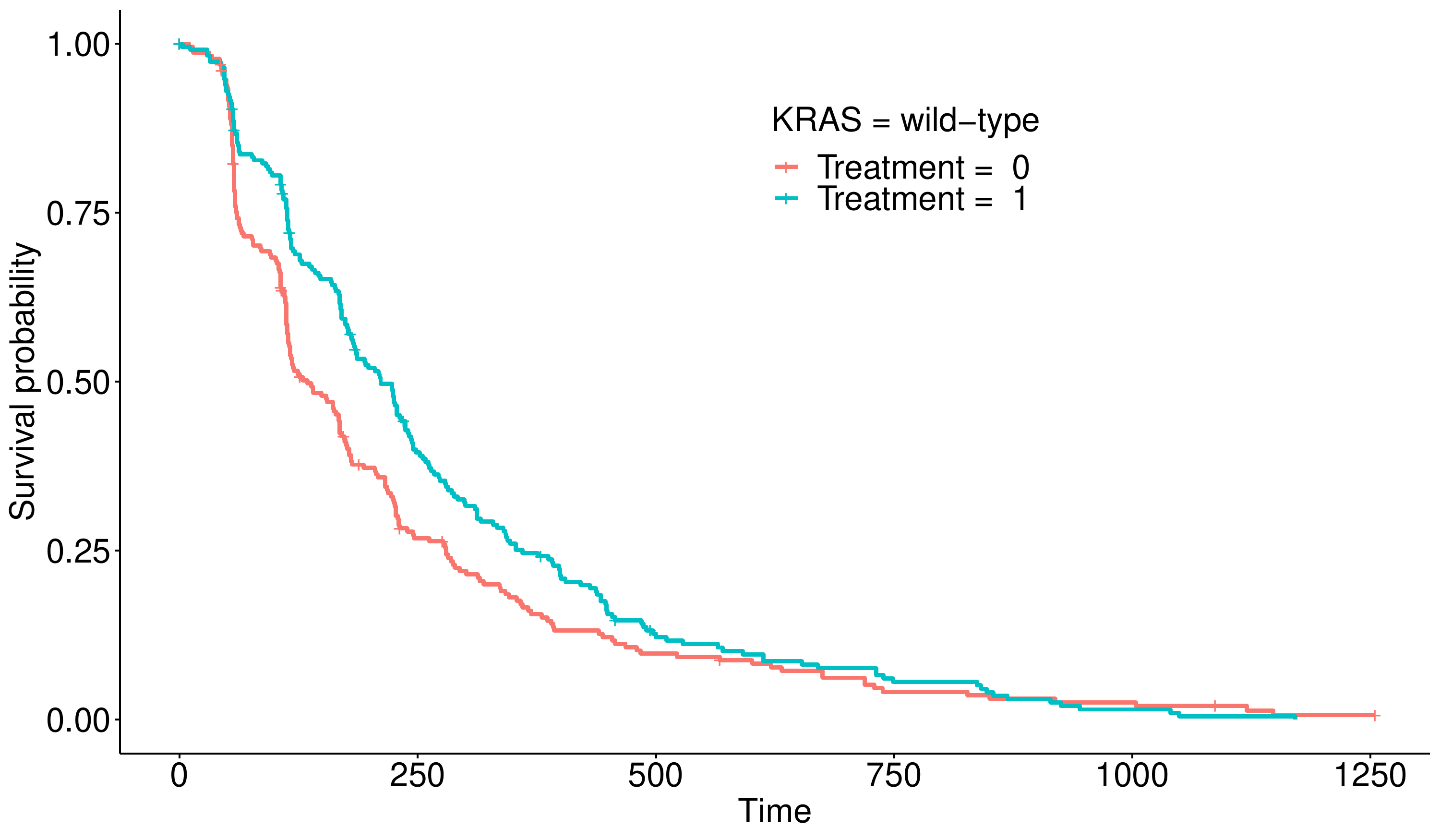}
\end{subfigure}\hfill
% \begin{subfigure}{0.5\textwidth}
% \caption{Prim tumor type}
% \label{fig:kp_tumor}
% \includegraphics[width = \textwidth]{Figures/Amgen/TUMCATCD_KP.pdf}
% \end{subfigure}\hfill
\begin{subfigure}{0.5\textwidth}
\caption{KRAS = mutant}
\label{supfig:kp_KRAS_mutant}
\includegraphics[width = \textwidth]{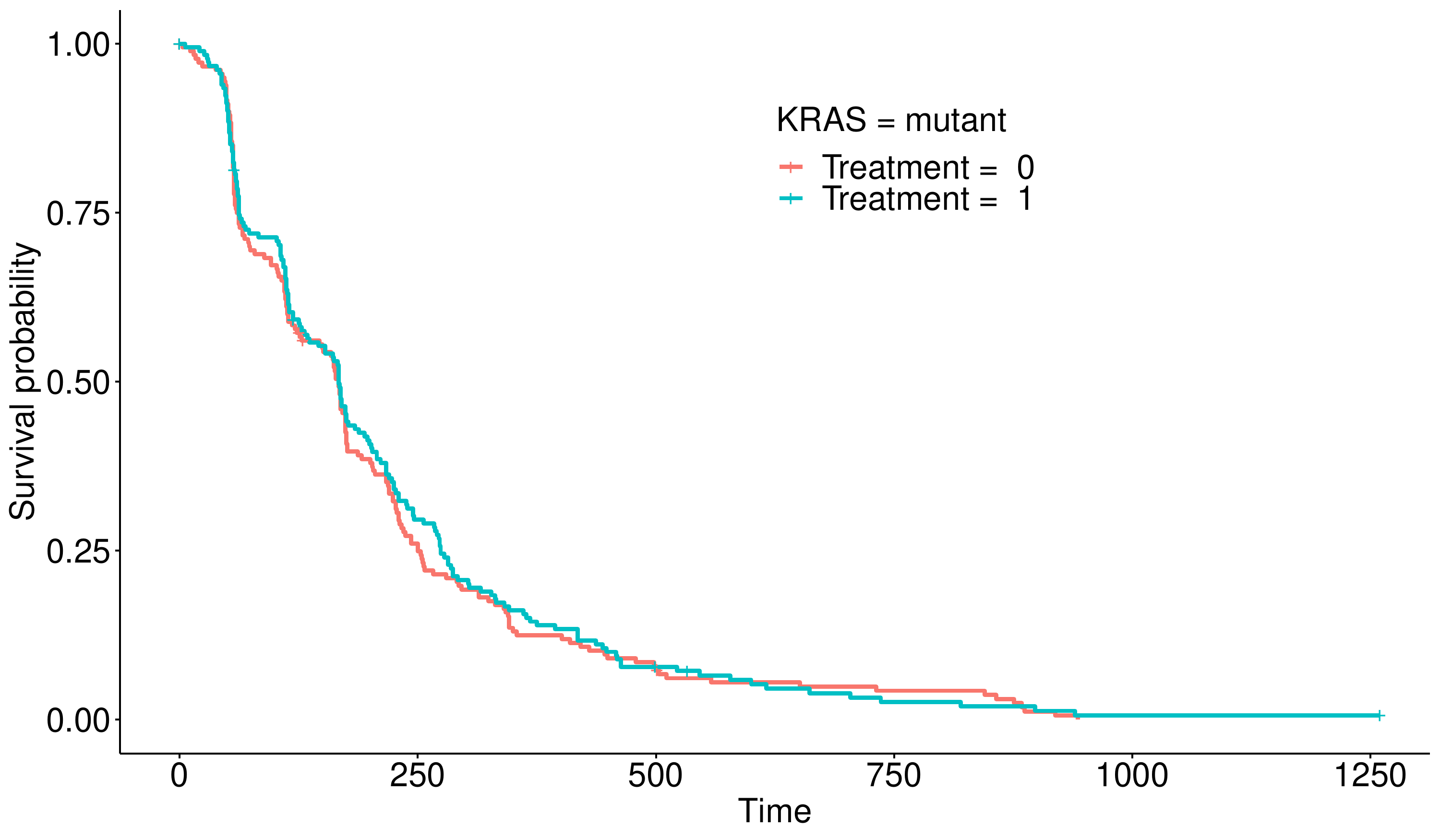}
\end{subfigure}
\end{figure}

\newpage

\clearpage

\end{document}